\documentclass[twocolumn]{aastex631}

\begin{document}

\title{CEERS Key Paper IV: A triality on the nature of HST-dark galaxies}


\author[0000-0003-4528-5639]{Pablo G. P\'erez-Gonz\'alez}
\affiliation{Centro de Astrobiolog\'{\i}a (CAB), CSIC-INTA, Ctra. de Ajalvir km 4, Torrej\'on de Ardoz, E-28850, Madrid, Spain}

\author[0000-0001-6813-875X]{Guillermo Barro}
\affiliation{Department of Physics, University of the Pacific, Stockton, CA 90340 USA}

\author[0000-0002-8053-8040]{Marianna Annunziatella}
\affiliation{Centro de Astrobiolog\'{\i}a (CAB), CSIC-INTA, Ctra. de Ajalvir km 4, Torrej\'on de Ardoz, E-28850, Madrid, Spain}

\author[0000-0001-6820-0015]{Luca Costantin}
\affiliation{Centro de Astrobiolog\'{\i}a (CAB), CSIC-INTA, Ctra. de Ajalvir km 4, Torrej\'on de Ardoz, E-28850, Madrid, Spain}

\author[0000-0002-8365-5525]{\'Angela Garc\'ia-Argum\'anez}
\affiliation{Departamento de Física de la Tierra y Astrofísica, Facultad de CC Físicas, Universidad Complutense de Madrid, E-28040, Madrid, Spain}
\affiliation{Instituto de Física de Partículas y del Cosmos IPARCOS, Facultad de CC Físicas, Universidad Complutense de Madrid, 28040 Madrid, Spain}

\author[0000-0001-8688-2443]{Elizabeth J.\ McGrath}
\affiliation{Department of Physics and Astronomy, Colby College, Waterville, ME 04901, USA}

\author[0000-0001-8115-5845]{Rosa M. M\'erida}
\affiliation{Centro de Astrobiolog\'{\i}a (CAB), CSIC-INTA, Ctra. de Ajalvir km 4, Torrej\'on de Ardoz, E-28850, Madrid, Spain}

\author[0000-0002-0786-7307]{Jorge A. Zavala}
\affiliation{National Astronomical Observatory of Japan, 2-21-1 Osawa, Mitaka, Tokyo 181-8588, Japan}

\author[0000-0002-7959-8783]{Pablo Arrabal Haro}
\affiliation{NSF's National Optical-IR Astronomy Research Laboratory, 950 N. Cherry Ave., Tucson, AZ 85719, USA}

\author[0000-0002-9921-9218]{Micaela B. Bagley}
\affiliation{Department of Astronomy, The University of Texas at Austin, Austin, TX, USA}

\author[0000-0001-8534-7502]{Bren E. Backhaus}
\affiliation{Department of Physics, 196 Auditorium Road, Unit 3046, University of Connecticut, Storrs, CT 06269}

\author[0000-0002-2517-6446]{Peter Behroozi}
\affiliation{Department of Astronomy and Steward Observatory, University of Arizona, Tucson, AZ 85721, USA}
\affiliation{Division of Science, National Astronomical Observatory of Japan, 2-21-1 Osawa, Mitaka, Tokyo 181-8588, Japan}

\author[0000-0002-5564-9873]{Eric F.\ Bell}
\affiliation{Department of Astronomy, University of Michigan, 1085 S. University Ave, Ann Arbor, MI 48109-1107, USA}

\author[0000-0003-0492-4924]{Laura Bisigello}
\affiliation{Dipartimento di Fisica e Astronomia "G.Galilei", Universit\'a di Padova, Via Marzolo 8, I-35131 Padova, Italy}
\affiliation{INAF--Osservatorio Astronomico di Padova, Vicolo dell'Osservatorio 5, I-35122, Padova, Italy}

\author[0000-0003-3441-903X]{V\'eronique Buat}
\affiliation{Aix Marseille Univ, CNRS, CNES, LAM Marseille, France}

\author[0000-0003-2536-1614]{Antonello Calabr{\`o}} 
\affiliation{Osservatorio Astronomico di Roma, via Frascati 33, Monte Porzio Catone, Italy}

\author[0000-0002-0930-6466]{Caitlin M. Casey}
\affiliation{Department of Astronomy, The University of Texas at Austin, Austin, TX, USA}

\author[0000-0001-7151-009X]{Nikko J. Cleri}
\affiliation{Department of Physics and Astronomy, Texas A\&M University, College Station, TX, 77843-4242 USA}
\affiliation{George P.\ and Cynthia Woods Mitchell Institute for Fundamental Physics and Astronomy, Texas A\&M University, College Station, TX, 77843-4242 USA}

\author[0000-0002-4343-0479]{Rosemary T. Coogan}
\affiliation{Universit{\'e} Paris-Saclay, Université Paris Cit{\'e}, CEA, CNRS, AIM, 91191, Gif-sur-Yvette, France}

\author[0000-0003-1371-6019]{M. C. Cooper}
\affiliation{Department of Physics \& Astronomy, University of California, Irvine, 4129 Reines Hall, Irvine, CA 92697, USA}

\author[0000-0002-3892-0190]{Asantha R. Cooray}
\affiliation{Department of Physics \& Astronomy, University of California, Irvine, 4129 Reines Hall, Irvine, CA 92697, USA}

\author[0000-0003-4174-0374]{Avishai Dekel}
\affil{Racah Institute of Physics, The Hebrew University of Jerusalem, Jerusalem 91904, Israel}

\author[0000-0001-5414-5131]{Mark Dickinson}
\affiliation{NSF's National Optical-IR Astronomy Research Laboratory, 950 N. Cherry Ave., Tucson, AZ 85719, USA}

\author[0000-0002-7631-647X]{David Elbaz}
\affil{Universit{\'e} Paris-Saclay, Université Paris Cit{\'e}, CEA, CNRS, AIM, 91191, Gif-sur-Yvette, France}

\author[0000-0001-7113-2738]{Henry C. Ferguson}
\affiliation{Space Telescope Science Institute, Baltimore, MD, USA}

\author[0000-0001-8519-1130]{Steven L. Finkelstein}
\affiliation{Department of Astronomy, The University of Texas at Austin, Austin, TX, USA}

\author[0000-0003-3820-2823]{Adriano Fontana}
\affiliation{INAF - Osservatorio Astronomico di Roma, via di Frascati 33, 00078 Monte Porzio Catone, Italy}

\author[0000-0002-3560-8599]{Maximilien Franco}
\affiliation{Department of Astronomy, The University of Texas at Austin, Austin, TX, USA}

\author[0000-0003-2098-9568]{Jonathan P. Gardner}
\affiliation{Astrophysics Science Division, NASA Goddard Space Flight Center, 8800 Greenbelt Rd, Greenbelt, MD 20771, USA}

\author[0000-0002-7831-8751]{Mauro Giavalisco}
\affiliation{University of Massachusetts Amherst, 710 North Pleasant Street, Amherst, MA 01003-9305, USA}

\author[0000-0002-4085-9165]{Carlos G{\'o}mez-Guijarro}
\affil{Universit{\'e} Paris-Saclay, Université Paris Cit{\'e}, CEA, CNRS, AIM, 91191, Gif-sur-Yvette, France}

\author[0000-0002-5688-0663]{Andrea Grazian}
\affiliation{INAF--Osservatorio Astronomico di Padova, Vicolo dell'Osservatorio 5, I-35122, Padova, Italy}

\author[0000-0001-9440-8872]{Norman A. Grogin}
\affiliation{Space Telescope Science Institute, Baltimore, MD, USA}

\author[0000-0002-4162-6523]{Yuchen Guo}
\affiliation{Department of Astronomy, The University of Texas at Austin, Austin, TX, USA}

\author[0000-0002-1416-8483]{Marc Huertas-Company}
\affil{Instituto de Astrof\'isica de Canarias, La Laguna, Tenerife, Spain}
\affil{Universidad de la Laguna, La Laguna, Tenerife, Spain}
\affil{Universit\'e Paris-Cit\'e, LERMA - Observatoire de Paris, PSL, Paris, France}

\author[0000-0002-1590-0568]{Shardha Jogee}
\affiliation{Department of Astronomy, The University of Texas at Austin, Austin, TX, USA}

\author[0000-0001-9187-3605]{Jeyhan S. Kartaltepe}
\affiliation{Laboratory for Multiwavelength Astrophysics, School of Physics and Astronomy, Rochester Institute of Technology, 84 Lomb Memorial Drive, Rochester, NY 14623, USA}

\author[0000-0001-8152-3943]{Lisa J. Kewley}
\affiliation{Center for Astrophysics, Harvard \& Smithsonian, 60 Garden Street, Cambridge, MA 02138, USA}

\author[0000-0002-5537-8110]{Allison Kirkpatrick}
\affiliation{Department of Physics and Astronomy, University of Kansas, Lawrence, KS 66045, USA}

\author[0000-0002-8360-3880]{Dale D. Kocevski}
\affiliation{Department of Physics and Astronomy, Colby College, Waterville, ME 04901, USA}

\author[0000-0002-6610-2048]{Anton M. Koekemoer}
\affiliation{Space Telescope Science Institute, 3700 San Martin Dr., Baltimore, MD 21218, USA}

\author[0000-0002-7530-8857]{Arianna S. Long}
\altaffiliation{NASA Hubble Fellow}
\affiliation{Department of Astronomy, The University of Texas at Austin, Austin, TX, USA}

\author[0000-0003-3130-5643]{Jennifer M. Lotz}
\affiliation{Gemini Observatory/NSF's National Optical-IR Astronomy Research Laboratory, 950 N. Cherry Ave., Tucson, AZ 85719, USA}

\author[0000-0003-1581-7825]{Ray A. Lucas}
\affiliation{Space Telescope Science Institute, 3700 San Martin Drive, Baltimore, MD 21218, USA}

\author[0000-0001-7503-8482]{Casey Papovich}
\affiliation{Department of Physics and Astronomy, Texas A\&M University, College Station, TX, 77843-4242 USA}
\affiliation{George P.\ and Cynthia Woods Mitchell Institute for Fundamental Physics and Astronomy, Texas A\&M University, College Station, TX, 77843-4242 USA}


\author[0000-0003-3382-5941]{Nor Pirzkal}
\affiliation{ESA/AURA Space Telescope Science Institute}

\author[0000-0002-5269-6527]{Swara Ravindranath}
\affiliation{Space Telescope Science Institute, 3700 San Martin Drive, Baltimore, MD 21218, USA}

\author[0000-0002-6748-6821]{Rachel S. Somerville}
\affiliation{Center for Computational Astrophysics, Flatiron Institute, 162 5th Avenue, New York, NY, 10010, USA}

\author[0000-0002-8224-4505]{Sandro Tacchella}
\affiliation{Kavli Institute for Cosmology, University of Cambridge, Madingley Road, Cambridge, CB3 0HA, UK}\affiliation{Cavendish Laboratory, University of Cambridge, 19 JJ Thomson Avenue, Cambridge, CB3 0HE, UK}

\author[0000-0002-1410-0470]{Jonathan R. Trump}
\affiliation{Department of Physics, 196 Auditorium Road, Unit 3046, University of Connecticut, Storrs, CT 06269, USA}

\author[0000-0002-9593-8274]{Weichen Wang}
\affiliation{Department of Physics and Astronomy, Johns Hopkins University, 3400 N. Charles Street, Baltimore, MD 21218, USA}

\author[0000-0003-3903-6935]{Stephen M.~Wilkins} %
\affiliation{Astronomy Centre, University of Sussex, Falmer, Brighton BN1 9QH, UK}
\affiliation{Institute of Space Sciences and Astronomy, University of Malta, Msida MSD 2080, Malta}

\author[0000-0003-3735-1931]{Stijn Wuyts}
\affiliation{Department of Physics, University of Bath, Claverton Down, Bath BA2 7AY, UK}

\author[0000-0001-8835-7722]{Guang Yang}
\affiliation{Kapteyn Astronomical Institute, University of Groningen, P.O. Box 800, 9700 AV Groningen, The Netherlands}
\affiliation{SRON Netherlands Institute for Space Research, Postbus 800, 9700 AV Groningen, The Netherlands}

\author[0000-0003-3466-035X]{L. Y. Aaron\ Yung}
\altaffiliation{NASA Postdoctoral Fellow}
\affiliation{Astrophysics Science Division, NASA Goddard Space Flight Center, 8800 Greenbelt Rd, Greenbelt, MD 20771, USA}


\begin{abstract}
The new capabilities that JWST offers in the near- and mid-infrared (IR) are used to investigate in unprecedented detail the nature of optical/near-IR faint, mid-IR bright sources, HST-dark galaxies among them. We gather JWST data from the CEERS survey in the Extended Groth Strip, jointly with HST data, and analyze spatially resolved optical-to-mid-IR spectral energy distributions (SEDs) to estimate photometric redshifts in two dimensions and stellar populations properties in a pixel-by-pixel basis for red galaxies detected by NIRCam. We select 138 galaxies with $F150W-F356W>1.5$~mag and $F356W<27.5$~mag. The nature of these sources is threefold: (1) 71\% are dusty star-forming galaxies (SFGs) at $2<z<6$ with $9<\log\mathrm{M\!_\star/M}_{\odot}<11$ and a variety of specific SFRs ($<1$ to $>100$~Gyr$^{-1}$); (2) 18\% are quiescent/dormant (i.e., subject to reignition/rejuvenation) galaxies (QG) at $3<z<5$, with $\log\mathrm{M\!_\star/M}_{\odot}\sim10$ and post-starburst mass-weighted ages (0.5-1.0 Gyr); (3) 11\% are strong young starbursts with indications of high-EW emission lines (typically, [OIII]+H$\beta$) at $6<z<7$ (XELG-$z6$) and  $\log\mathrm{M\!_\star/M}_{\odot}\sim9.5$. The sample is dominated by disk-like  galaxies with a remarkable compactness for XELG-$z6$ (effective radii smaller than 0.4 kpc). Large attenuations in SFGs, $2<\mathrm{A(V)}<5$ mag, are found within 1.5 times the effective radius, approximately 2 kpc, while QGs present $\mathrm{A(V)}\sim0.2$~mag. Our SED-fitting technique reproduces the expected dust emission luminosities of IR-bright and sub-millimeter galaxies. This study implies high levels of star formation activity between $z\sim20$ and $z\sim10$, where virtually 100\% of our galaxies had already formed $10^8$ M$_{\odot}$, 60\% had assembled $10^9$ M$_{\odot}$, and 10\% $10^{10}$ M$_{\odot}$ ({\it in situ} or {\it ex situ}).
\end{abstract}


\section{Introduction} \label{sec:intro}

Every time a new space observatory is launched with improved capabilities at longer and longer wavelengths, the Universe reveals the existence of surprisingly massive and/or highly metal-enriched and/or dusty galaxies at higher and higher redshifts, where some of us naïve astronomers are surprised to discover such evolved systems. This was the case for IRAS \citep[e.g.,][]{1996ARA&A..34..749S}, ISO \citep[e.g.,][]{2001ApJ...556..562C}, {\it Spitzer} \citep[e.g.,][]{2005ApJ...630...82P,2006ApJ...644..792R}, {\it Herschel} \citep[e.g.,][]{2011A&A...533A.119E,2013MNRAS.432...23G}, but also from ground infrared observatories \citep[e.g.,][]{1988ApJ...331L..77E,2003ApJ...587L..79F,2004ApJ...617..746D,2005ApJ...622..772C,2008ApJ...673L.127D}.

One of the last episodes of this recursive story happened with the comparison of data coming from the most powerful mid-infrared (mid-IR) space telescope existing before 2022, {\it Spitzer}, and the most advanced optical space telescope to date, {\it Hubble}. The combination of their capabilities revealed a population of red galaxies, nearly undetected or even escaping the sensitivity limits in the near-infrared (near-IR) provided by the WFC3 instrument onboard HST (and even dimmer in the optical), but relatively bright in the mid-IR, well detected by the IRAC camera onboard {\it Spitzer} \citep{2011ApJ...742L..13H,2012ApJ...750L..20C,2016ApJ...816...84W,2019ApJ...876..135A,2021ApJ...922..114S,2022arXiv220707125K}.

Among the population of mid-IR bright, near-IR/optically faint galaxies, some are missed by HST due to depth limitations and they are thus called {\it HST-dark} sources, although the name is misleading due to the variety of depths of HST/WFC3 observations on the sky. The nature of most  mid-IR selected systems, and especially HST-dark sources, is still uncertain. This is mainly because their stellar emission could only be detected robustly by IRAC (and their dust content by ALMA, NOEMA, or SCUBA, see references below), their spectral energy distributions are thus poorly constrained with just a few flux data points, and the vast majority of them are too faint for spectrographs (operating at any wavelength from the UV to the sub-millimeter range; but see \citealt{2012MNRAS.427.1066S}, \citealt{2019ApJ...884..154W}, and \citealt{2020A&A...642A.155Z}) to analyze their nature in detail. 

However, the discovery of these intrinsically red sources directly point to an elusive high redshift (probably $z>2-3$) population of either massive galaxies experiencing a very dusty star-forming event or quiescently evolving from very early cosmic times. This is supported by the large fraction of detections by {\it Spitzer}/MIPS, {\it Herschel}/PACS and SPIRE \citep{
2011ApJ...742L..13H,2012ApJ...750L..20C,2019ApJ...876..135A}, and at sub-millimeter and radio wavelengths  
\citep{2014ApJ...788..125S,2018A&A...620A.152F,2019ApJ...878...73Y,2019ApJ...884..154W,2021A&A...648A...8W,2021ApJ...909...23T,2022ApJ...925...23M,2022arXiv221003135X}. In addition, a minor fraction of HST-dark galaxies could be linked to emission-line galaxies presenting high equivalent widths and thus being prominent in the mid-IR when the lines enter the IRAC bands (cf. Figure~12 in \citealt{2019ApJ...876..135A} and references therein). 

The cosmological importance of HST-dark galaxies to improve our knowledge about galaxy evolution is not negligible. The existence of a numerous population of massive galaxies already in place or even evolving passively at $z>2-3$, and even up to $z\sim6$ (and beyond, see \citealt{2022ApJ...927..170T}, \citealt{2022arXiv220714733B}, \citealt{2022arXiv220712446L}, \citealt{2022arXiv220801630N}, \citealt{2022arXiv220814999E},) is very difficult to reproduce by state-of-the-art galaxy evolution models (see discussion in \citealt{2019ApJ...876..135A}, and recent discussion on JWST results in \citealt{2022arXiv220810479L}), which also suffer from their limitations in area to understand the nature of these samples presenting relatively small volume densities. 

In this paper we want to benefit from the huge jump offered by the brand new JWST and its near- and mid-infrared unique capabilities to understand the nature of this interesting  population of galaxies whose nature and relevance is still debatable, HST-dark galaxies among them, near-infrared faint, mid-infrared sources in general.

Indeed, with the advent of JWST, new mid-IR bright, near-IR faint galaxies have been identified and started to be characterized \citep{2022arXiv220714733B,2022arXiv220800986C,2022arXiv220712446L,2022arXiv220801630N}. These works have confirmed the high-redshift nature of HST-dark galaxies, their relevance for compiling a complete stellar mass census in the first 1-2~Gyr of the lifetime of the Universe, and their challenging numbers and properties for galaxy evolution models. Disagreement with theoretical predictions has also been (preliminary) found with the large numbers of very high-redshift ($z>10$) galaxy candidates being found in relatively shallow JWST/NIRCam data \citep[e.g.,][]{2022arXiv220709436C,2022arXiv220712356D,2022arXiv220712474F,2022arXiv220801612H,2022arXiv220709434N,2022arXiv220813582O,2022arXiv220802825R,2022arXiv220711217A}, implying high levels of star formation activity in the very early Universe (even though these candidates must still be confirmed and separated from contaminants such as dusty galaxies at $2<z<5$, see \citealt{2022arXiv220801816Z}).

In this paper we take advantage of the new JWST data for studying galaxies that are red in the near- to mid-IR color (some of them HST-dark), benefiting from 2 different aspects. First, we  exploit the deeper flux limits provided by JWST and extend the selection of red mid-IR detected galaxies carried out by previously published works, mainly using {\it Spitzer}/IRAC and HST/WFC3 data. And second, we also exploit the exquisite spatial resolution provided by JWST and HST in a unprecedentedly wide spectral range covering from the observed optical wavelengths (around 0.6~$\mu$m) to the observed mid-IR ($\sim4.4$~$\mu$m). We remark that this spectral range has been used extensively to study the integrated light of galaxies at cosmological distances up to $z\sim10$, and we are now able to use the same type of spectral energy distributions (SEDs) but with much better spatial resolution in the entire (and very common in the last 20 years since {\it Spitzer} was launched) optical-to-mid-IR spectral window thanks to JWST.

With this methodology, our goals are twofold. First, we aim to robustly characterize the redshift, stellar mass and evolutionary stage of HST-dark galaxies, and  mid-to-near-IR red galaxies in general. Within this goal, we are especially interested in identifying the first massive galaxies ever formed and, particularly, evolved galaxies that quenched 0.5-1.0~Gyr after a starburst. Such systems are classically called \textit{quiescent} galaxies. We note that rather than the term {\it quiescent}, it may be more adequate at early epochs to talk about {\it dormant} galaxies, since the Universe itself is not much older than those ages, there is no time to be quiescent (in the same way used to talk about other galaxy populations at lower redshifts, such as {\it red nuggets}; \citealt{2020ApJ...889L...3C},
\citealt{2020A&A...638L..11T},
\citealt{2021IAUS..359..431F},
\citealt{2021IAUS..359..441L},
\citealt{2022arXiv220804601L}) but these early galaxies may reignite and rejuvenate in their future evolution \citep{2008ApJ...687...50P,2014ApJ...796...35F,2022arXiv221003832W,2022ApJ...926..134T}.

Our second goal is to benefit from the $8\times$ jump in spatial resolution provided by JWST in the mid-IR (with respect to {\it Spitzer}), and $\times 1-2$ in the near-infrared (with respect to HST). With those enhanced capabilities, we will study the 2-dimensional distribution of the stellar populations in HST-dark galaxies, searching for the location of dusty starbursts, clues into their mass-growth and structural assembly channels, and, more general, information about the first episodes in the star formation history of massive galaxies occurring at the earliest cosmic epochs.

This paper is organized as follows. Section~\ref{sec:data} presents the observations used in our study. Section~\ref{sec:selection} discusses the selection of near-IR faint, mid-IR bright galaxies and the expectations for their nature. In Section~\ref{sec:photometry}, we describe our photometric measurements. Next, Sections~\ref{sec:photoz} presents the methodology to analyze our sample using spatially resolved spectral energy distributions to obtain photometric redshifts and stellar population properties. Section~\ref{sec:analysis} presents our results about the integrated properties of our sample as well as the radial gradients of the stellar populations they harbor and the morphologies of our galaxies. Finally, Section~\ref{sec:summary} summarizes our results and conclusions.

Throughout the paper, we assume a flat cosmology with $\mathrm{\Omega_M\, =\, 0.3,\, \Omega_{\Lambda}\, =\, 0.7}$, and a Hubble constant $\mathrm{H_0\, =\, 70\, km\,s^{-1} Mpc^{-1}}$. We use AB magnitudes \citep{1983ApJ...266..713O}. All stellar mass and SFR estimations assume a universal \citet{2003PASP..115..763C} IMF.

\section{Data}\label{sec:data}

In this paper, we use the JWST imaging data acquired within the Cosmic Evolution Early Released Science (CEERS) project \citep{2017jwst.prop.1345F} in June 2022. This dataset comprises 4 NIRCam pointings (1, 2, 3, and 6, as named in the APT observing strategy file) in the Extended Groth Strip (EGS, \citealt{2007ApJ...660L...1D}), observed in filters $F115W$, $F150W$, $F200W$, $F277W$, $F356W$, $F410M$, and $F444W$. The total surveyed area is 38.8~arcmin$^2$. The 5$\sigma$ depths in each filter for point-like sources using 0.15\arcsec\, circular apertures are 29.24, 29.08, 29.23, 30.08, 30.13, 29.74, 29.50 mag, respectively, 28.96, 28.84, 29.02, 29.74, 29.77, 29.34, 29.09 mag after applying aperture corrections. The 5$\sigma$ limits for individual pixels (important for our 2-dimensional study of the stellar populations in high redshift galaxies) are 31.61, 31.45, 31.61, 32.45, 32.50, 32.11, 31.85 mag for the filters mentioned above, respectively.

The NIRCam data were calibrated with the {\it jwst} pipeline version 1.7.2, reference files in pmap version 0989 (which includes detector-to-detector matched, improved absolute photometric calibration; see Bagley et al., in prep.). Apart from the standard pipeline stages, we performed careful alignment of the 3 images available for each pointing and filter before stacking them. We also applied a background homogenization algorithm prior to obtaining the final mosaics. The whole dataset was registered to the same World Coordinate System reference frame (based on Gaia DR1.2; \citealt{2016A&A...595A...1G}, \citealt{2016A&A...595A...2G}) and put to the same pixel scale, namely, 0.03 arcsec/pixel.

Complementary to the JWST observations, we also used the Hubble Space Telescope HST data provided by the AEGIS \citep{2007ApJ...660L...1D} and CANDELS  \citep{2011ApJS..197...35G,2011ApJS..197...36K} collaborations in the EGS region. The data were reduced again and drizzled to the same pixel size of the JWST observations, 0.03 arcsec/pixel, by the CEERS team (version v1.9, see Bagley et al., in prep.). We added to our analysis those bands with better spatial resolution than the worst obtained by JWST, in filter $F444W$ (see Section~\ref{sec:photometry}), namely $F606W$ and $F814W$ from the Advanced Camera for Surveys, i.e., we exclude WFC3 observations from our analysis. Depths for these images are 29.29 and 29.05~mag, respectively, for point-like sources in a 0.15\arcsec\, circular aperture, accounting for negligible aperture corrections. The 5$\sigma$ limits for individual pixels are 31.66 and 31.42 mag, respectively.

We cross-correlated our sample with the IR catalogs for {\it Spitzer} MIPS, and {\it Herschel} PACS and SPIRE presented in \citet{2019ApJS..243...22B}, constructed with a deconvolution methodology described in \citet[][see also \citealt{2016MNRAS.459.1626R}, \citealt{2019MNRAS.485..586R}]{2010A&A...518L..15P}. The 5$\sigma$ depths for these data are 45~$\mu$Jy and 3.5~mJy for MIPS 24~$\mu$m and 70~$\mu$m, 8.7 and 13.1~mJy for PACS 100 and 160~$\mu$m, and 14.7, 17.3, and 17.9~mJy for SPIRE 250, 350, and 500~$\mu$m.

At longer wavelengths, we also used the sub-millimeter catalog presented in \citet{2017MNRAS.464.3369Z,2018MNRAS.475.5585Z}, which reaches a 5$\sigma$ central depth of 0.2~mJy at 850~$\mu m$ with SCUBA-2 as well as our NOEMA 1.1~mm follow-up project on a sub-sample of these SCUBA-2 galaxies (proposal W20CK; see also \citealt{2022arXiv220801816Z} and Ciesla et al. in prep.). To search for radio galaxies, we used the VLA/EGS 20 cm (1.4 GHz) survey described by \citet{2007ApJ...660L..77I}. In addition, we searched for counterparts for our galaxies in the X-ray catalog presented in \citet{2015ApJS..220...10N}.

Finally, we used the spectroscopic redshifts compiled by \citet{2017ApJS..229...32S}, adding new sources published by the MOSFIRE Deep Evolution Field (MOSDEF) team \citep{2015ApJS..218...15K,2015ApJ...806..259R} in their 2021 data release. The spectroscopic sample in the 4 NIRCam fields include 466 galaxies at a median and quartiles redshift $z=0.81_{0.55}^{1.41}$, with 17\% at $z>2$, all below $z=3.7$.

\section{Sample selection}
\label{sec:selection}

In this section, we describe how using a $F150W-F356W$ vs$.$ $F356W$ color-magnitude diagram we select a sample of red galaxies in the near-to-mid-infrared color, HST-dark systems among them. To motivate the discussion in the following sections, we also describe in some detail the redshifts and types of galaxies expected for a selection based on different thresholds in color-magnitude.

\subsection{Pre-JWST searches for mid-IR bright, optical/near-IR faint galaxies}
\label{sec:color_selection}

Previous works have shown that searching for massive, dusty and
quiescent galaxies at $z\gtrsim3$ using color-magnitude diagrams is a
very effective technique
\citep{2009A&A...500..705M,2011ApJ...742L..13H,2012ApJ...750L..20C,2013ApJ...777L..19L,2014ApJ...794...68N,2016ApJ...816...84W,2019ApJ...876..135A,2015ApJ...803...11S,2019ApJ...876..135A}. Briefly, these methods use a long baseline optical/NIR
color to identify galaxies with very red SEDs, which would be caused
either by strong Balmer breaks/D4000 spectral features or large,
dust-driven UV attenuations, which are more common in evolved, massive
galaxies. Mid-to-near IR large colors might also be present in
galaxies with strong (high equivalent width) emission lines
\citep[see,
e.g.,][]{2014ApJ...784...58S,2015ApJ...801..122S,2016ApJ...823..143R,2019ApJ...876..135A}.
Prior to JWST, these methods typically relied on the HST/$F160W$
filter for the blue band and either {\it Spitzer}/IRAC $[3.6]$ or [4.5] channels for the red (and selection) band.  The resulting samples
find galaxies with relatively bright IRAC magnitudes ([3.6] \& or
[4.5] $\leq$ 24.5~ mag) that are often very faint, even undetected, in the $H$-band, implying
$F160W>26-27$~mag (depending on the observed field). Thus, they are commonly referred to as IRAC-bright,
HST-dark (or -drops) galaxies
\citep{2012ApJ...750L..20C,2015ApJ...803...11S,2019ApJ...876..135A}. 

The new and extremely deep near- and mid-IR first datasets from JWST reaches more than 2 magnitudes deeper in $F150W$ and $F356W$ than the previous HST and {\it Spitzer} photometry and, consequently, it has become easier to identify and characterize HST-dark galaxies both IR-bright and much fainter
\citep{2022arXiv220714733B,2022arXiv220814999E,2022arXiv220801630N,2022arXiv220712446L}.

Here we aim to study the properties of optical/near-IR faint galaxies selected with a similar method to those in previous papers but taking advantage of the new JWST capabilities. Therefore, we use the following thresholds
in color-magnitude : $F150W-F356W>1.5$~mag, $F356W<27.5$~mag. The red color is similar to the values used in previous works \citep[e.g.,][]{2019ApJ...876..135A,2016ApJ...816...84W}, while the much deeper limiting magnitude in $F356W$,
compared to IRAC depths, [3.6]$\sim 24.5$~mag, implies that we can potentially find massive galaxies 
($\log\mathrm{M\!_\star/M}_{\odot} >$ 11) up to $z=8$ without including too many lower redshift galaxies with $\log\mathrm{M\!_\star/M}_{\odot} <$
10, therefore keeping the overall sample focused on massive galaxies. 

In order to understand better its heterogeneous nature, we divide the color-selected sample into two subsets,
with magnitudes brighter and fainter than $F150W=26$~mag, labeled HST-faint and HST-dark, respectively. The latter are a proxy for HST/WFC3 dropouts  that were previously only detected in IRAC and/or IR/radio surveys, or completely unknown. 
The HST-faint sample, which consists mainly of HST/$F160W$ detected galaxies 
identified before in CANDELS and 3D-HST catalogs \citep{2014ApJS..214...24S,2017ApJS..229...32S}, 
will be used for completeness, reference and comparison to illustrate how the formal definition
of an HST/WFC3-dropout (which depends on the depth of the survey and changes from paper to paper)
affects the median redshifts and stellar population properties of an HST-dark sample (see next section). We do not consider here any galaxy brighter than $F150W<23$~mag, which are typically massive galaxies at $z\lesssim2$ and have been well-characterized by HST and ground-based telescopes, even with spectroscopy. 

\subsection{Understanding the selection of different galaxy types in the color-diagram}
\label{color_magnitude_diagram}

\begin{figure*}[htp!]
\centering
\epsscale{1.15}
\plotone{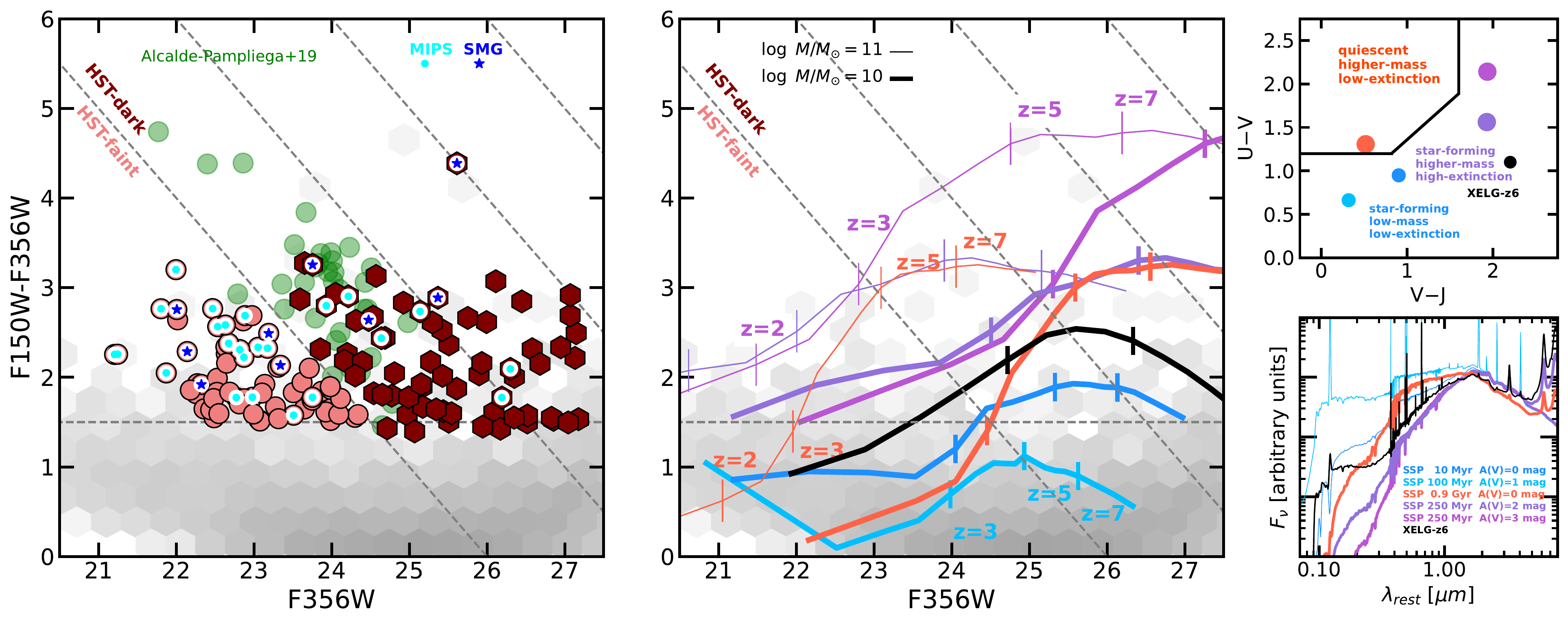}
\caption{The left panel shows a $F150W-F356W$ vs$.$ $F356W$ color-magnitude
diagram. The overall population of galaxies in the
CEERS sample are shown in a gray-scale density map (cut to $F356W<28$~mag). The markers show the color-selected, $F150W-F356W>1.5$~mag (horizontal dashed gray line) sample divided into HST-faint (light circles) and HST-dark (dark hexagons) galaxies with magnitudes
brighter and fainter than $F150W=26$~mag, respectively. Diagonal dashed gray lines at 26, 28, 30 and 32 mag and shown. Nearly
70\% of the HST-dark galaxies are undetected in previous HST/WFC3
based surveys ($\sim$90\% of those with $F150W>27$~mag). Cyan circles and blue stars indicate galaxies within our sample detected in the far-IR and radio/sub-mm wavelengths, and green circles show the sample of HST-dark galaxies from \citet{2019ApJ...876..135A}. The middle panel shows the same plot with colored
lines presenting the redshift dependence (marked with vertical lines) in color-magnitude for 6 different galaxy types with stellar masses $\log\mathrm{M\!_\star/M}_{\odot}=11$ and 10,
thin and thick lines, respectively.  The typical SEDs and rest-frame
$UVJ$ colors of each type are indicated in the bottom and top
panels on the right. The templates correspond to star-forming galaxies with low (blue tones) and high (magenta/purple) levels of obscuration, a quiescent galaxy template (red) and a extreme starburst spectrum with high equivalent width lines (black).}
\label{fig:color_magnitude_selection}
\end{figure*}

Figure~\ref{fig:color_magnitude_selection} shows our selection
method. The left panel depicts the $F150W-F356W$ vs$.$ $F356W$
color-magnitude plot for the galaxies in the CEERS sample using a gray-scale hexdiagram to indicate the regions with a higher number density. The selection of sources and integrated photometry used in this plot have been performed with a hot$+$cold Sextractor run \citep{1996A&AS..117..393B}, using \citet{1980ApJS...43..305K} apertures. The details of the catalog will be presented in Finkelstein et al. (2023, in prep.).

The color threshold used to select the sample is
indicated with a dashed grey line, and the resulting sample of
HST-faint and -dark galaxies is shown with circles and hexagons (light and dark colors, respectively). For comparison purposes, we show the sample of
HST-dark galaxies from \citet{2019ApJ...876..135A}, selected in
the CANDELS/GOODS region, with green circles. The overall distribution of
those galaxies agrees very well with the location of our HST-dark sample at relatively bright $F356W\lesssim25$~mag, except for the three bright and very red galaxies in the \citet{2019ApJ...876..135A} sample, which have no equivalent in our JWST-based selection.

To illustrate in more detail the types of galaxies and redshift ranges selected by the color-magnitude diagram, and also highlight the advantages and short-comings of this method, the left panel of Figure~\ref{fig:color_magnitude_selection} shows the tracks in color-magnitude of 6 different galaxy templates as a function of redshift. These galaxies are chosen to be (broadly) representative of the typical types in different regions of the $UVJ$ diagram \citep[top-right
panel;][]{2009ApJ...691.1879W,2010ApJ...719.1715W,2012ApJ...748L..27P,2019ApJ...880L...9L,2021ApJ...922L..32Z}, i.e., they range from blue,
low-mass and low-attenuation galaxies (bottom-left region of the $UVJ$ diagram; blue colors) to
redder, more massive, dusty star-forming galaxies (top-right; purple
colors) and quiescent galaxies (top-left; red/orange colors). The average SEDs for these galaxies are shown in the bottom-right panel of Figure~\ref{fig:color_magnitude_selection}. The SEDs are normalized at $\sim$1.6~$\mu$m rest-frame to highlight the reddening of the UV/optical emission with increasing age and/or dust-attenuation (from blue to purple templates). 

In Figure~\ref{fig:color_magnitude_selection}, the thin/thick line widths color-redshift
tracks show the effect of scaling the templates to low and high
stellar masses, $\log\mathrm{M\!_\star/M}_{\odot}=11$ and 10,
respectively, which are roughly traced by the $F356W$ magnitude (with
some degeneracies and scatter). The low mass/extinction templates are
shown only for $\log\mathrm{M\!_\star/M}_{\odot}=10$, since large dust contents typically exist at the high-mass end.

Overall, the color-redshift tracks illustrate that red colors
$F150W-F356W\gtrsim1.5$~mag and bright $F356W$ magnitudes provide an excellent proxy to identify massive, dusty star-forming galaxies at redshifts above $z\gtrsim2$, and massive, quiescent galaxies at slightly higher redshifts above $z\gtrsim3$. We note also that some of
the bluest quiescent galaxies ($U-V\sim1.3$~mag, orange color-track)
are still slightly bluer than $F150W-F356W=1.5$~mag at $z=3$, and were missed by color thresholds of previous HST$+${\it Spitzer} based papers. For this paper, considering composite stellar
population models with some additional residual recent star formation and not adding much mass (and thus not affecting the $F356W$ flux significantly), we use a color cut $F150W-F356W=1.5$~mag.

The color-redshift tracks also show that using a limiting magnitude down to $F356W\sim27$~mag includes many more galaxies at higher redshifts
and/or lower masses than previous IRAC-bright searches, which being
capped at $[3.6]\sim24.5$~mag (and in relatively large sky areas),
often picked the most massive galaxies
($\log\mathrm{M\!_\star/M}_{\odot} \sim$ 11) at redshifts of
$z\gtrsim3-4$ \citep{2016ApJ...816...84W,2019ApJ...876..135A}. Indeed, the $\log\mathrm{M\!_\star/M}_{\odot} = 10$ color-tracks (thick
lines) indicate that the selection would be roughly complete down to
that mass for quiescent galaxies up to $z\lesssim8$ (red/orange) and
even for the most dusty galaxies up to z$\sim$6
(purple/magenta). Similarly, at the high-mass end
($\log\mathrm{M\!_\star/M}_{\odot} >$ 11; thin lines), the
color-magnitude selection would identify all the quiescent or dusty
galaxies up to $z\sim6$. We note that the top-right corner of the figure, where we
would expect the most massive dusty galaxies at $z>5$ is nearly empty. This is likely
due to the relatively low number density of such sources, and very
massive galaxies in general (e.g., \citealt{2013ApJ...777...18M}; \citealt{2015ApJ...803...11S}) combined
with the small area surveyed by the current CEERS pointings
($\sim5\times$ smaller than for example the CANDELS/GOODS regions). We
further discuss these implications in section 6.2.

One potential concern for the mass-completeness would be the existence
of massive galaxies at $z>5-7$ with less canonical (composite) SEDs,
such us the ones identified by \citet{2022arXiv220712446L}. Those
galaxies have mirror-flipped-L-shaped SEDs (V-shaped in f$_{\lambda}$) with red colors in the rest-frame near-IR
($V-J\sim2$~mag) and blue colors in the rest-frame UV similar to those
of our blue templates (black tracks and marker in Figure~\ref{fig:color_magnitude_selection}). 
The strong UV-upturn and relatively red optical SED suggest that these are experiencing a 
recent burst of unobscured star formation but also have a pre-existing more evolved or dust-enshrouded stellar population.
However, the optical emission could be partially contaminated by the presence of high equivalent-width emission 
lines, which makes the estimate of the stellar mass more uncertain \citep{2022arXiv220814999E}.
The color-tracks for these galaxies decline very
quickly down to $F150W-F356W\sim1$~mag at $z\gtrsim5$ instead of
plateauing at around $F150W-F356W\sim2.5$~mag slightly under the quiescent and
dusty tracks. As we will demonstrate in the following sections, we have several of this type of galaxies in our sample thanks to our color cut at $F150W-F356W=1.5$~mag, but still some more active systems could exist and be missed by our selection. If those galaxies are common at the highest redshifts, they will be systematically missed by typical color-magnitude
selections unless we decrease the color cut (at the expense of sample
contamination by less extreme systems).

\subsection{Overall properties of the color-selected sample: HST-faint and -dark}

Our color-magnitude selected sample is composed by 138 galaxies, whose median/quartiles magnitude and colors are $\langle F356W\rangle=24.2^{25.6}_{23.0}$~mag and $\langle F150W-F356W\rangle=1.9^{2.3}_{1.6}$~mag. Half of the sample, 50\% (36\%),  69 (49) galaxies, are HST-dark, $F150W>26(27)$~mag, the rest would qualify as HST-faint. 

Among the HST-dark subsample, the fraction of sources included in HST catalogs of the EGS area \citep{2014ApJS..214...24S,2017ApJS..229...32S} is less than 28\% and all of these galaxies are brighter than
$F160W=27$~mag. This number goes up to 35\% if we include an IRAC selection \citep{2011ApJS..193...30B}. Among the HST-faint subsample, 85\% of galaxies was previously cataloged by HST-selected and/or IRAC-selected surveys. 

\subsection{Counterparts in far-IR, sub-millimeter and X-ray catalogs}
\label{sec:firmirxray}

Out of the 138 galaxies in our sample, 32 of them (23\%) can be identified with MIPS 24~$\mu m$ sources in the catalog presented in \citet{2019ApJS..243...22B}, searching in a 1.5\arcsec\, radius region, 75\% of them being in the HST-faint sample. One sixth of them has MIPS 70~$\mu m$ measurements, all with $SNR<2$. A visual inspection of the MIPS24 sources reveals that 11 of them (75\% in the HST-faint sample) are very bright and isolated MIPS emitters, with an average flux of 67~$\mu$Jy detected at the 9$\sigma$ level and average redshift $\langle 
z\rangle=2.8$. The rest of the counterparts, 21 galaxies (65\% in the HST-faint sample), are quite faint and/or possibly blended with nearby objects, presenting and average flux of 50~$\mu$Jy detected at the 7$\sigma$ level and average redshift $\langle z\rangle=3.2$.

{\it Herschel} detections (all above 4$\sigma$) by PACS at 100~$\mu m$ were found for 7 galaxies (5\% of the sample), average flux 3.1~mJy detected on average at the 7$\sigma$ level, with 2 of them in the HST-dark sample. For PACS 160~$\mu m$, we found 6 galaxies with $SNR>4$ measurements, average flux 7.5~mJy detected on average at the 7$\sigma$ level, 1 in the HST-dark sample. For SPIRE bands, we found 3 possible counterparts with $SNR>4$, average flux 13~mJy detected at, on average, 4$\sigma$ level, 1 in the HST-dark sample.

A total of 10 galaxies in our sample (7\% of the total) are likely associated with SCUBA-2 sources reported in \citet[][we note that this survey does not cover our whole area]{2017MNRAS.464.3369Z}. Nine of them are unequivocally associated with the dusty galaxies via 1.1~mm  high-angular resolution observations obtained with NOEMA (Zavala et al. 2022; Ciesla et al. in preparation), while the remaining four sources lie within the 1.5~arcsec of the 850~$\mu$m position.  An additional 3 sources  were found in the VLA 20~cm maps from \citet{2007ApJ...660L..77I}, within a 1.5~arcsec radius.

Finally, we also cross-correlated our sources with the X-ray catalog presented in \citet{2015ApJS..220...10N} finding 2 (3) association(s) within 0.5\arcsec\, (1.6\arcsec). 

The IDs, coordinates, and basic information about our sample are given in Table~\ref{tab:selection}.

\section{Photometry in 2 dimensions}
\label{sec:photometry}

\begin{figure*}
    \includegraphics[clip, trim=1.3cm 3.2cm 1.5cm 12.5cm,scale=0.99]{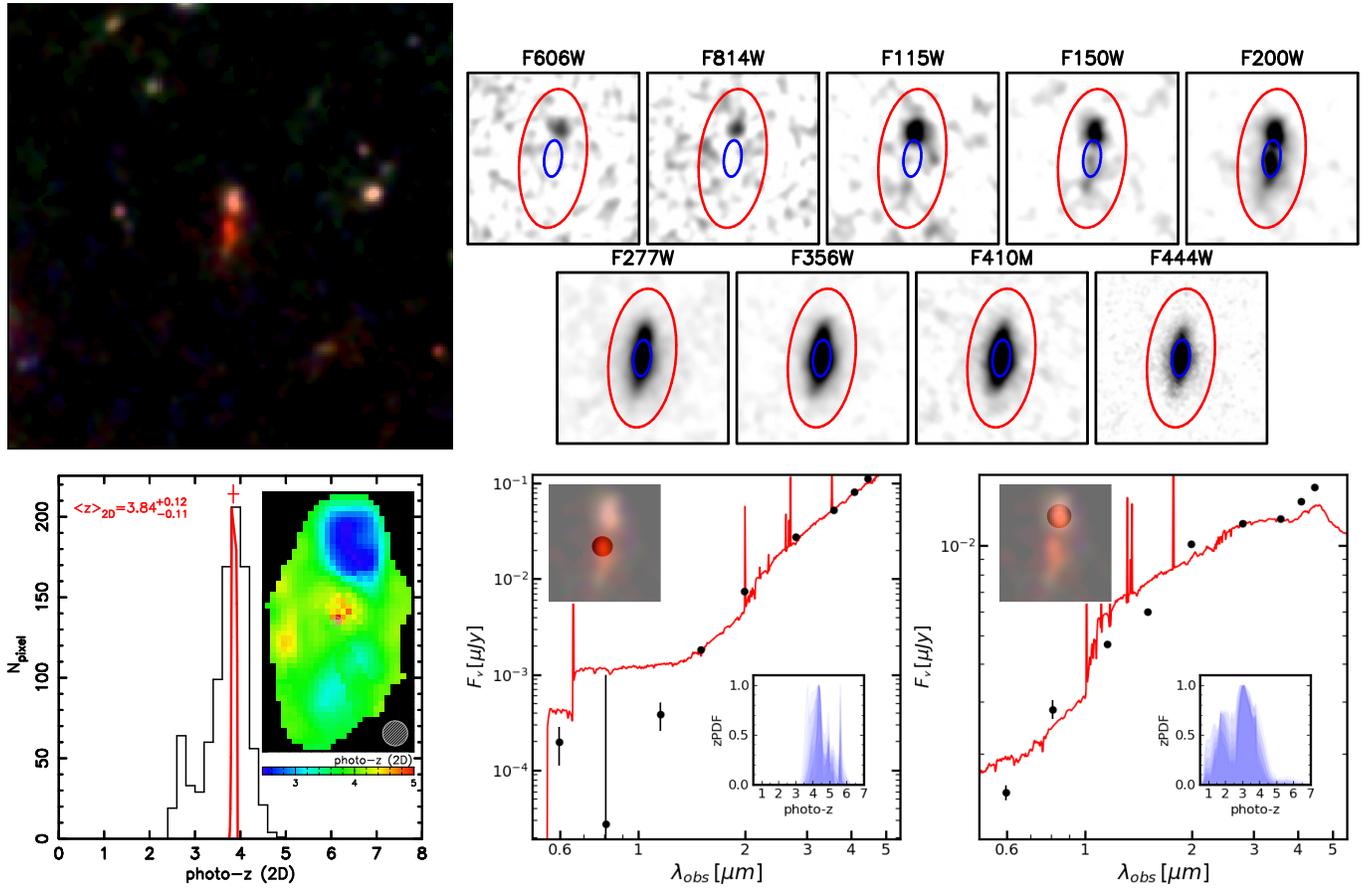}
    \caption{Example of a two-dimensional photometric redshift determination for one of our extended sources (identified as 2 galaxies at different redshifts, see text for more details). The upper left panel shows an RGB $8\arcsec\times8\arcsec$ postage stamp of nircam2-7122, using NIRCam filters $F115W$, $F150W$, and $F200W$. The top right thumbnails show individual images of the source in all HST/ACS and JWST/NIRCam filters used in this work. The images have been PSF-matched to the $F444W$ image. The red ellipse marks the aperture used for the integrated photometric measurements and the blue ellipse depicts the aperture containing half of the total flux in $F356W$. On the bottom-left, we show the histogram of most-probable photometric redshifts for all pixels with enough SNR (see text for details) within the integrated aperture. The combined redshift probability distribution function (zPDF) for all these pixels is shown in red, while the most probable redshift and its uncertainty for this source are marked and written in red. The inset shows the photometric redshift map. On the bottom-right, we show the SEDs of two circular apertures positioned on the central region of the source and the northern part, respectively (see circular regions marked in the RGB image insets). In red, we show the sum of all photometric redshift templates fitting the corresponding pixels (offsetted to the average redshift of the region for clarity). The bottom-right insets in these SED plots show the individual probability distribution functions for all pixels within the aperture, plotted with a transparency parameter.}
    \label{fig:2Dphotoz}
\end{figure*}

Taking advantage of the unprecedented spatial resolution combined with a wide wavelength coverage and depth provided by JWST and HST, in this paper we base our analysis in 2-dimensional spectral energy distributions (2D-SEDs) rather than the more common study of the integrated emission of galaxies. We identify 3 benefits from this approach:
\begin{itemize}
\item[(1)] The spectral analysis is similar to what has been extensively done before in the literature based on HST and {\it Spitzer}/IRAC data for complete galaxies, i.e., choosing a 2D approach does not imply losing photometric bands now that we have JWST data.
\item[(2)] Smaller parts of a galaxy (especially if it is significantly extended) should have less complex star formation histories (SFHs), so a 2D approach is expected to help in diminishing the effect of stellar population synthesis degeneracies, especially for high redshift sources where the stellar population age is relatively more limited. More specifically, our 2D approach can produce relatively complex SFH for an entire galaxy (similar to ``non-parametric" approaches; see, e.g., \citealt{2012MNRAS.421.2002P}, \citealt{2019ApJ...876....3L}, \citealt{2021ApJS..254...22J}, \citealt{2022arXiv220804325J}) based on simple parametrizations (e.g., an exponentially decaying SFH) for smaller parts of it.
\item[(3)] Independent estimations of the photometric redshift should allow to obtain more accurate results, especially if the galaxies present stellar population gradients, as well as help with the segmentation of images. 
\end{itemize}

On the con side of our 2D approach, we can mention: 
\begin{itemize}
\item[(1)] Squeezing the 2D stellar population synthesis procedure to very small regions of a galaxy can be detrimental since the derived properties can be affected by stochasticity in the IMF and/or correlations of properties in physically connected regions (e.g., linked to escape of ionizing photons from one region of the galaxy to another).
\item[(2)] The low surface brightness component of galaxies cannot be studied with the same spatial resolution as the ``core" of the galaxy.
\item[(3)] The combination of results from individual pixels or resolution elements to recover the properties of the galaxy as a whole might not be straightforward (e.g., in the determination of the redshift of the galaxy based on regions with very different SNR photometry). 
\item[(4)] The faintest and smallest galaxies would now allow for a 2D approach.
\item[(5)] The pixel-by-pixel approach needs well-determined PSFs and matching kernels as well as accurately aligned images, details which might not be as important when working with photometry using large apertures.
\end{itemize}

Overall, the advantages in the analysis of our galaxies are significant, so we follow this route and perform 2-dimensional analysis of the emission of galaxies, which we describe next. We invite the reader to see an evaluation of the method for $z>1$ galaxies in \citet{2022arXiv220714062G} and similar approaches based on HST data \citep{1999MNRAS.303..641A,2012ApJ...753..114W}. This paper is focused on presenting an interesting population of galaxies that JWST has revealed and allowed to study in detail, and will discuss the optimization of procedures for 2D SED-fitting in more detail in future work, when more comprehensive datasets and data calibrations are available.

The JWST and HST images described in Section~\ref{sec:data} were registered to the same World Coordinate System including a local solution for each galaxy in the sample, in order to ensure an accurate alignment on a pixel-by-pixel basis. To do that, we considered a square region around each source and measured centroids for all galaxies in the CANDELS catalog within that region for all images in the different filters. Based on the cross-correlation of those sources with respect to a reference band, $F444W$ in our case, we realigned and remapped all images to a common grid. The accuracy of this alignment is better than 0.001\arcsec, on average, with a scatter smaller than 0.006\arcsec (one fifth of the pixel). These statistics are independent of the band and are not affected by more complex morphologies of galaxies observed in the bluest bands, in part because the relative WCS alignment of different JWST (and HST) bands is extremely accurate (see Bagley et al., in prep.).

After aligning the images, we PSF-matched them using $F444W$ as the reference (see example in Figure~\ref{fig:2Dphotoz}). This is the band with the most extended PSF in our dataset, FWHM 0.16\arcsec (based on the actual PSF constructed with CEERS data). Using all known stars provided by the CANDELS catalog and fainter stars identified with JWST colors, we built PSFs for each band and kernels using {\it photutils} to match them to the $F444W$ PSF.

We then measured photometry on a pixel-by-pixel basis using the locally-registered, PSF-matched JWST and HST images. We built SEDs for each pixel and kept for analysis those which had more than 3 bands with fluxes measured at the 3$\sigma$ level or better. For bands with low-significance fluxes, we assumed 5$\sigma$ upper limits in all the subsequent analysis, including photometric redshift estimation and stellar population synthesis modeling. Considering 3$\sigma$ upper limits instead had a very small impact on photometric redshifts, with an average difference between estimations $\langle\Delta z/(1+z)\rangle=0.02$ and 5 galaxies out of the full sample of 138 sources changing redshift by more than $\Delta z=0.2$. Fixing the redshift, the effect of using 3$\sigma$ upper limits in the stellar population synthesis is negligible since the fits are mostly dominated by actually measured fluxes.

\begin{figure*}
    \begin{center}
    \includegraphics[clip, trim=4.9cm 6.2cm 0.3cm 9.0cm,scale=1.1]{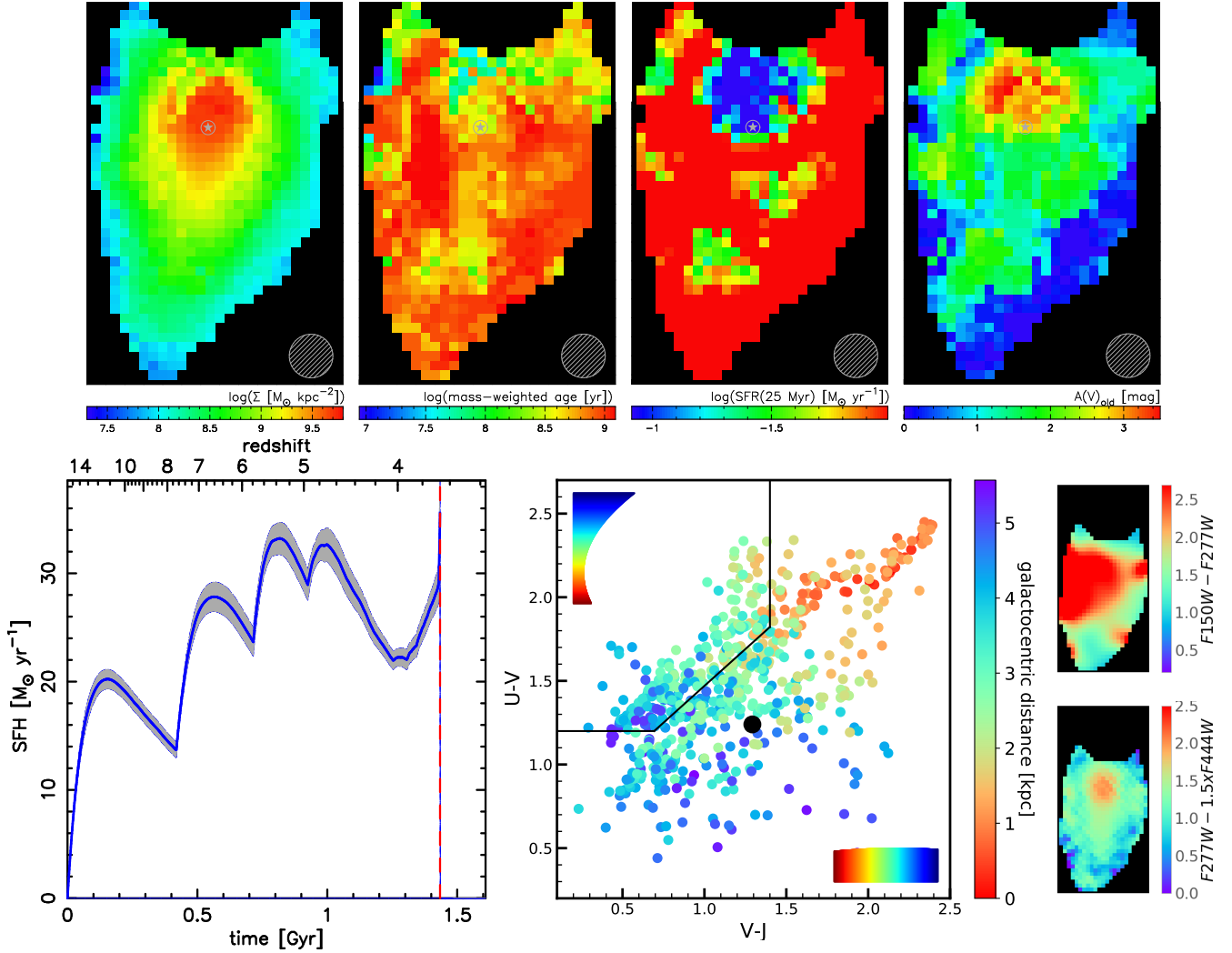}
    \caption{Results of our pixel-by-pixel SED modeling procedure for the example source presented in Figure~\ref{fig:2Dphotoz}. In this figure, we just keep the Southern component of the source, which we identify as a distinct galaxy that meets our selection criteria and lies at a different redshift than the Northern region of the source. The top 4 panels show maps of the stellar mass surface density, the mass-weighted age, the SFR averaged over the last 25~Myr of the SFH of the galaxy, and the dust attenuation in the $V$-band. The centroid of the galaxy calculated with the stellar mass map is marked with a gray star. The striped circle depicts the FWHM of the data. On the bottom-left, we show the integrated SFH and its uncertainty obtained by adding the results for all pixels, after applying an aperture correction based on the average flux ratio between the sum of all pixels used in the 2D analysis and the integrated aperture. On the bottom-center panel, we show a $UVJ$ diagram for all pixels used in our analysis, color-coded by distance to the galaxy centroid. Average uncertainties in both axis as a function of radius are plotted in the top-left (uncertainties in $V-J$) and bottom-right (in $U-V$) corners of the panel. The position of the integrated colors of the galaxy is marked with a black circle. On the bottom-right panel, we show color maps using the closest observed filters to the $UVJ$ rest-frame colors considering the redshift of the galaxy ($z=3.7$), namely $F150W-F277W$ for $U-V$ and $F277W-1.5\times F444W$ for $V-J$ (the multiplicative factor being applied to account for the extrapolation to probe the $J$-band).}
    \label{fig:SED}
    \end{center}
\end{figure*}

We also measured integrated photometry based on a \citet{1980ApJS...43..305K} aperture (marked in red in Figure~\ref{fig:2Dphotoz}). We note that the latter covers an area which is larger than the sum of the individual pixels for which we were able to build SEDs. The typical aperture correction from the sum of pixels to the Kron aperture is 7\% for our sample.

The median number of pixels with usable SEDs per galaxy in our sample is 472 (approximately 20 resolution elements, defined as the area of the FWHM), with 85\% (65\%) of the sample having more than 100 (150) pixels ($\sim$4/7 resolution elements). We note that the pixels in our images correspond to physical sizes between 250 and 170~pc, and the $F444W$ FWHM to 1.3 and 1~kpc, for the redshift range covered by our sample ($2\lesssim z\lesssim7$).

\section{Photometric redshifts and stellar population synthesis modeling in 2 dimensions}
\label{sec:photoz}

The SEDs described in Section~\ref{sec:photometry} were used to estimate photometric redshift and stellar populations properties on a pixel-by-pixel basis, and also for the integrated aperture. We describe the procedures in the following subsections.

\subsection{Estimation of photometric redshifts}

Photometric redshifts for each pixel were estimated with a modified version of the {\sc eazy} code \citep{2008ApJ...686.1503B}. The modification consisted in allowing the template fitting algorithm to use (5$\sigma$) upper limits as an input, not allowing any fit to provide brighter fluxes (achieved by penalizing the $\chi^2$ calculation) than those limits and excluding the band in the $\chi^2$ calculation for templates providing lower fluxes (M\'erida et al. 2022, submitted). The code was run disabling the template error feature (which penalized IRAC bands as a default, i.e., NIRCam LW fluxes would also be affected) and not using any prior. We used v1.3 templates, which include a dusty galaxy and a high-EW emission-line galaxy spectrum.

After obtaining photometric redshifts for each pixel, we combined the probability distribution functions (zPDF) for all pixels belonging to a given galaxy using the procedure described in \citet{2013ApJ...775...93D}. Briefly, different zPDFs are combined allowing for catastrophic fits as well as some degree of interdependency among the different estimations, parametrized with a quantity called $\alpha$, which can take values from 1 to the number of combined zPDFs. We note that, in our case, the estimation of photometric redshifts for nearby pixels should be correlated, since the FWHM of the PSF-matched data is 0.16\arcsec\, (5 pixels), but for most galaxies we counted with completely independent estimations given their large extension. We made some tests on the parameter used to account for this interdependence of estimations and found no significant variations in the final most-probable redshifts provided for each galaxy for $\alpha>2$.

The procedure of photometric redshift estimation in 2 dimensions is exemplified in Figure~\ref{fig:2Dphotoz} with a peculiar source, as explained next. We show one of the selected galaxies, nircam2-7122, which seems to have 2 stellar clumps in the JWST data. The Northern part, orange in the RGB image, was included in HST catalogs \citep[e.g., in][]{2017ApJS..229...32S}. At wavelengths redder than 2~$\mu$m, another clump starts to be visible to the South, a region that gets brighter at longer wavelengths while the Northern knot gets relatively dimmer (and almost disappears by 4~$\mu m$). Our 2D analysis indicates that the Northern (already known) galaxy lies at a lower redshift, as revealed by the photo-z map, which counts with 4-5 resolution elements for the Northern galaxy and 6-7 times more for the new JWST source. SEDs and photo-z fits are shown in Figure~\ref{fig:2Dphotoz} for the centers of the 2 distinct galaxies. A similar analysis was performed in the rest of the sample, just keeping the regions which present consistent photometric redshifts in the study of the stellar populations presented in the following sections. 

Our approach to analyze the photo-z maps was that in order to consider a possible contamination from an interloper at a different redshift, the region must have a size of at least 2 resolution elements (defined in terms of the FWHM). This allows to get rid of noise, visible in some zones of the photo-z map presented in Figure~\ref{fig:2Dphotoz}. In that plot, we clearly identify a different galaxy or knot with this criterion to the North (but not in other zones). We must also consider that the map is presented in terms of the most probable photo-z, but each pixel counts with a complete zPDF. 

With this in mind, in order to assign redshifts to different regions of a given galaxy, we considered a probability threshold to decide when 2 zPDFs (constructed with several pixels as explained above) provide incompatible redshifts. This assumes that a zPDF would catch degeneracies in the photo-z determination linked to stellar population gradients throughout the galaxy (or belonging to different galaxies). We considered as separate galaxies those regions where the combined zPDF implied a probability smaller than 10\% of the region being at the redshift of the rest of the galaxy.

We remark that the identification of redshift interlopers in this work implies that some regions of a given source were removed from the subsequent analysis (as in Figure~\ref{fig:2dsps}). This means that the total masses could be underestimated if the removed zone is indeed part of the galaxy, but the SFHs (or other properties such as SFRs, sSFR, A(V), etc...) would be robust for the kept galaxy area.

The evaluation and calibration of the 2D photometric redshift determination, i.e., how to combine the information from different pixels to obtain the best redshift for a given galaxy as well as help with the segmentation of objects, would need more extended datasets in terms of area and availability of accurate measurements at higher redshifts than what current spectroscopic samples provide (and would be extended in the coming months with JWST observations). 

As a first evaluation, we run the procedure through all the galaxies with spectroscopic redshifts in the 4 available CEERS pointings (see Section~\ref{sec:data}), just using the 9 filters selected for our study of HST-dark/faint galaxies. These galaxies sum up 1.1 million pixels, for which we calculated photometric redshifts. We obtained $\sigma_{\mathrm{NMAD}}$ values (as defined in \citealt{2008ApJ...686.1503B}) a factor of 2 better (0.035 vs$.$ 0.018) than using the integrated SEDs. 

\subsection{Stellar population synthesis modeling}

\begin{figure*}[t]
\centering
\epsscale{1.15}
\plotone{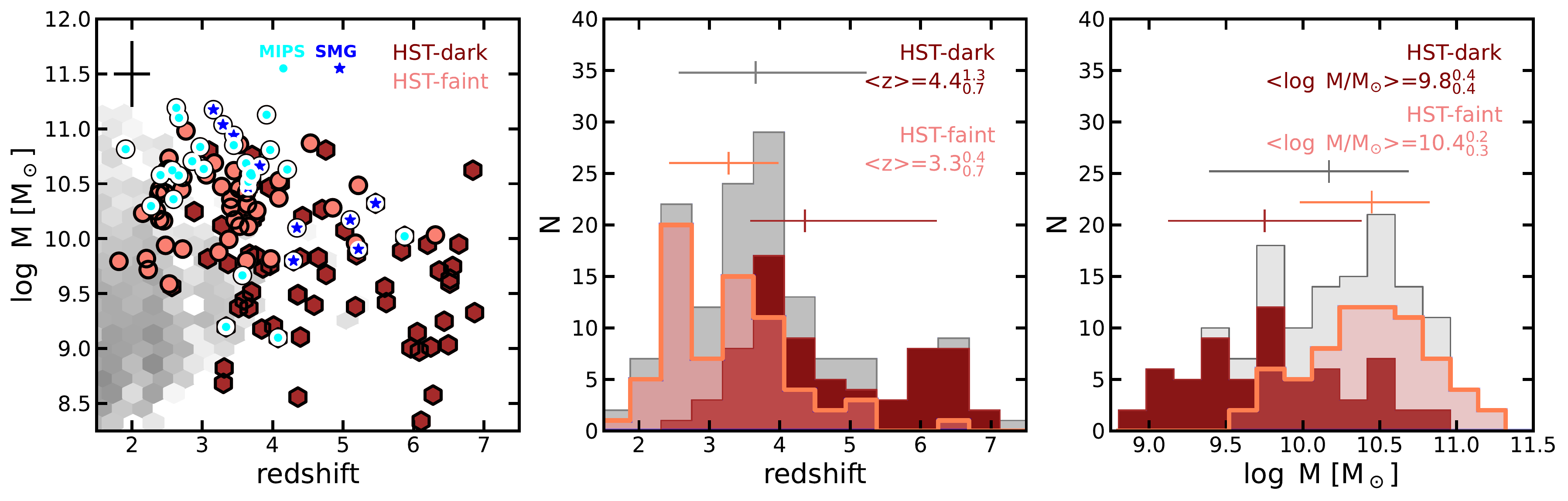}
\caption{The left panel shows the stellar mass versus photometric redshift 
diagram for all the galaxies in the CANDELS catalog (gray-scale density
map) and the color-magnitude selected sample divided into HST-dark
(dark hexagons) and HST-faint (light circles). MIPS24 emitters are marked with a cyan circle, sub-mm and radio galaxies with a blue star. The central and right panels show the
photometric redshift and stellar mass distributions for the HST-dark and HST-faint  galaxies (with the same color scheme) and the whole sample (gray histogram).  Some statistical information (median and quartiles) for the HST-faint and HST-dark subsamples are provided.}
\label{fig:zphot_mass_histograms}
\end{figure*}

Once we obtained a most-probable redshift combining the estimations for all pixels belonging to each galaxy, we fixed its value and analyzed the stellar populations. We fitted the pixel-by-pixel and integrated SEDs with the synthesizer code \citep{2003MNRAS.338..508P,2008ApJ...675..234P} assuming a delayed-exponential as the SFH, with timescale values $\tau$ between 100~Myr and 5~Gyr, ages between 1 Myr and the age of the Universe at the redshift of the source, all discrete metallicities provided by the \citet{2003MNRAS.344.1000B} models, and a \citet{2000ApJ...533..682C} attenuation law with $V$-band extinction values, $A(V)$ between 0 and 5~mag. We assume a universal initial mass function (IMF) that follows the \citet{2003PASP..115..763C} functional form. Nebular continuum and emission lines were added to the models as described in \citet{2008ApJ...675..234P}. A Monte~Carlo method was carried out in order to obtain uncertainties and account for degeneracies (see \citealt{2016MNRAS.457.3743D}). We warn the reader that this kind of methodology would not account for systematic uncertainties such as those arising from the parametrization of the SFH, metallicity evolution or the complexity of dust effects (in general, and as a function of position in the galaxy).

Figure~\ref{fig:SED} shows the results of the stellar population synthesis (SPS) modeling for the galaxy presented in Figure~\ref{fig:2Dphotoz}, after removing the Northern component which is revealed to lie at a different redshift by our 2D photo-z method. All relevant quantities have been corrected by the aperture correction mentioned in Section~\ref{sec:photometry}. The galaxy presents a knot with significant amounts of dust, $A(V)\sim3$~mag, separated from the stellar mass centroid, and (integrated) SFR around 10~M$_\odot\,\mathrm{yr}^{-1}$. The rest of the galaxy presents a less active and low-dust content disky structure (see also $UVJ$ diagram per pixel). The SFH of the galaxy shows 3 main episodes of star formation, one beyond $z\sim10$, another (extended) at $z\sim5.5$, and the recent burst in the center. The galaxy is not detected by MIPS, PACS, SPIRE, or (sub-)mm wavelengths (nor in X-ray) at the 3$\sigma$ level or better. This galaxy is transiting through a post-starburst phase with some residual dusty star formation near the nuclear region, presenting mass-weighted ages of up to a few hundred Myr. 

The 2D $UVJ$ diagram on the bottom right of Figure~\ref{fig:SED} shows that the full galaxy would be located in an intermediate region between blue, low-dust content, star-forming galaxies and dusty starbursts. The pixels near the mass-weighted center of the galaxy lie towards the dust-enshrouded zone, while outer pixels beyond the effective radius tend to lie in the blue cloud and post-starburst/quiescent wedge. In order to demonstrate the reliability of this spatially-resolved $UVJ$ diagram, we show color maps directly constructed with the JWST bands which roughly probe the $UVJ$ bands, namely, $F150W$, $F277W$, and $F444W$. For the latter, given that we would need some extrapolation to probe the $J$-band, we multiply it by a 1.5$\times$ factor, so we roughly match in the color map the $V-J$ values in the color-color plot.

Our 2D-SPS methodology was compared in \citet{2022arXiv220714062G} with fits of the integrated emission of simulated galaxies in Illustris using SFHs including a single burst and two star formation events, each following a delayed exponentially-decaying SFH. We demonstrated that this more classical approach could underestimate the mass-weighted ages by factors of a few (cf. Figure 8 in \citealt{2022arXiv220714062G}). The paper also showed that our 2D-SPS approach could reproduce the Illustris SFHs with 20-30\% accuracies in physical properties such as mass weighted ages and formation times. For this paper, we compared our 2D-SPS-based stellar masses with those obtained obtained from single-burst delayed exponentially-decaying SFHs fitting the integrated emission, obtaining a very small systematic offset, $\Delta\log\mathrm{M\!_\star}=0.02$~dex, and a scatter that ranges from 0.05~dex at $\log\mathrm{M\!_\star/M}_{\odot}>10$ to 0.3~dex for smaller masses. A detailed comparison of modeling procedures including 2D-SPS, integrated photometry with parametric and non-parametric SFHs will be presented in other papers (e.g., Garc\'{\i}a-Argum\'anez et al., in prep).

\section{Analysis of the integrated and bi-dimensional stellar population properties}
\label{sec:analysis}

In this section, we analyze the
photometric redshifts, stellar masses, morphology, and both the integrated and spatially resolved current vs past star formation activity of the 138
galaxies selected as ``HST-dark" and ``HST-faint" using the mid-to-near IR color-magnitude diagram.

\subsection{Redshifts and stellar masses}

The left panel of Figure~\ref{fig:zphot_mass_histograms} shows the
photometric redshift versus stellar mass diagram for galaxies in the
HST WFC3/$F160W$-selected CANDELS catalog (gray scale density map) overlapping
with the CEERS region, and the galaxy sample selected by our
JWST color-magnitude criterion.  We also distinguish between JWST galaxies 
that are brighter and fainter than $F150W=26$~mag, namely HST-faint and -dark galaxies
(light and dark colors, hexagons and circles), respectively. At first glance, the main difference
in the distribution of galaxies selected with HST and the new JWST-selected sample is that the latter
reveals a population of red, massive ($\log\mathrm{M\!_\star/M}_{\odot}=9-11$)
galaxies at $z\gtrsim2$ many of which were previously undetected (HST-dark) even
in the deepest HST/WFC3 surveys such as CANDELS. 


In more detail, the HST-faint sample selects predominantly massive
galaxies ($\log\mathrm{M\!_\star/M}_{\odot}\gtrsim10$) over a
redshift range around $z\sim3$. This is consistent with the distribution of points for
HST-faint galaxies in the color-magnitude diagram presented in Figure~\ref{fig:color_magnitude_selection}. As
illustrated by the template color-tracks, similarly massive galaxies
at redshifts $z\lesssim2$ are not in the HST-faint sample because they have
bluer colors ($F150W-F356W<1.5$~mag) and/or brighter $F150W$ magnitudes, while
massive galaxies at higher redshifts tend to be redder and fainter, and thus fall
into the selection region of the HST-dark sample. 

Consequently, the
HST-dark sample exhibits a higher median redshift $z\sim4.4$ and a
broader distribution that extends up to $z\sim7$. The redshift
overlap between HST -faint and -dark samples at $z\sim3.5$ (central
panel of Figure~\ref{fig:zphot_mass_histograms}) is a direct
consequence of the magnitude limit used to define the two sets
($F150W=26$~mag). If we choose a fainter magnitude, many of the HST-dark
galaxies at higher-z will move to the HST-faint sample increasing its
median redshift and lowering the median redshift of the HST-dark
sample. We note that while $F150W=26$~mag is an adequate limit to identify
WFC3 dropouts in the shallower CANDELS mosaics in EGS, the deeper
WFC3 imaging in GOODS/UDF fields (see. e.g., \citealt{2013ApJS..207...24G}, \citealt{2019ApJS..243...22B}) means
that HST-dark galaxies in those fields will be fainter in $F160W$ (by
0.5 to 1~mag) and will have, on average, higher redshifts (as presented exactly in those fields in, e.g., \citealt{2019ApJ...876..135A}). 

The redshift histogram for HST-dark galaxies reveals a peak around $z\sim6.5$. The sources in this peak are peculiar, as we will also show in the coming sections. Here we emphasize that these sources typically present a blue $F356W-F410M$ color, on average $-0.3\pm0.2$~mag, compared with an average of $+0.2\pm0.2$~mag for the rest of the sample. Some of them are also brighter in the $F410M$ compared to both $F356W$ and $F444W$.

In terms of the stellar mass distribution (right panel of
Figure~\ref{fig:zphot_mass_histograms}), the HST-dark sample has a
lower average stellar mass than the HST-faint sample, $\log
\mathrm{M\!_\star/M}_{\odot}\sim9.8$ vs$.$ $\log\mathrm{M\!_\star/M}_{\odot}\sim10.4$, although both
distributions have galaxies with masses up to $\log
\mathrm{M\!_\star/M}_{\odot}\sim11.0$. The HST-dark sample also extends to lower stellar masses than the
HST-faint sample. This is primarily due to the faint limiting magnitude
of the selection in $F356W$, which, to first order, correlates with the
stellar mass of the galaxies. As illustrated by the template
color-tracks in Figure~\ref{fig:color_magnitude_selection}, some of
the galaxies in the HST-dark sample with magnitudes around
$F356W\sim27$~mag will have $\log\mathrm{M\!_\star/M}_{\odot}\lesssim10$ at redshifts
ranging from $z=3$ to 7.

\begin{figure*}   
    \includegraphics[width=18cm]{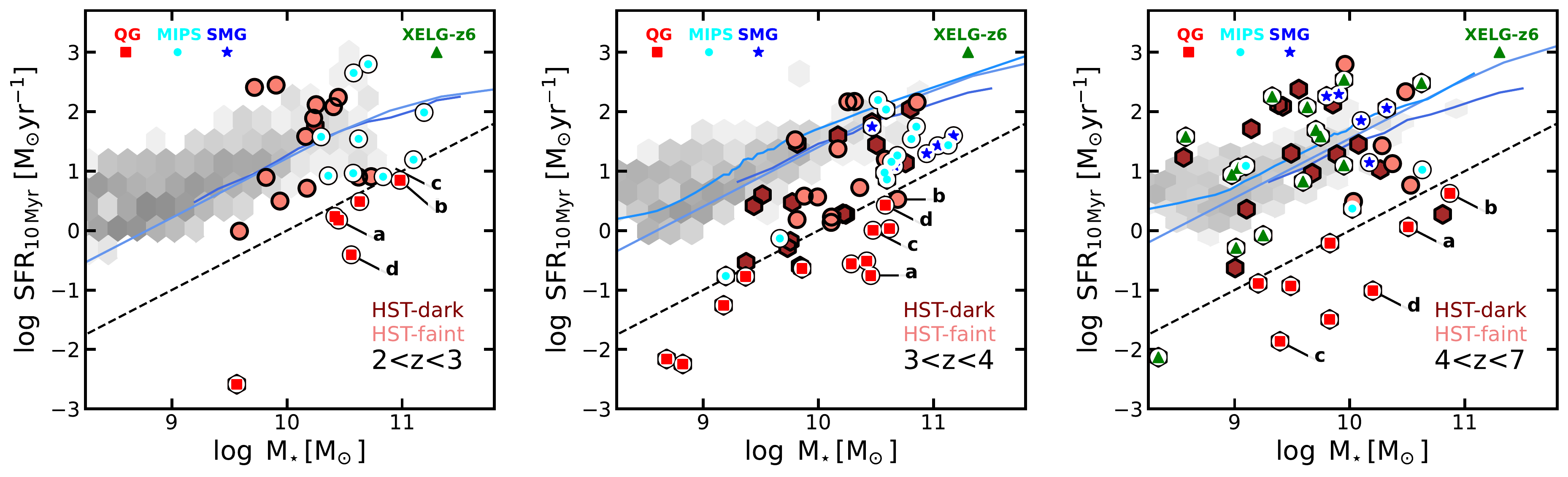}\\
    \includegraphics[width=18.cm,angle=0]{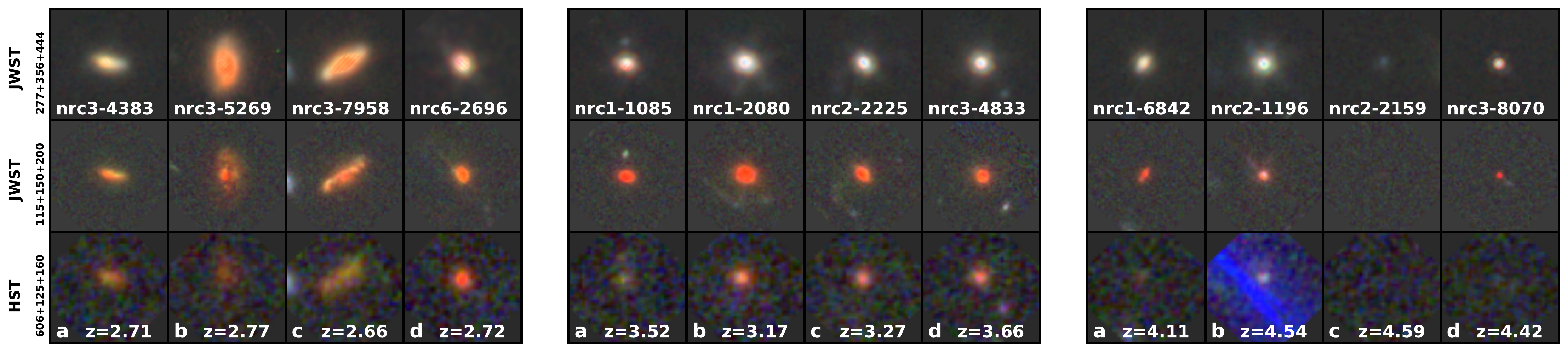}
    \caption{Main sequence (SFR vs$.$ stellar mass, SFMS) plots for our sample of galaxies with red near-to-mid IR colors, compared to the CANDELS sample. From left to right, each panel shows the SFMS in different redshift bins, selected to exemplify: the range dominated by HST-faint galaxies ($F150W<26$~mag), $2<z<3$; HST-dark sources ($F150W>26$~mag), $4<z<7$; and a transition range. In each panel our sample is shown with hexagons (HST-dark) and circles (HST-faint). MIPS24 emitters are marked with a cyan circle, sub-mm and radio galaxies with a blue star, quiescent galaxies with a red square, and XELG-$z6$ with green triangles. The hexmap shows the distribution of galaxies in the CANDELS catalog \citep{2017ApJS..229...32S}, with SFRs estimated with the method explained in \citet{2019ApJS..243...22B}. The colored lines depict the SFMS at $z=2-7$ from  \citet{2015A&A...575A..74S}, \citet{2017ApJ...847...76S}, and \citet{2019ApJS..243...22B}. The cutouts on the bottom show the 5\arcsec$\times$5\arcsec images of the quiescent galaxies marked with letters in the SFMS, also providing IDs and redshifts. From bottom-to-top, the images are color composites of 3 HST bands, and 3 JWST bands, short and long wavelengths, respectively.}
    \label{fig:ms}
\end{figure*}

Another consequence of having a fainter limiting magnitude is that the
average stellar mass of our HST-dark sample is also lower than those
of pre-JWST HST-dark studies which selected galaxies using {\it
Spitzer}/IRAC and thus were restricted to much brighter limiting
magnitudes of [3.6]$\sim$24.0-24.5~mag. Therefore, these samples had
virtually no galaxies with $\log\mathrm{M\!_\star/M}_{\odot}\lesssim10$. For example,
\citet{2016ApJ...816...84W} or \citet{2019ApJ...876..135A} find average
stellar masses of the order of $\log\mathrm{M\!_\star/M}_{\odot}=10.5$ and $10.8$,
respectively, for their HST-dark, IRAC bright samples.  A different effect
that can contribute to the larger stellar masses of these IRAC-based
studies is that they cover a larger area than the current CEERS survey
by up to a factor $\times$4-5 and, therefore, they are more likely
to find a higher fraction of rare galaxies with very large stellar
masses. Furthermore, the number of those
galaxies detected over the relatively small area surveyed so far by
CEERS is more susceptible to cosmic variance.

\subsection{Recent/past star formation and colors}

In this section we study the star formation properties and rest-frame
colors of our sample in the context of the star formation main sequence
and the $UVJ$ diagram. Based on these analysis, we will divide the sample
into star formation activity subsets that will be explored in the
upcoming sections to analyze their stacked properties (e.g.,
star formation or mass-assembly histories). In particular, we will classify 
the sample in 3 classes: quiescent sources, star-forming galaxies, many of them detected at far-infrared and sub-mm wavelengths (note that the coverage of the CEERS footprint by sub-millimeter surveys is very inhomogeneous, so this is not a complete sample), and peculiar galaxies at $z\sim6$. We refer the reader to subsection~\ref{sec:classify} for more details.

\subsubsection{Recent star formation properties and the star-forming main sequence}
\label{sec:sSFR}

Figure~\ref{fig:ms} shows the SFR vs stellar mass plot for our sample, known as the star formation main sequence (SFMS) diagram. In order to further  understand the properties of our sample, which derive directly from the selection method, we divide the sample in several redshift bins in the following plots. The selected redshift bins are: one dominated by HST-faint galaxies, $2<z<3$, one dominated by HST-dark galaxies, $4<z<7$, and an intermediate range with similar number of galaxies from the 2 subsamples, $3<z<4$. SFRs have been derived by calculating the average value from the SFHs of each galaxy in different time-intervals, ranging from the last 5 to the last 100 Myr (namely, 5, 10, 25, 50, 75, and 100~Myr). In this figure, we use the SFR defined as that averaged over the last 10 Myr. Similar trends are observed in plots using other definitions of the SFR.

In Figure~\ref{fig:ms}, we see our sample forming a band in the SFR-stellar mass plane around the SFMS, most galaxies would qualify as star-forming galaxies and a small fraction presents very small specific SFRs (a 0.1~Gyr$^{-1}$ limit is marked in the plot), expediting the classification as quiescent/dormant galaxy. Although our SFR and stellar mass estimations do follow a sequence, many of the galaxies lie below the "classical" SFMS (this is specially clear when using SFRs averaged on the last 75-100~Myr), typically obtained by estimating stellar masses from integrated-aperture SED-fitting and SFRs from observed rest-frame UV luminosities corrected for attenuation with recipes based on UV-slopes, or with SFR tracers linked to dust emission. At first sight, this kind of behavior (which has also been observed when comparing the observed SFMS with galaxy formation simulations such as Illustris, see \citealt{2015MNRAS.447.3548S,2019MNRAS.485.4817D}) might be interpreted as the SED-fitting analysis underestimating the SFRs due to high dust contents. 

\begin{figure}
\includegraphics[width=9.cm,angle=0]{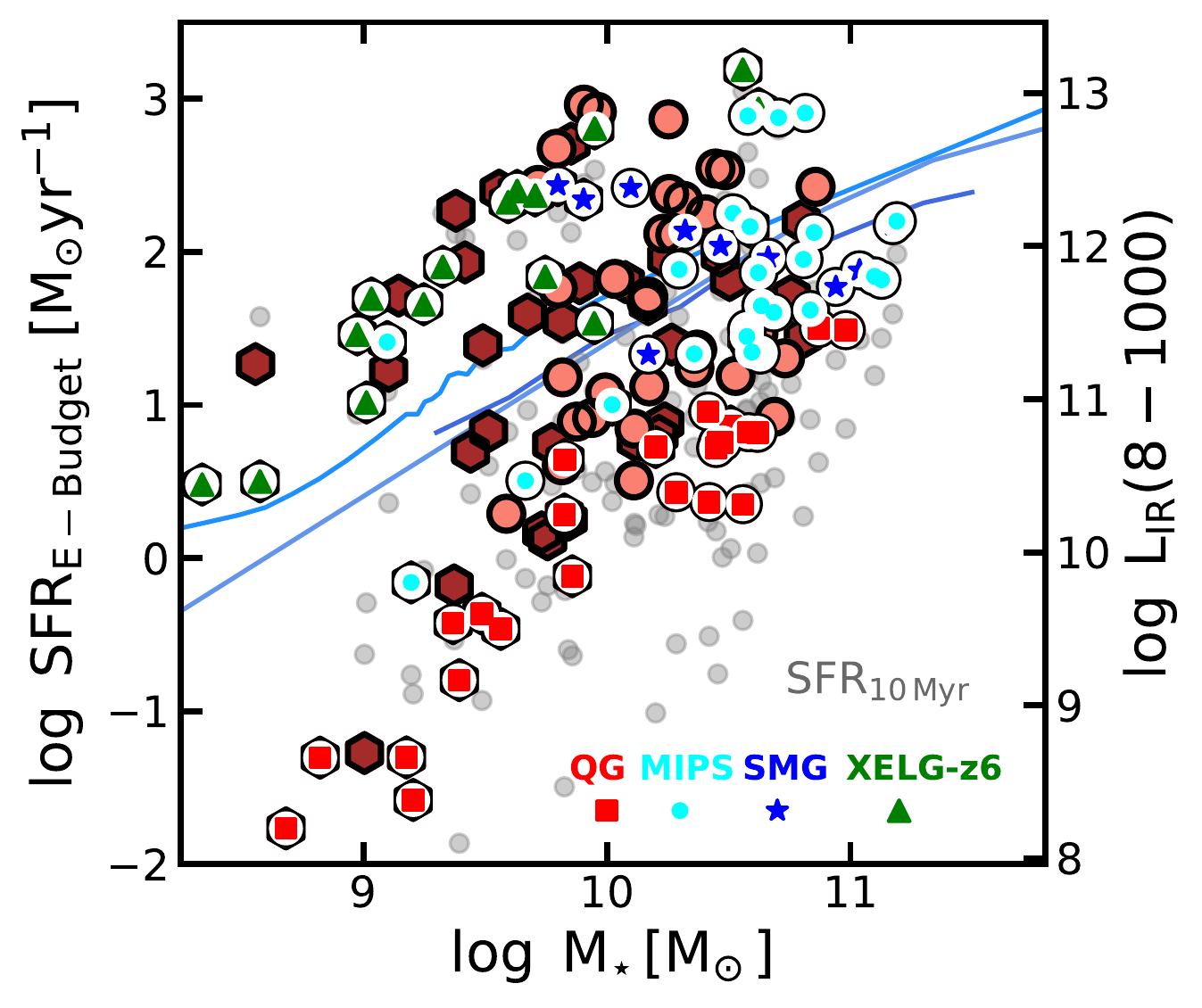}
\caption{Star formation main sequence plot for our sample of mid-IR bright, near-IR faint galaxies constructed with SFRs derived from our 2D stellar population synthesis analysis through the calculation of the dust-absorbed stellar emission. The energy extincted by dust in the UV/optical is assumed to be reemitted in the mid- to far-IR, from 8 to 1000~$\mu$m. The luminosity in this wavelength range is then converted to SFR using \citet{1998ARA&A..36..189K} factor. Symbols in gray show the SFRs obtained from the average of the integrated SFHs in the last 20~Myr. Blue curves depict literature main sequence relationships (same as in Figure~\ref{fig:ms}).}
\label{fig:sfms_energybudget}
\end{figure}

To probe the possible existence of highly embedded, optically-thick star-forming knots that cannot be characterized with rest-frame UV-to-near-IR SEDs, and test how the 2D approach handles this classical problem, we discuss the effects of dust on the SFMS plot in Figure~\ref{fig:sfms_energybudget}. For this purpose, we calculate the total energy absorbed by dust according to our pixel-by-pixel SPS analysis, adding up the contributions of all pixels for each galaxy to calculate the stellar dust-absorbed luminosity. Assuming an energy balance, this luminosity should be converted into dust emission in the mid-to-far IR. We then translate the dust-absorbed stellar luminosity to an IR-based SFR applying the calibration in \citet{1998ARA&A..36..189K}. We remark that this exercise should provide results which are completely comparable to the regular technique used to construct the SFMS plot at the high-mass end. For low-mass galaxies, with lower dust contents, SFRs are typically based on rest-frame UV luminosities, maybe added to absorbed emission with hybrid SFR tracers (see \citealt{2005ApJ...619L..51B}, \citealt{2007ApJ...666..870C}) or dust-corrected using UV-slopes (with more or less complicated recipes, see \citealt{2019ApJS..243...22B}). Therefore, for our galaxies, which span a relatively wide range of masses, the SFRs calculated with an energy-budget technique might be underestimated at the low stellar mass end if significant amounts of non-obscured star formation exists.

\begin{figure*}
    \includegraphics[width=18.cm,angle=0]{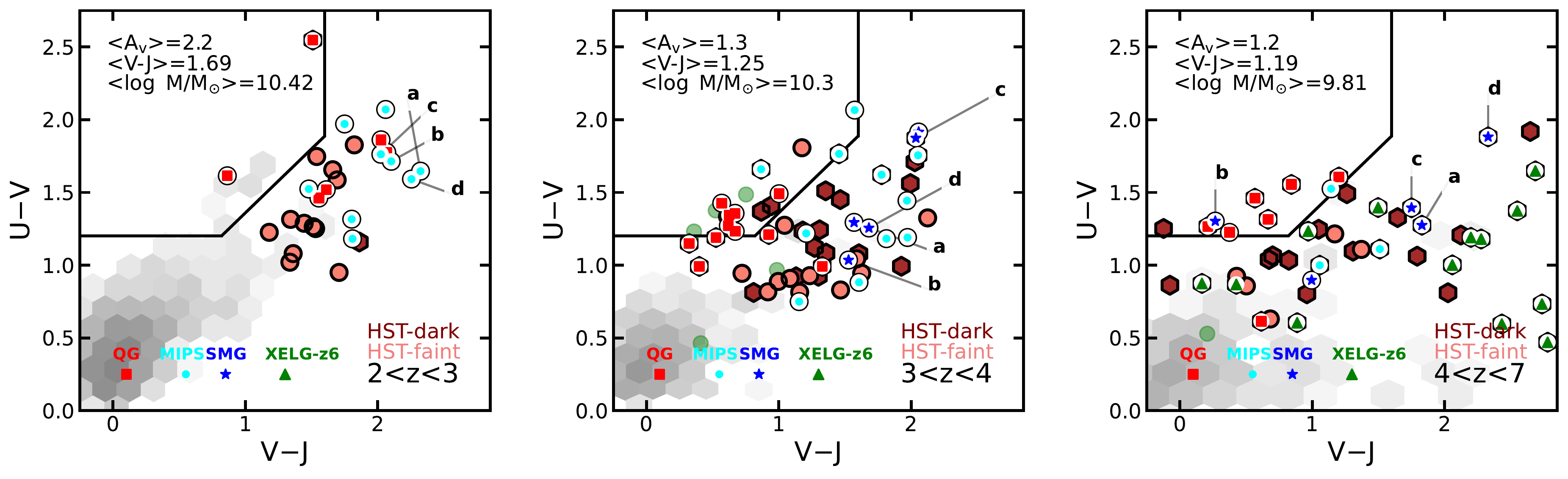}\\
    \includegraphics[width=18.cm,angle=0]{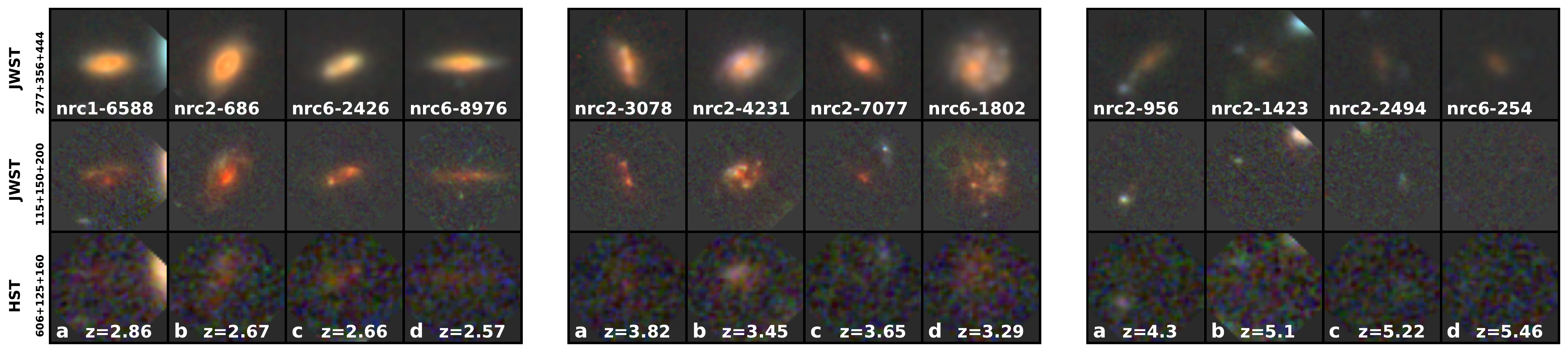}
    \caption{Rest-frame $U-V$ vs $V-J$ diagram at different redshifts for the galaxies in the CEERS sample (color gray-scale) and the color-selected sample divided in HST-faint and -dark (light and dark colors). MIPS24 emitters are marked with a cyan circle, sub-mm and radio galaxies with a blue star, quiescent galaxies by sSFR with a red square, and XELG-$z6$ with green triangles. The green circles show quiescent galaxies at $z>3$ from \citet{2022arXiv220800986C}. HST-dark galaxies make up the bulk of the sample at $z>4$ while HST-faint dominate in number at $\langle z\rangle=2.5$, and there is an even split between them at $\langle z\rangle=3.5$. The average color, extinction and stellar mass are indicated in the top left corner of each plot. Overall color-selected galaxies are predominantly massive, dusty or quiescent galaxies in agreement
with the goal of the selection criteria in Figure~\ref{fig:color_magnitude_selection}. Furthermore, there is a clear downward trend in extinction and stellar mass ranging from $\langle A_{V}\rangle\sim2.2$~mag and $\langle\log \mathrm{M\!_\star/M}_{\odot}\rangle\sim10.5$ at $z<3$ to $\langle A_{V}\rangle\sim1.2$~mag and $\langle\log \mathrm{M\!_\star/M}_{\odot}\rangle=10$ at $z>4$. The cutouts on the bottom show the 5\arcsec$\times$5\arcsec images of the MIPS and sub-mm/radio galaxies marked with letters in the $UVJ$ diagram, also providing IDs and redshifts. From bottom-to-top, the images are color composites of 3 HST bands, and 3 JWST bands, short and long wavelengths, respectively.}
    \label{fig:UVJ}
\end{figure*}

Figure~\ref{fig:sfms_energybudget} demonstrates that the 2D-based, global dust attenuations for the galaxies in our sample do reproduce the dust content and emission observed in galaxies and used for the construction of the SFMS. Even though the 5-to-100~Myr time-averaged SFRs fall below the MS, the fair comparison with observations behaves much better. On average, and down to $\log
\mathrm{M\!_\star/M}_{\odot}\sim9.7$, our energy-budget based SFRs reproduce the SFMS, with the median sSFR lying within 0.1~dex of the literature curves. However, the scatter of our SFMS is $2\times$ larger than the typical 0.2-0.3~dex reported in the literature. Whether this behavior is intrinsic or related to observational uncertainties is beyond the scope of this paper. 

In Figure~\ref{fig:sfms_energybudget} (and the right panel of Figure~\ref{fig:ms}), we mark the galaxies in the redshift peak at $z\sim6$ reported in the previous section (green triangles). All these sources present very large SFRs for their stellar mass, well above the SFMS, consequently qualifying as starburst galaxies. The SFRs derived with the energy budget calculation are also large, indicating the presence of significant amounts of dust. We remark here that the starburst galaxies seem to follow a separate sequence, very similar to the bimodality reported in  \citet{2022ApJ...930..128R}. We cannot claim any robust interpretation for this effect due to the small number of galaxies we have in our sample, but our modeling points out to systematic effects linked to age (rather, for example, than for extinction), since most of the galaxies located far away from the SFMS present very young bursts, many of them belonging to the XELG-$z6$ type. This would be consistent with the presence of strong emission lines, most probably [OIII]$+$H$\beta$, in these starburst systems, similar to the sources reported in \citet{2022arXiv220814999E}, \citet{2022arXiv221108255M} and \citet{2022arXiv221205072L}, but they also present relatively high dust contents, with average attenuations around $A(V)=2$~mag. Indeed, \citet{2022arXiv221108255M} reports not negligible amounts of dust in some of the (NIRCam-grism selected) spectroscopically confirmed [OIII] emitters.

We conclude that our 2D-SPS method based on JWST$+$HST data does provide robust dust estimations at spatial resolutions around 200~pc, and that SFRs from SED-fitting should be compared with SFRs from classical tracers with caution.

\subsubsection{Rest-frame $UVJ$ colors}


Figure~\ref{fig:UVJ} shows the overall distribution of all the CEERS galaxies
(gray-scale density map) and the color-selected sample, divided into HST-faint and -dark in the rest-frame $UVJ$ diagrams at different redshifts. The distribution of rest-frame colors is roughly consistent with what we see in the SFMS plot: the bulk of the sample are massive, dusty star-forming galaxies with a small fraction quiescent galaxies starting at $z\gtrsim3$. The majority of the galaxies identified as quiescent by sSFR (red squares) are located in the quiescent region of the $UVJ$ diagram, showing that the two methods are quite consistent. 

At $2<z<3$ nearly all the color-selected galaxies are HST-faint, dusty
and star-forming, with high average extinctions $\langle A_{V}\rangle=2.2$~mag and
relatively large stellar masses of $\langle\log\mathrm{M\!_\star/M}_{\odot}\rangle=10.4$.  This is
consistent with the color-tracks in
Figure~\ref{fig:color_magnitude_selection} that indicate that only
dusty galaxies are redder than the selection threshold at $z<3$. I.e.,
quiescent galaxies at these redshifts do not make the color cut.
Similarly, since these galaxies are also relatively massive, they tend
to be on the brighter end of both the $F356W$ and $F150W$ magnitudes and,
consequently, the majority are HST-faint.  Interestingly, many of
these galaxies are located in the very high dust-obscuration region of
the $UVJ$ diagram with rest-frame colors exceeding $V-J>2.0$~mag. Previous
works have pointed out that this dusty corner of the $UVJ$ diagram seems
to exhibit a higher fraction of objects with late-type morphologies
(or low \citealt{1968adga.book.....S} indices) and closer to edge-on inclinations which can
increase the reddening along the line-of-sight due to dust within the
disks which lead to values of $A_{V}\gtrsim2$~mag (e.g., \citealt{2012ApJ...748L..27P}; \citealt{2018ApJ...869..161W}).

At $3<z<4$, there is roughly an even split between HST-dark and -faint
galaxies. The average stellar mass is roughly the same, but the
average extinction is much lower, $\langle A_{V}\rangle=1.4$~mag, than
at $z\sim2.5$. Similarly, the average $V-J\sim1.5$~mag is smaller and
fewer galaxies are found in the dusty corner $V-J>2$~mag. Notable
exceptions are the sub-mm/radio galaxies with $V-J>1.5$~mag. The lower
average attenuation and $V-J$ color is likely driven by the inclusion
of a larger number of quiescent galaxies ($\sim17\%$), which enter the
color selection at $z>3$. We note also that at $z\sim3.5$ some of
the younger, perhaps recently quenched, quiescent galaxies with ages
$\sim1$~Gyr have bluer $U-V\sim1.3$~mag colors, and some of them will
not make the color threshold of the selection, $F150W-F356W>1.5$~mag,
even at $z=3$ (orange template in
Figure~\ref{fig:color_magnitude_selection}).  To emphasize this point,
in Figure~\ref{fig:UVJ} we show 6 out of the 17 galaxies in the recent
\citet{2022arXiv220800986C} paper which finds massive quiescent
galaxies at $z>3$ in the CEERS field. 3 out of these 6 galaxies (green
circles), are in our $UVJ$ quiescent region but are not selected by
our color criterion, as they are too blue in $F150W-F356W$. The other
11 galaxies with redder colors and higher redshifts are included in
our sample and 9 of them are also classified as quiescent by our
criteria.

At $4<z<7$, nearly all selected galaxies are HST-dark. The average
extinction and color continue their downward trend relative to the
lower redshift bins with $\langle A_{V}\rangle=1.2$~mag, $\langle
V-J\rangle = 1.2$~mag, but there is also a noticeable decrease in the
average stellar mass to
$\langle\log\mathrm{M\!_\star/M}_{\odot}\rangle=10$.  Since stellar
mass and dust are strongly correlated (e.g.,
\citealt{2018ApJ...858..100F}) a decrease in the average mass will
also lower the attenuation. The lower average mass is caused by a
decrease in the number of very massive (and dusty) galaxies (e.g.,
$\log \mathrm{M\!_\star/M}_{\odot}\gtrsim10.8-11$) in our sample at
redshift $z\gtrsim4$ (see Figure~4). This is most likely caused by a
combination of the steep decline in the number density of massive
galaxies with redshift and our small surveyed area. For example,
\citet{2019ApJ...876..135A} (see also \citealt{2013ApJ...777...18M};
\citealt{2015ApJ...803...11S}) finds densities of
$\log(\mathrm{N})\sim-4.75$ or $-5.00 $~[Mpc$^{-3}$] at $3<z<6$, even
after accouning for HST-dark galaxies. In the small area covered by
CEERS, that density translates to less than $\sim$1 galaxy (we find
none), and even at $\log \mathrm{M\!_\star/M}_{\odot}>10$ the number
of galaxies is expected to be of the order of $\sim20$ based on
pre-JWST studies (e.g., extrapolating \citealt{2013ApJ...777...18M},
$\log(\mathrm{N})\sim-3.75$~[Mpc$^{-3}$] at $z\sim3.5$).
Consequently, more massive, dusty star-forming galaxies like those identified in sub-mm
surveys \citep{2019Natur.572..211W} will likely be found in larger
area surveys. In fact, 2 (of the 5) sub-mm NOEMA galaxies at this
redshift are among the most massive and higher extinction galaxies. In
particular, galaxy a) was also recently discussed in
\citet{2022arXiv220801816Z} as a source that might be misidentified
with very high-z ($z\gtrsim10$) candidates.

Aside from the average trends, the fraction of quiescent galaxies is
a bit smaller compared to the previous redshift bin, $\sim$10\%, but, we note an
increase in the overall scatter such that there are more galaxies with
more extreme $V-J$ colors (e.g. $V-J>2$ or $V-J<0.25$). The majority of
these galaxies are all at $z>6$, in the redshift peak reported in previous sections. They also have peculiar SEDs, in terms of $F356W-F410M$ color, and SEDs which are red  at $\lambda>2\,\mu$m but get bluer at shorter wavelengths, similar to the SEDs of the galaxies presented in \citet{2022arXiv220712446L}. At those redshifts the $V-J$ color is not probed by the observed
SED and therefore it is extrapolated from the best-fit SED
template. Most of these galaxies are typically a combination
of a dusty stellar population (hence the red $V-J$) plus a low
mass, un-extinguished burst that causes the blue UV-rest-frame color and
a high, recent SFR (see SFH section).  The true stellar properties of
these galaxies are unclear and still debated, most remarkably whether the red color is
caused in predominant part by strong emission lines. Our 2D SED fits reveal strong (i.e., high-EW) [OIII]$\lambda4959,5007$ and H$\beta$ emission boosting the $F356W$ flux (and sometimes the $F140M$ emission), linked to a very young starburst on top of a an older population. For this reason, we will identify these sources as extreme emission-like galaxies at $z\sim6$ (XELG-$z6$, hereafter). They are a
relatively numerous population, in fact the dominate our sample at
$z>6$ and we might still be missing some, since their colors decrease
with redshift, ranging from $F150W-F356W=3$~mag down 
to our limit 1.5~mag (see postage stamps in Figure~\ref{fig:mass-size}). In fact, the majority of galaxies from
 \citet{2022arXiv220712446L} at $z>7$ are bluer and fainter at $F356W$, so we only have 3 in common with their
sample. Given the selection of our sample, we are biased towards galaxies with emission lines entering the $F356W$ passband, but if similar high-EW ELGs exist at lower redshifts, the lines could lie on other filters and affect the interpretation of JWST data. This is the case for one of the objects discussed in \citet{2022arXiv220801816Z}, for which $z=4.6$  and $z=16.7$ \citep{2022arXiv220712356D} redshift solutions are obtained, depending strongly on the usage of templates with high-EW emission lines. In our 2D study, that galaxy, nircam2-2159 is assigned with a redshift $z=4.59$, with all its 74 pixels preferring the low redshift solution over the $z\sim16$ value. Apart from that, and interestingly, all XELGs-$z6$ share identical
morphologies/appearances (see Section~\ref{sec:structure}).

\subsubsection{Selecting subsets of galaxies based on star formation activity}
\label{sec:classify}

Based on Figures~\ref{fig:ms} to \ref{fig:UVJ}, in the following sections we will divide our sample in 3 different galaxy types according to their star formation activity. We identify 25 galaxies as quiescent/dormant (QG) for which the SFMS plot provides $sSFR<0.1$~Gyr$^{-1}$, 17 of them inside the quiescent wedge in the $UVJ$ diagram. For the identification of QGs, we forced that the galaxies were below the quoted sSFR level for all the SFR timescales mentioned in Section~\ref{sec:sSFR} (ranging from 5 to 100~Myr), and  the energy-budget sSFR. All other galaxies are considered as star-forming galaxies (SFGs), except the special case of $z>6$ sources, which we already introduced, XELG-$z6$. For QGs and SFGs, in some plots we will also consider 2 redshift regimes, $z<4$ and $z>4$ for SFGs, and $z<3.5$ and $z>3.5$ for QGs, which divide the samples in roughly equal numbers (and help with the interpretation of the results).

\subsection{Structural properties}
\label{sec:structure}
\subsubsection{Visual appearances}

Figures~\ref{fig:ms} and \ref{fig:UVJ} show the cutouts of several interesting galaxies in our sample (HST-dark and -faint) at different redshifts. There are 3 cutouts per galaxy showing the color composite images based on 3 HST bands and 3 JWST bands, at short and long wavelengths. The purpose of these images is to not only illustrate the appearances of the galaxies as seen by JWST but also the substantial differences with respect to the HST images for galaxies which are faint and even dropouts in $F160W$.

At $2<z<3$ some of the very dusty galaxies at $V-J\gtrsim2$ in Figure~\ref{fig:UVJ} do
indeed look like nearly edge-on disks with signs of dust lanes
visible in the NIRCam short-wavelength RGB images, while a smoother
morphology is observed in the long-wavelength bands.  Moreover, the
comparison between the JWST and HST images for these galaxies reveals
very striking differences, with the latter missing, in many cases, a
significant fraction of the outskirts of the galaxies and,
consequently, underestimating their true sizes (see next section on
parametric morphologies). 

At higher redshift, $3<z<4$, the short-wavelength JWST bands start to probe rest-frame UV of these
(red) dusty and quiescent galaxies, and consequently they can miss part of the
structure in the outer parts of the galaxies.  On the other hand, the
higher spatial resolution of the short-wavelength bands appears to identify one or more
clumps in some galaxies which are not clearly seen in the more smooth
appearance of the long-wavelength JWST bands. The rest-frame UV-dimming is even more extreme for the shallower HST bands which, for
some galaxies, can only detect a handful of pixels even in those previously
identified in the CANDELS catalog. 

Lastly, at $4<z<7$ galaxies become much fainter in the JWST short-wavelength bands 
and, in many cases, there is hardly anything to see on the images, with the exception of some quiescent galaxies that are still bright in $F200W$. In the long-wavelength bands the galaxies are clearly
detected but also faint and, in most cases, there are fewer distinct
structural features (i.e., disks, or clear differences
between the inner/outer parts of the galaxy) relative to what is seen at lower redshifts.

\subsubsection{Morphological parametric measurements}

\begin{figure*}   
    \includegraphics[width=18cm]{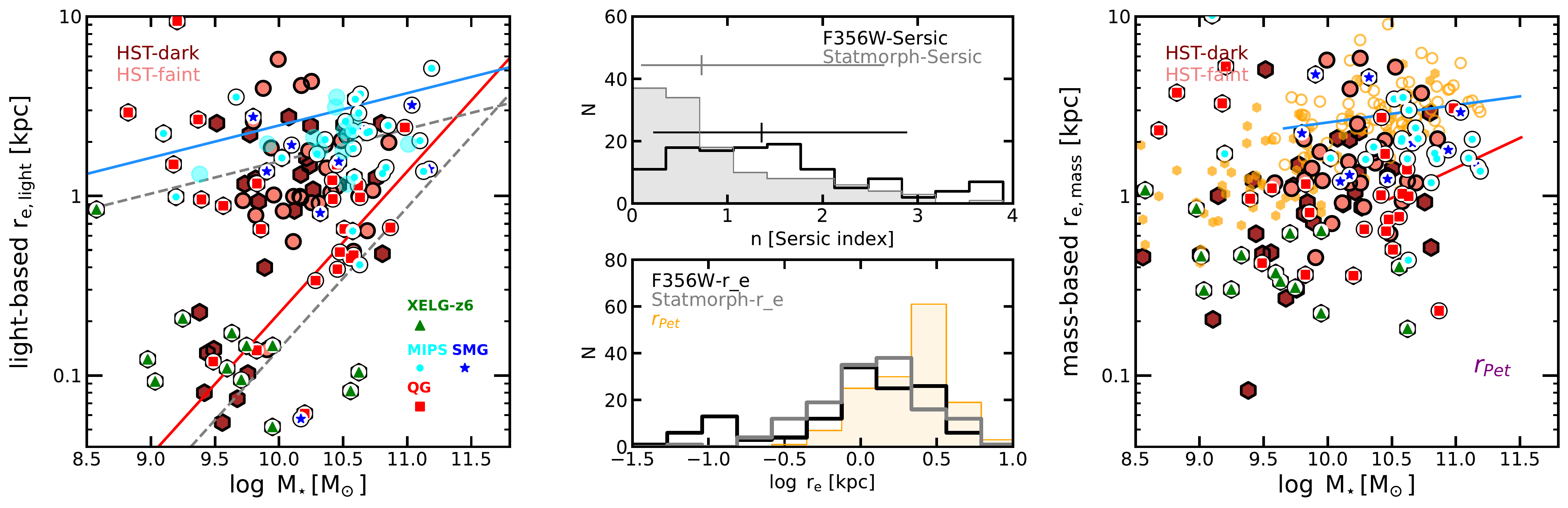} \\
    \includegraphics[width=18cm]{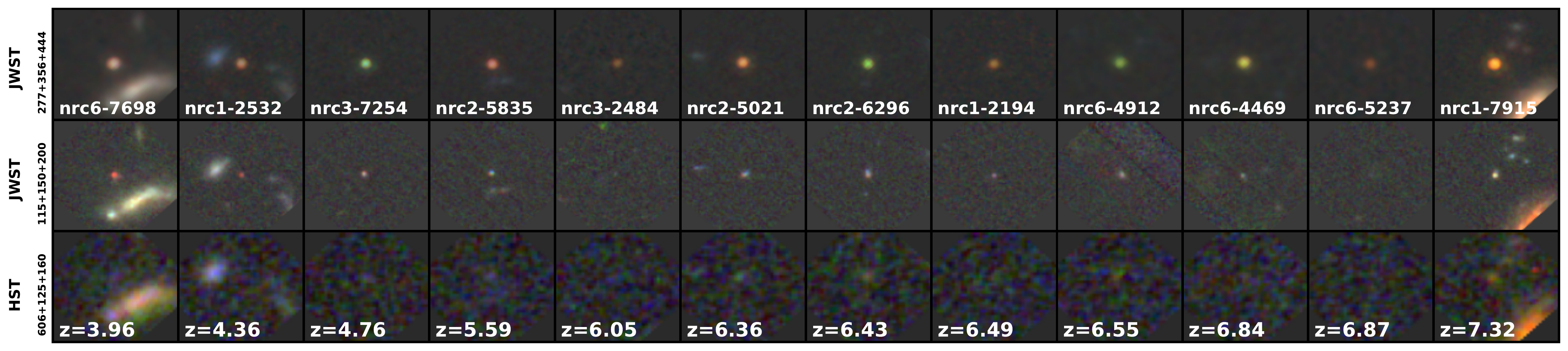} \\
    \caption{Morphological properties of our sample of near-IR faint, mid-IR bright galaxies. On the left panel, we show the mass-size plot for HST-dark galaxies (hexagons) and HST-faint sources (circles), compared to the CEERS sample (gray shaded density map). Sizes are represented by the effective radius of the Sérsic profile fit performed with the GALFIT software \citep{2002AJ....124..266P} on $F356W$ images. MIPS24 emitters are marked with a cyan circle, sub-mm and radio galaxies with a blue star, quiescent galaxies by sSFR with a red square, and XELG-$z6$ with green triangles. We depict the scaling relationships for star-forming and quiescent galaxies at z=2.75 (blue and red, respectively) from \citet{2014ApJ...788...28V}. In the middle panels, histograms for the Sérsic index (top) and effective radius (bottom) are provided. We show estimations based on $F356W$ images (light) as explained before, and on mass maps analyzed with the statmorph software \citep{2019MNRAS.483.4140R}. The middle bottom panel also show the distributions of \citet{1976ApJ...209L...1P} radii in mass maps. On the right panel, the mass-size relationship is again shown, this time based on mass maps analyses performed with statmorph. Orange points show Petrosian radii. The blue and red lines show the (r$_{e, mass}$ based) mass-size relations for star-forming galaxies and quiescent galaxies from \citet{2019ApJ...885L..22S}. The cutouts on the bottom show the 5\arcsec$\times$5\arcsec images of XELG-$z6$ galaxies and other similar objects (not identified explicitly as such) at redshifts ranging from $5<z<7$, also providing IDs and redshifts. From bottom-to-top, the images are color composites of 3 HST bands, and 3 JWST bands, short and long wavelengths, respectively.}
    \label{fig:mass-size}
\end{figure*}

Figure~\ref{fig:mass-size} shows the mass-size relation for the color-selected sample
and the bulk of the CEERS sample (density map).  The effective radii
($r_{e}$), measured along the major axis, and \citet{1968adga.book.....S} indices ($n$) were
determined from the $F150W$, $F200W$, and $F356W$ images using {\tt GALFIT} v3.0.5 \citep{2002AJ....124..266P}.
The code was run on the background-subtracted images, with the sky background held fixed at zero during fitting. Image cutouts which were fed to {\tt GALFIT} have dimensions 2.5 times the Kron radius. The ERR array, which includes background sky, Poisson, and read noise, was used as the input noise map. Empirical PSFs were constructed using stars in all CEERS pointings. All galaxies in the image cutout within 3 magnitudes of the primary source were fit simultaneously.  All other sources were masked out during fitting. Initial values for position, magnitude, effective radius, axis ratio, and position angle were set from the input photometry catalog.  In addition, the following constraints were applied while fitting to keep values for neighboring sources within reasonable bounds: S\'ersic index $0.2\le n \le 8.0$, effective radius $0.3 \le r_e \le 400\rm{~pixels}$, axis ratio  $0.01 \le q \le 1$, magnitude $\pm3$ mag from the initial value, and position $\pm3$ pixels from the initial value. The radii in Figure~\ref{fig:mass-size} are based on the reddest band ($F356W$) since many of the HST-dark
galaxies are quite faint in the bluer bands. The blue and red lines
show the location of the mass-size scaling relations for star-forming
and quiescent galaxies at $z=2.75$ based on HST/WFC3 data from \citet{2014ApJ...788...28V}.

Overall, the bulk of the sample appears to follow the mass-size relation
for star-forming galaxies, which is consistent with most of the sample
being classified as star-forming by either the sSFR or rest-frame
color criteria. More precisely, the color-selected sample has slightly
lower average sizes than the star-forming relation at $z=2.75$, i.e.,
they follow a relation with a lower zero-point. We compare the $F356W$
and $F150W$ radii for the galaxies well-detected in the bluest band (mostly
HST-faint galaxies) and we find that the sizes are systematically
smaller in $F356W$ by $r_{e,F150W}/r_{e,F356W}=1.7$. This agrees with the results of \citet{2014ApJ...788...28V} who also found smaller sizes
at longer wavelengths when comparing HST/ACS and WFC3 measurements. If
we account for this systematic effect by lowering the mass-size relation
for star-forming galaxies at $z=2.75$ (blue line) by 0.23~dex (dashed grey line), it
agrees well with the distribution of our sample.  This is not
surprising for the HST-faint galaxies with an average redshift of
$\langle z\rangle=3.2$, but it suggests that the HST-dark galaxies at higher
redshifts, $\langle z\rangle=4.2$, are not significantly smaller. I.e., we do not
find a strong size evolution for HST-dark galaxies and, in fact, many of them are quite large (see right panel of Figure and discussion below), as recently pointed out by \citet{2022arXiv220801630N} (blue circles in Figure~\ref{fig:mass-size}). We also find that the axis-ratio ($q$) distribution peaks at $q\sim0.6$ and ranges from $q=0.2$ to 0.8, as expected for a sample of mainly disks galaxies, and we verify that the objects in
\citet{2022arXiv220801630N} are indeed among those with the lowest axis-ratios.

As discussed in previous sections, the overall fraction of quiescent
galaxies in the sample is relatively small ($\sim$15\%). Among them,
the most massive, $\log \mathrm{M\!_\star/M}_{\odot}>10$, appear to
follow relatively closely the mass-size relation of quiescent galaxies
at $z=2.75$ whereas the least massive, $\log
\mathrm{M\!_\star/M}_{\odot}<10$, deviate from the relation towards
larger masses. Nonetheless, most of these tend to have, on average,
smaller sizes than star-forming galaxies of the same mass, as we would
expect.

The most significant difference between the HST-faint and -dark
samples is that the latter contains a population of galaxies with very
small sizes $r_{e,F356W}<0.25$~kpc and relatively small masses of
$\log\mathrm{M\!_\star/M}_{\odot}\lesssim9.5$. Roughly $\sim$60\% of
these galaxies are XELG-$z6$, which reinforces the idea
that they are indeed compact starbursts with strong emission lines
coming from small a nuclear region. But similar galaxies in terms of compact morphology and rest-frame UV colors, but at $z=4-6$, are also found in our sample. Many of these galaxies have GALFIT
fits with S\'ersic index that either $n>3$ (spheroidal) or
$n<0.2$. The latter suggest that the size estimates are more
uncertain, likely because the small sizes and, in many cases fainter
magnitudes, are close to the spatial-resolution limit of the F356W
images. Nonetheless, the visual inspection of the residuals does not
reveal any problems and their visual appearances are indeed very
small, featureless and compact (see cutouts in
Figure~\ref{fig:mass-size}). Interestingly, 3 of these peculiar
galaxies are quiescent by sSFR or $UVJ$ colors, and 2 more have low
sSFRs and are adjacent to the quiescent region of the $UVJ$ diagram
(i.e., post-SB like or dormant just starting a reignition).

The right panel of Figure~\ref{fig:mass-size} shows the mass-size relationship but this time obtained with statmorph \citep{2019MNRAS.483.4140R} measurements performed on the mass maps. We provide several measurements of the size of our galaxies based on stellar mass 2D distributions, namely, the Petrosian radius (which informs about the total extent of the sources down to the surface brightness detection limit), and the effective radius (commonly used in this kind of mass-size plot). As mentioned in Section~\ref{sec:photometry}, most of our galaxies are quite extended, with Petrosian radius extending from 1 to 10 kpc. The typical effective radius of our sample is 2~kpc, with HST-dark galaxies including a broader range of sizes and extending to smaller sizes than the HST-faint sample. The former follow a correlation which is steeper than what was obtained for $z\sim2.5$ by \citet{2019ApJ...885L..22S} for star-forming galaxies, and very similar to that for quiescent systems. 

Overall the mass-size distribution of our sample in light and mass are quite similar. The most prominent differences are a larger scatter in the mass-based estimations and the larger sizes found in the mass maps for the very compact objects mentioned in the previous paragraph. Those galaxies have Sérsic indices below $n=1$. To aid in the discussion of light-to-mass morphological estimations, the middle panel of Figure~\ref{fig:mass-size}  shows the distribution of S\'ersic index
values, which agrees with the results discussed above. First, we remark that the median (quartiles) ratio between effective and Petrosian radii is $1.8_{1.2}^{2.5}$. Second, the majority of the
color-selected galaxies are disk-like, both for the HST-faint and
-dark subsamples. The HST-dark galaxies have a tail of small S\'ersic index
values which are precisely the very small galaxies mentioned
above. The average S\'ersic index of the HST-dark is marginally higher and
extends to higher values than for the HST-faint galaxies. This is
expected because most of the quiescent galaxies are in that
sub-sample.

\subsection{Star Formation Histories}

\begin{figure*}   
    \includegraphics[clip, trim=0.8cm 0.cm 1.5cm 0.2cm,width=8.6cm]{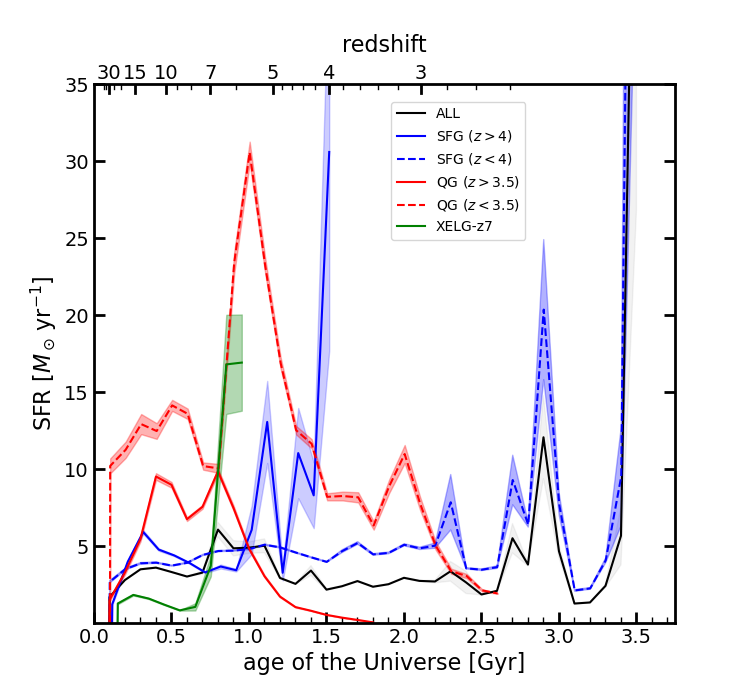}
    \includegraphics[clip, trim=2.6cm 0.1cm 13.2cm 0.7cm,width=9.0cm]{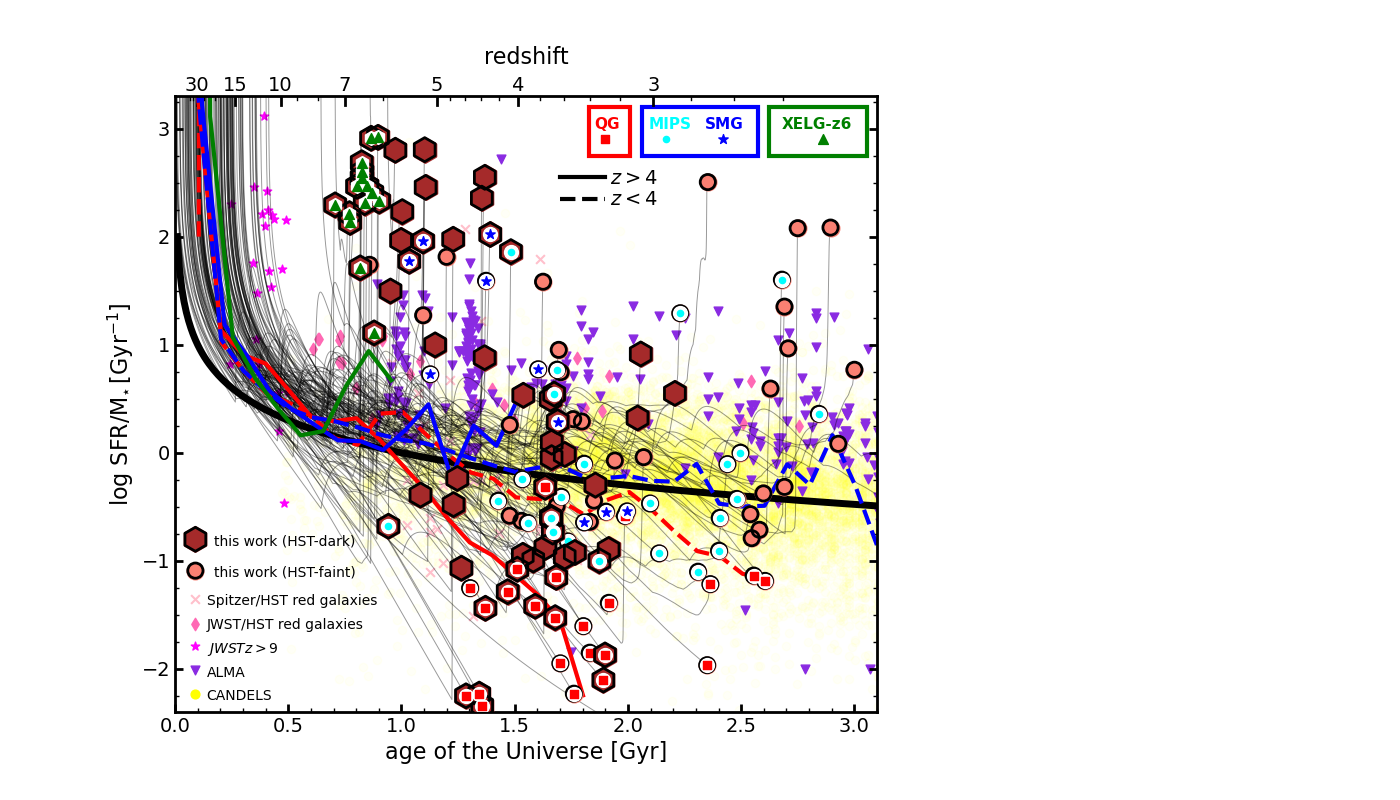}
    \caption{Left: Average and scatter of the SFHs (in terms of the age of the Universe and corresponding redshift) for all the galaxies in our sample (black), quiescent galaxies (red), and star-forming galaxies (blue). The latter types have been divided in two subsamples based on their redshifts, roughly with equal number of sources in each subgroup. We also depict the SFH for the XELGs-$z6$ type (green). Right: Gray lines show the evolution of the specific SFRs of our galaxies according to the 2D-based SFHs. The black-border symbols show the positions of our galaxies at the epoch of observation, hexagons for HST-dark and circles for HST-faint galaxies. Quiescent galaxies in our sample are marked with red squares, dusty star-forming galaxies detected by MIPS and (sub-)mm surveys with cyan circles and blue stars, respectively, and XELG-$z6$ sources with a green triangle. The black thick line corresponds to inverse specific SFRs (i.e, doubling times) equal to the age of the Universe. The average tracks for types of galaxies shown in the left panel are also shown. Comparison samples are depicted in color, using literature estimations of the redshift, stellar mass, and SFR. Yellow points correspond to the CANDELS catalog. Purple pointing-down triangles depict ALMA-detected galaxies \citep{2016ApJ...820...83S,2018MNRAS.481.1631B,2020ApJ...901...74T,2022A&A...664A..73B,2022ApJ...928...31S}. Orange crosses show near-IR faint, mid-IR bright sources (i.e., red galaxies) discovered in the {\it Spitzer}/HST era \citep{2011ApJ...742L..13H,2016ApJ...816...84W,2019ApJ...876..135A}. Magenta diamonds depict the high-z red (and tentatively massive) galaxies discovered by JWST in comparison with HST data \citep{2022arXiv220714733B,2022arXiv220800986C,2022arXiv220712446L,2022arXiv220801630N}. And pink stars correspond to $z>9$ JWST-selected galaxies \citep{2022arXiv220709434N,2022arXiv220712474F,2022arXiv220802825R,2022arXiv220711379S,2022MNRAS.511.4464A}.}
    \label{fig:sfh_ssfr}
\end{figure*}

Figure~\ref{fig:sfh_ssfr} shows the average global SFHs derived for the galaxies in our sample divided by star formation activity, namely, QGs, SFGs and XELGs-$z6$. Global SFHs were computed by summing up the measurements from all pixels in a galaxy (each one fitted with a delayed exponentially decaying parametrization). This SFH was scaled up to reproduce the total stellar mass of the object by multiplying by the average (using all photometric bands) flux aperture correction obtained from the comparison of the sum of pixels arriving to our 2D stellar population analysis and the integrated-light elliptical aperture. After that, we transformed all SFHs to the common time frame of the age of the Universe. Then, we averaged the functions in 100~Myr time intervals using all galaxies lying at a redshift below that corresponding to the given Universe age. In order to make the average SFH curves more meaningful (in terms of combining coetaneous galaxies with similar natures), we divided the SFG and QG samples in two regimes, one at lower redshifts, $z<4$ and $z<3.5$, respectively, and one at higher, with roughly the same number of galaxies in each bin. 

The average SFH of our sample (black line) is roughly constant, at a 4~M$_\odot$\,yr$^{-1}$ level, from $z\sim30$ down to $z\sim3$, when star formation bursts occur peaking 3-10 times above the more steady previous state. This translates to forming around $10^{10}\,\mathrm{M}_\odot$ of stars above $z>3$, in 2~Gyr, and an extra 30\% of that mass in more recent starbursts. Our sample is thus dominated by (dusty) a population of star-forming galaxies which smoothly (when averaging their properties) assemble their mass until there is an intense star-forming event with high dust content.

If we restrict the analysis to SFGs, a similar SFH is observed for the $z<4$ sample (which dominates the sample), with a slightly larger average level. For $z>4$ SFGs, we see a steady state at a similar level, 5~M$_\odot$\,yr$^{-1}$, followed by a starburst at $z=4-6$, with peak SFRs 6 times larger than the previous plateau. We note that these starbursts are averaged out when considering the full sample, which implies that they must be short and stochastic from galaxy to galaxy.

The combined interpretation for the SFHs we derive for SFGs point to a relatively constant SFR for around 1~Gyr broken by a first major starburst event beyond $z=4-6$ (maybe only in some galaxies), and another one below $z<3$. The SMGs in our sample are caught in the first burst, implying that the first stage with less extreme SFRs and lasting for 1~Gyr was capable of producing enough amounts of metals to feed the dusty starburst revealed by the sub-mm/mm observations.

The left panel of Figure~\ref{fig:sfh_ssfr} also shows the SFH for the QGs in our sample, again divided by redshift ($z<3.5$ and $z>3.5$ to have similar numbers of galaxies in each subsample). For the $z<3.5$ QGs, after some constant activity at a 10-15~M$_\odot$\,yr$^{-1}$ level, larger than the value observed for other galaxy types, we see a very prominent star formation peak, reaching 5 times larger SFRs values, around 30~M$_\odot$\,yr$^{-1}$, occurring at $z=5-7$, at a similar epoch when high-z SFGs present their first peak. Then the SFH decays very rapidly (a factor of 4 in 500~Myr) and establishes at the 5-10~M$_\odot$\,yr$^{-1}$ level for 1 more Gyr.

For high-z QGs the SFH peak occurs even before, $z\sim7-10$, it is not as marked, decays slightly more rapidly (300-400~Myr) and reaches almost null activity afterwards. We note that the high-z and low-z subsamples of QGs have different median masses (see Table~\ref{tab:stats}), which explain the average levels of the SFH in Figure~\ref{fig:sfh_ssfr} (also true for other subsamples).

Finally, we also show in this panel the SFH for XELG-$z6$, which start at a very low level and increase very rapidly by a factor of 8 at $z\sim6-7$, similarly to the behavior observed for $z<3.5$ QGs and at a very similar epoch.

The right panel of Figure~\ref{fig:sfh_ssfr} shows the evolution of galaxies in a sSFR vs age of the Universe plot. We remark that the tracks followed by galaxies roughly align with the line representing a mass-doubling time equal to the age of the Universe (shown in black). This means that, on average, the SFH of galaxies transit through a steady star formation phase (the constant SFR region in the left panel) for several hundreds of Myr (even up to 1~Gyr). Indeed, the CANDELS sample (shown in yellow) nicely concentrates around that line. At $z\gtrsim10$, however, our SFHs imply very quick mass-doubling times, which points out to a very active phase in the young Universe, consistent with the relatively high number of very high-z sources being detected in JWST surveys, several times above expectations (cf. Figure~5 in \citealt{2022arXiv220712474F} and pink stars in our plot). We must warn the reader that the behavior of the SFHs at very high-z can be affected by the very non-linear relationship between time and redshift.  By $z\sim7$ we start distinguishing the individual galaxies, which move to a starburst phase, a quiescent state, or remain in the mass-doubling line.

\begin{figure*}   
    \includegraphics[clip, trim=3.2cm 0.0cm 0.0cm 0.0cm,width=20cm]{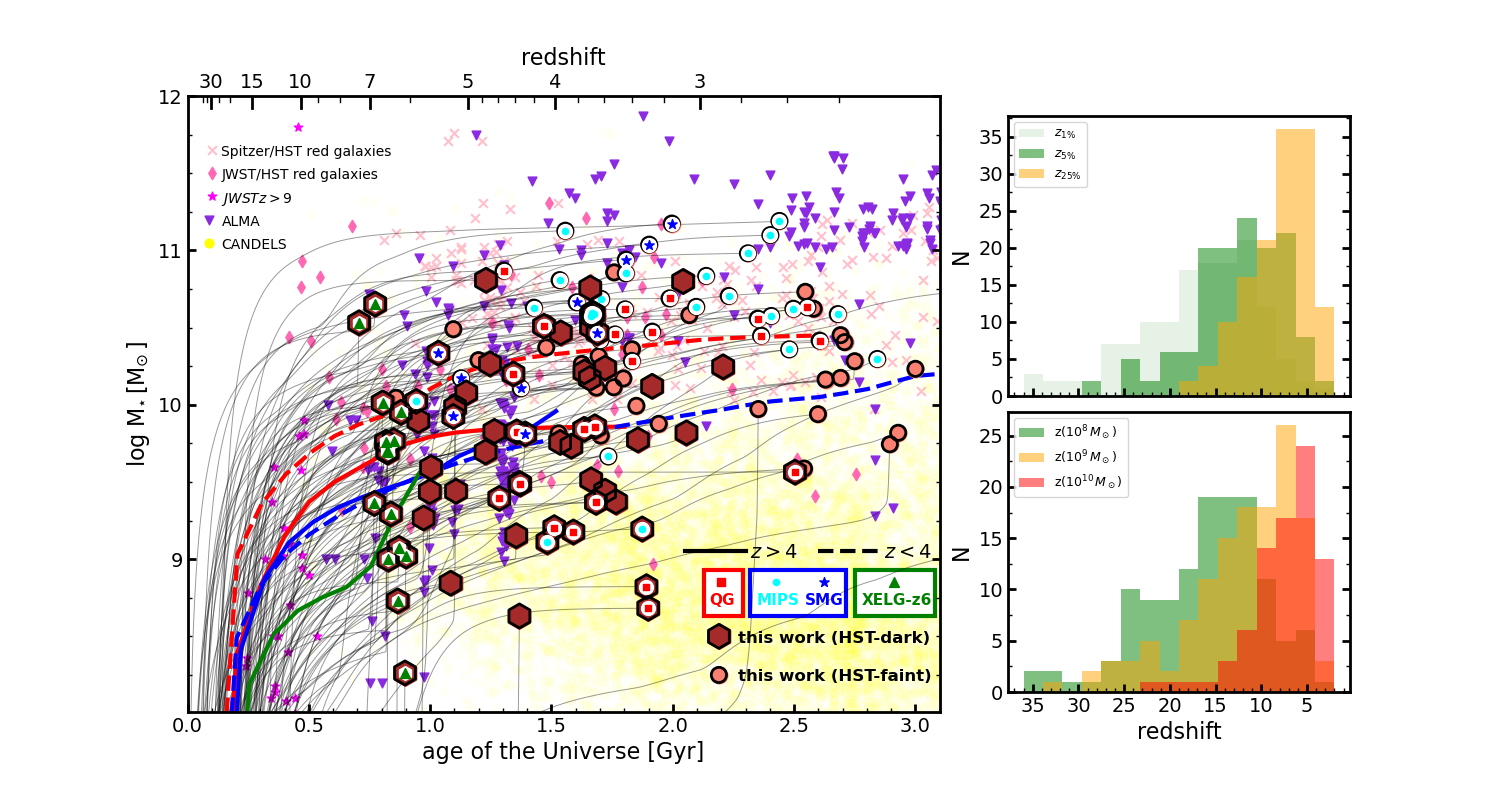}
    \caption{Left: Gray lines show the evolution of the stellar mass of our galaxies according to the 2D-based SFHs. The black-border symbols show the positions of our galaxies at the epoch of observation, hexagons for HST-dark and circles for HST-faint galaxies. Quiescent galaxies in our sample are marked with a red square, dusty star-forming galaxies detected by MIPS and (sub-)mm surveys  with cyan circles and blue stars, respectively, and XELG-$z6$ sources with a green triangle. The average tracks for quiescent galaxies, dusty starburst (both types divided by redshift, with lower/higher redshifts depicted with dashed/continuous lines), and XELG-$z6$ are also shown. Comparison samples are also depicted, using literature estimations of the redshift and the stellar mass, and with the same colors mentioned in Figure~\ref{fig:sfh_ssfr}. On the top-right panel, we show the distribution of redshifts where our galaxies have formed 1, 5, and 25\% of their observed stellar mass. On the bottom-right panel, we show the histograms of redshifts where our galaxies have already formed 10$^{8,9,10}$~M$_{\odot}$ in stellar mass. }
    \label{fig:masstracks}
\end{figure*}

The right panel of Figure~\ref{fig:sfh_ssfr} also shows some interesting galaxy samples to help understand the nature of our sources. Apart from the general CANDELS sample shown in yellow, which follows the mass-doubling equal to Universe age line and the SFH tracks plotted in the figure, we also depict the galaxies detected with HST vs IRAC colors and presented in papers such as \citet{2016ApJ...816...84W} or \citet{2019ApJ...876..135A}, with the SFR and mass (quite uncertain due to data limitations prior to JWST) estimation they were able to assign. Most of these galaxies present high specific SFRs from redshifts $z=2-6$, similar to some of our dusty starbursts. Another less numerous sub-population presents specific SFRs compatible with quiescent systems, extending even at higher redshifts than our quiescent galaxy candidates.

Another interesting sample of galaxies we include in the right panel of Figure~\ref{fig:sfh_ssfr} are ALMA sources (see caption for references). Especially interesting are those at $4<z<6$, which present similar specific SFR values to some of our $z>4$ SFGs (some confirmed SMGs) and XELG-$z6$. The SMGs in our sample and the ALMA sources concentrate around the epoch where our QGs experienced their last intense starburst event before quenching.

Finally, we also include on the right panel of Figure~\ref{fig:sfh_ssfr} the high redshift massive galaxies reported by \citet{2022arXiv220712446L} and the $z>9$ galaxy candidates presented by several groups based on the first analyses of the first JWST datasets \citep{2022arXiv220709434N,2022arXiv220712474F}. In both cases, some of our SFHs would be able to reproduce their properties, but the bulk of our galaxies would be linked to progenitors presenting smaller specific SFRs than these high-z sources, implying also that we are just starting to probe the tip of the iceberg of $z>9$ populations.  

Figure~\ref{fig:masstracks} shows, on the left panel, the evolution of the stellar masses of our galaxies as a function of the Universe age, jointly with the same comparison samples depicted in the previous figure. Virtually all of our sources experience a very quick mass assembly at $z\gtrsim10$, followed by a less pronounced mass increment, very similar for all galaxies except for very few of them quickly increasing their mass by more than 1~dex below $z=7$, the XELG-$z6$.

The HST-faint and HST-dark galaxies selected in this paper correspond to a high mass end of the distribution of masses and redshifts of the CANDELS sample. Many of our galaxies have stellar masses similar to those of HST-dark systems selected with IRAC vs$.$ HST data (pink crosses), but those samples seem to be biased towards massive systems, $\log\mathrm{M\!_\star/M}_{\odot}>10$. This might be linked to limitations in the area of our CEERS survey, not comparable (yet) to the datasets used by those previous works. It might also be linked to confusion problems in the IRAC observations, since many of our galaxies are indeed surrounded by neighbors and only the JWST data have now been able to separate them in the mid-IR. And, obviously, IRAC samples are biased towards brighter objects, compared to the new JWST data.

Compared to ALMA sources, our sample present similar stellar masses at $z>4$. Probably the selections are converging in this epoch. We also remark that some of the ALMA sources at $7<z<10$ could well be progenitors of our lower redshift galaxies, because they lie within the locus occupied by our SFHs.

It is worth noticing that the sample of XELG-$z6$ presents an average track implying that they are quickly assembling a significant fraction of their stellar mass at $z\sim7$.

The last 2 samples we compare with are \citet{2022arXiv220712446L} massive high-z galaxies and $z>9$ JWST galaxy candidates \citep{2022arXiv220709434N,2022arXiv220712474F}. The former significantly deviates from our SFHs, they would seem to correspond more, should their stellar mass and redshift be accurate, to progenitors of very massive HST-dark galaxies which we have not yet detected with JWST, maybe due to survey area limitations. In fact, for the galaxies in that paper that we have in our sample, we obtain similar redshifts but up to 10$\times$ smaller stellar masses.
The masses and redshifts for the $z>9$ samples compare well with the tracks followed by many of our SFHs, although they typically lie around 1~dex below the average SFH for our samples, implying that we are not seeing comparable populations. Furthermore, $z>9$ detections would be more compatible with the $\log\mathrm{M\!_\star/M}_{\odot}<10$, $3<z<5$ galaxies in the CANDELS or our sample (including quiescent objects). 

The right panels of Figure~\ref{fig:masstracks} present the histograms of formation redshifts of our sample. On the bottom panel, we show when our red galaxies reached masses $10^{8-10}$~M$_\odot$. Typically, $10^{8}$~M$_\odot$ of stars were assembled by $z\sim15$, with a large scatter. Our galaxies formed their first $10^{9}$~M$_\odot$ by $z\sim12$ (on average), and reached masses as high as $10^{10}$~M$_\odot$ at $z=7$. Complementary, we also show the histograms of formation redshifts for 5\%, 10\%, 25\% of the current stellar mass of the galaxies in our sample. Overall, our SFHs, especially those for quiescent galaxies, imply a very active Universe beyond $z>10$ and up to $z\sim25$, with some quite massive galaxies ($\log\mathrm{M\!_\star/M}_{\odot}>10$) at even $z=15-20$ (or, quite probably, several progenitors to merge later). We remark that all this star formation activity can occur {\it in situ}  or {\it ex situ} in distinct progenitors at high redshifts.

\begin{figure*}   
    \includegraphics[clip, trim=3.1cm 3.3cm 3.7cm 5.0cm,width=18cm]{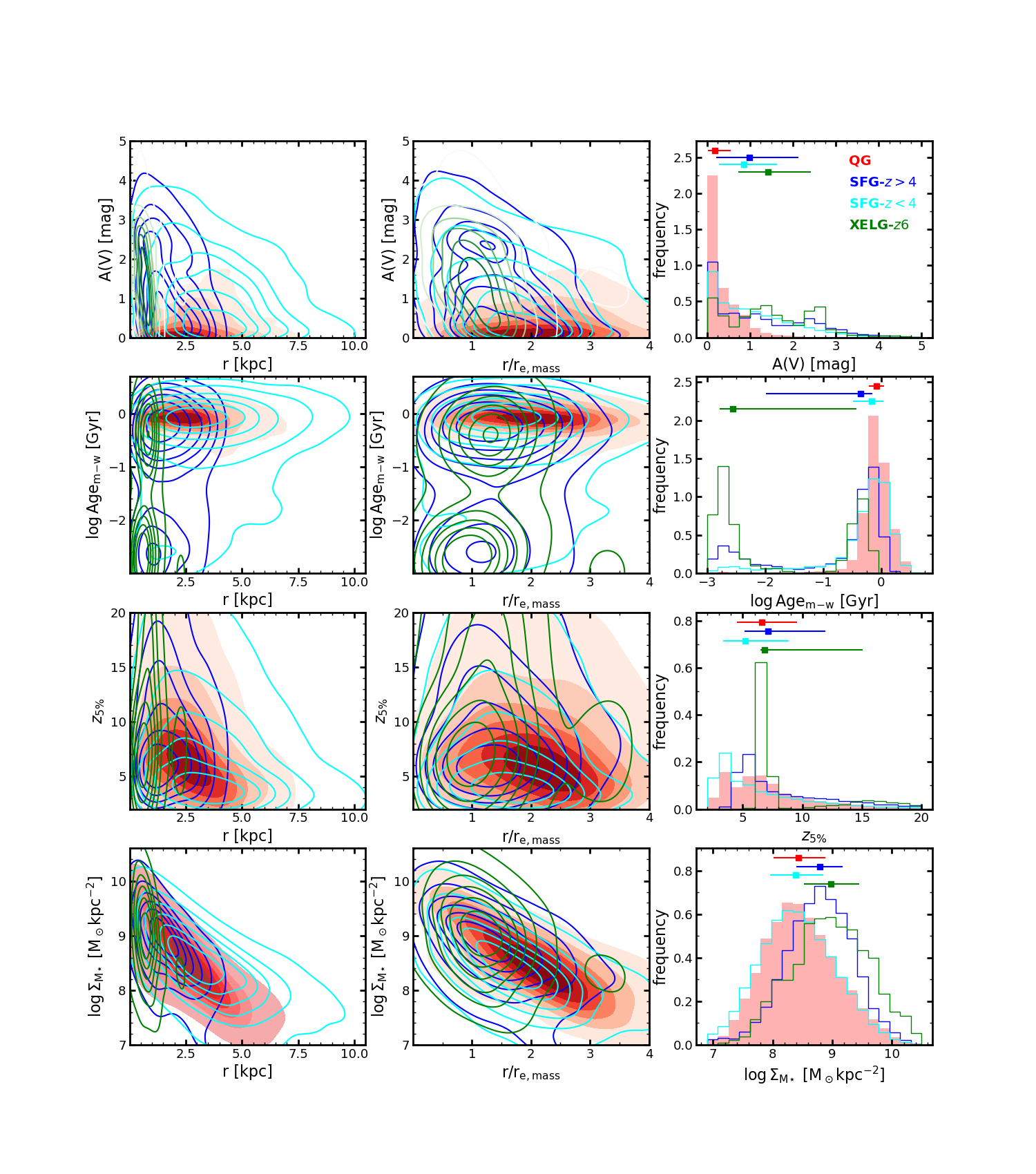}
    \caption{Radial distribution of the main stellar population properties for our sample of HST-faint and HST-dark galaxies, divided into 4 samples: quiescent galaxies (QGs, red), star-forming galaxies at $z<4$ and $z>4$ (in cyan and blue) and XELG-$z6$ sources (green). From top to bottom, we show attenuations in the $V$-band, mass-weighted ages, redshift at which each pixel formed 5\% of its current stellar mass, and stellar mass surface density. In the left column we plot these properties as a function of absolute galactocentric distances in kpc. In the middle-column panels, we normalize the distances using the effective radius calculated in the stellar mass maps. On the right column, we show histograms of all properties, jointly with median and quartiles.}
    \label{fig:2dsps}
\end{figure*}

\subsection{Spatially resolved analysis of the stellar populations}

In this section, we focus on the analysis of how the red galaxies in our sample have assembled, investigating the stellar mass density 2D distribution, the ages of stellar populations present in different parts of the galaxy, and the location of the dusty starbursts that many of our sources present. 

Figure~\ref{fig:2dsps} shows the distribution of the stellar population properties for all pixels in our analysis as a function of galactocentric distance (in absolute units and relative to the effective radius) and for different galaxy subsamples. Median values and quartiles are marked and provided in Table~\ref{tab:stats}. We note three interesting results from this figure.

First, focusing on the top-right panel, the $z>4$ SFGs (including high-z SMGs) subsample (depicted in blue) is identified with the highest dust attenuations, typically $\mathrm{A(V)}>2$~mag. For $z<4$ SFGs, there is a tail of systems harboring regions (pixels) with similarly large dust contents, many of them belong to the MIPS24 emitters in our sample. We remark that the sub-mm and mm surveys in the CEERS field are heterogeneous in area coverage and depth, so our $z>4$ SFG sample could still include additional SMGs. Distinctively, quiescent galaxies are composed by regions with $\mathrm{A(V)}<1$~mag dust attenuations, the median being around 0.3~mag. The regions with the highest attenuations are preferentially located in the inner 2~kpc, or $2\times r_{e, M}$, with a significant gradient towards larger radii, similar to what has been reported at lower redshifts \citep[see][]{2017MNRAS.469.4063W,2022arXiv220912954M}, but with larger nuclear attenuations in the case of SMGs \citep[compared to main sequence galaxies at $z\sim2$; see, e.g.,][]{2018ApJ...859...56T}. XELG-$z6$ present a quite flat attenuation distribution, which extends from unobscured to A(V)$=$3~mag, i.e., the starburst we see as an emission line is also dusty, but these are very compact sources, as we mentioned earlier.

The second result we highlight is that all the galaxies in our sample present evolved stellar populations in terms of mass-weighted age. The frequency plots for QGs and SFGs peak at similar ages, around 1~Gyr, although the distribution for quiescent galaxies is skewed to older values, the contrary can be said for SFGs. The distribution of mass-weighted ages for regions in $z>4$ SFGs is different to that for $z<4$ SFGs, the former tend to have more regions with ages between 1 and 10~Myr. Most of these very young regions are in the inner 2.5~kpc, within $1.5\times r_{e, M}$. Even though the histograms of mass-weighted ages for QGs and SFGs (especially those at the lowest redshifts) peak at similar values, the latter starting forming earlier, there is a significant number of pixels (dominating in number but not in mass) which had already formed 5\% of their current mass at $5<z<10$. This contrasts with the properties of XELG-$z6$, most of their pixels are assembling their mass quickly at $z=6-7$.

The last result we extract from Figure~\ref{fig:2dsps} refers to the stellar mass surface density. QGs and $z<4$ SFGs present very similar distributions at the dense end, at values higher than  $10^{8.5}$~M$_\odot$\,kpc$^{-2}$. Relatively, the regions with the highest densities, above $10^{9.5}$~M$_\odot$\,kpc$^{-2}$, are more common in quiescent galaxies and at distances within 2~kpc or one effective radius. We see a tail at the high-density end for $z>4$ SFGs and XELG-$z6$, with larger values than those observed by QGs, but this is probably an artifact introduced by the different redshifts and spatial resolutions of the subsample. The two higher redshift subsamples would be more comparable, and XELG-$z6$ present more pixels with higher densities, compared to $z>4$ SFGs.

\section{Summary and conclusions}
\label{sec:summary}

We have presented a novel approach to the study of galaxies at cosmological distances with the combination of JWST and HST data. Using their superb sensitivities and spatial resolution along a wide spectral region covering the optical, near- and mid-infrared, we analyze spectral energy distributions in a pixel-by-pixel basis for a sample of 138 mid-IR selected ($F356W<27.5$~mag), HST/near-IR faint ($F160W>24$~mag), red ($F150W-F356W>1.5$~mag) galaxies, some of them qualifying as what have been commonly known as HST-dark galaxies, but also extending previous surveys of optical/near-IR dark galaxies to fainter mid-IR magnitudes. Our innovative technique provides several independent estimations for the photometric redshift of each individual galaxy, which can also be used to help with the segmentation of JWST images. In addition, we carry out a 2-dimensional analysis of the stellar populations in this type of galaxies, with promising results about the identification of highly dust-obscured star-forming knots within galaxies, a typical limitation of stellar population synthesis modeling of broad-band data. This method reveals for the first time the integrated and spatially resolved properties of HST-dark and JWST-only galaxies, and improve those of previously detected HST-faint galaxies.

Our JWST-based color-magnitude selection reveals the existence of evolved, relatively massive galaxies with redshifts $2\lesssim z\lesssim7$, with a triality of natures. (1) Massive, dusty star-forming galaxies (SFGs) with redshifts ranging from $z\sim2$ up to $z\sim6$, with dust emission detected in mid-to-far-IR preferentially below $z<3$, and at (sub-)millimeter, and/or radio wavelengths at higher redshifts. This type dominates our sample, they add up 71\% of all our selected red galaxies. (2) Quiescent or dormant (i.e., subject to reignition at later epochs, QGs) galaxies between $z=3$ and $z=5$, accounting for 18\% of our sample. And the remaining 11\% of galaxies forming (3) a population of very compact galaxies with red colors in the rest-frame optical, blue $F356W-F410W$ colors probably arising from high-EW [OIII]$\lambda4959,5007$ and H$\beta$ emission-lines (partly explaining their selection by our $F150W-F356W$ color cut), and a UV upturn, presenting very high specific SFRs and lying at $6<z<7$ (and possibly at redshifts as low as $z\sim5$), and thus being identified as extreme emission-line galaxies (of starburst or AGN origin), XELG-$z6$.

The existence of all these types of galaxies implies high levels of star formation activity in redshifts as high to $z=10-25$, where most of our galaxies had already formed ({\it in situ} or {\it ex situ}) $10^8$~M$_\odot$, more than half $10^9$~M$_\odot$, and a few up to $10^{10}$~M$_\odot$,.

The SFGs in our selection present different properties at $z<4$ and $z>4$. The lower-redshift  galaxies are larger than the higher redshift sources, with relatively less area experiencing a dusty (up to $\mathrm{A(V)}=5$~mag) starburst, and they also expand to higher stellar masses.  Therefore, they are more evolved systems in terms of stellar populations and morphology, which is dominated by a disk with dusty star-forming knots concentrating in the inner 2~kpc. In terms of the SFH, all SFGs present secular epochs with SFRs around 5-10~M$_\odot$\,yr$^{-1}$ broken by a starburst producing 100~Myr-averaged SFRs 8-10 times larger than those values.

The QGs in our sample are relatively young, as expected by the age of the Universe at the time of observation, and are caught in a post-starburst phase around 1~Gyr after a star formation event with up to 10 times higher SFRs than the earliest evolutionary stages. These starburst occur at  $5<z<7$ for $z<3.5$ QGS, where we have some dusty SFGs with instantaneous SFRs above 100~M\,yr$^{-1}$.

High redshift ($z>4$) SFGs could be linked to low-redshift  ($z<3.5$) QGs in terms of SFH, sizes, and stellar mass density profiles. The dusty SFGs could also be linked to XELG-$z6$ galaxies regarding the peak of star formation, but the secular SFR level is lower for the latter, which might imply that they do not have a progenitor-descendant relationship or mergers play a significant role at $z>7$ to increase the secular level for low-z QGs.  

\begin{acknowledgements}
We would like to thank Prof. Giulia Rodighiero for carefully and very constructively reviewing the manuscript. PGP-G acknowledges support  from  Spanish  Ministerio  de  Ciencia e Innovaci\'on MCIN/AEI/10.13039/501100011033 through grant PGC2018-093499-B-I00. \'AGA acknowledges the support of the Universidad Complutense de Madrid through the predoctoral grant CT17/17-CT18/17. This work has made use of the Rainbow Cosmological Surveys Database, which is operated by the Centro de Astrobiología (CAB), CSIC-INTA, partnered with the University of California Observatories at Santa Cruz (UCO/Lick, UCSC). This work is based on observations carried out under project number W20CK with the IRAM NOEMA Interferometer. IRAM is supported by INSU/CNRS (France), MPG (Germany) and IGN (Spain).
\end{acknowledgements}

\software{astropy \citep{2022ApJ...935..167A}, EAZY \citep{2008ApJ...686.1503B}, GALFIT  \citep{2002AJ....124..266P}, matplotlib \citep{2007CSE.....9...90H}, NumPy \citep{2011CSE....13b..22V}, photutils \citep{2022zndo...6825092B}, PZETA (\citealt{2005ApJ...630...82P}, \citealt{2008ApJ...675..234P}), Rainbow pipeline (\citealt{2005ApJ...630...82P}, \citealt{2008ApJ...675..234P}, \citealt{2011ApJS..193...30B}), SciPy \citep{2020NatMe..17..261V}, SExtractor \citep{1996A&AS..117..393B}, Synthesizer (\citealt{2005ApJ...630...82P}, \citealt{2008ApJ...675..234P}}).

The data used in this paper come from the CEERS project and are available and documented at \url{ceers.github.io/dr05.html}
and at MAST via \dataset[10.17909/z7p0-8481]{\doi{10.17909/z7p0-8481}}. See \citet{2022arXiv221102495B} for details. 

\startlongtable
\begin{deluxetable*}{lcccccrcl}
\centerwidetable 
\tabletypesize{\scriptsize}
\tablehead{\colhead{ID} & \colhead{RA (J2000)} & \colhead{DEC (J2000)} & \colhead{$F356W$} & \colhead{$F150W-F356W$} & \colhead{z} & \colhead{$\log\mathrm{M\!_\star}$} & \colhead{$\log\mathrm{SFR(10~Myr)}$} & \colhead{comment}\\
 & [degrees] & [degrees] & [mag] & [mag] & & [M$_\odot$] & [M$_\odot$\,yr$^{-1}$] & }
\startdata
nircam1-349    &  214.95571202 &  +52.98342646 & 27.31$\pm$0.06 & $>$ 2.77 & $ 6.24_{-0.02}^{+0.03}$ &  9.01$\pm$0.09 & -0.29$\pm$0.10 & XELG-$z6$\\
nircam1-1085   &  214.95787499 &  +52.98030148 & 23.31$\pm$0.03 &  1.91$\pm$0.04 & $ 3.52_{-0.06}^{+0.04}$ & 10.46$\pm$0.03 & -0.76$\pm$0.03 & QG\\
nircam1-1623   &  214.92577403 &  +52.95443735 & 24.79$\pm$0.03 &  2.11$\pm$0.05 & $ 3.07_{-0.13}^{+0.06}$ &  9.82$\pm$0.03 &  1.47$\pm$0.06 & SFG\\
nircam1-1744   &  215.01127115 &  +53.01358970 & 24.79$\pm$0.03 &  1.78$\pm$0.04 & $ 5.17_{-0.00}^{+0.00}$ &  9.96$\pm$0.06 &  2.79$\pm$0.05 & SFG\\
nircam1-1752   &  215.01221476 &  +53.01469376 & 24.72$\pm$0.03 &  1.56$\pm$0.04 & $ 3.62_{-0.13}^{+0.03}$ &  9.80$\pm$0.03 &  1.53$\pm$0.03 & SFG\\
nircam1-1867   &  214.99840491 &  +53.00461910 & 25.95$\pm$0.03 &  2.03$\pm$0.06 & $ 6.52_{-0.08}^{+0.03}$ &  9.63$\pm$0.10 &  2.07$\pm$0.04 & XELG-$z6$\\
nircam1-2080   &  214.98181546 &  +52.99123394 & 22.32$\pm$0.03 &  1.72$\pm$0.04 & $ 3.17_{-0.24}^{+0.18}$ & 10.69$\pm$0.01 &  0.52$\pm$0.06 & QG\\
nircam1-2194   &  214.95683242 &  +52.97315320 & 26.52$\pm$0.04 &  1.26$\pm$0.06 & $ 6.49_{-0.11}^{+1.00}$ &  9.03$\pm$0.26 &  1.05$\pm$0.01 & XELG-$z6$\\
nircam1-2532   &  214.95083827 &  +52.96686330 & 26.26$\pm$0.04 &  2.29$\pm$0.06 & $ 4.36_{-0.18}^{+0.13}$ &  9.49$\pm$0.07 &  1.30$\pm$0.02 & SFG\\
nircam1-2800   &  214.94918535 &  +52.96414043 & 26.20$\pm$0.04 &  1.64$\pm$0.08 & $ 6.08_{-0.04}^{+0.08}$ &  8.98$\pm$0.15 &  0.94$\pm$0.04 & XELG-$z6$\\
nircam1-3117   &  214.94696940 &  +52.96026712 & 23.11$\pm$0.03 &  1.36$\pm$0.04 & $ 2.40_{-0.12}^{+0.12}$ & 10.17$\pm$0.02 &  0.71$\pm$0.03 & SFG\\
nircam1-3454   &  214.98870870 &  +52.98862085 & 24.58$\pm$0.03 &  1.83$\pm$0.04 & $ 3.38_{-0.08}^{+0.07}$ &  9.99$\pm$0.03 &  0.56$\pm$0.01 & SFG\\
nircam1-4762   &  214.99777939 &  +52.98612084 & 22.91$\pm$0.03 &  2.65$\pm$0.04 & $ 2.72_{-0.11}^{+0.07}$ &  9.91$\pm$0.02 &  2.45$\pm$0.05 & SFG\\
nircam1-5091   &  214.99703630 &  +52.98372380 & 25.08$\pm$0.03 &  1.28$\pm$0.05 & $ 3.86_{-0.61}^{+0.33}$ &  9.73$\pm$0.04 & -0.29$\pm$0.01 & SFG\\
nircam1-5244   &  214.99782799 &  +52.98284270 & 21.96$\pm$0.03 &  1.78$\pm$0.04 & $ 2.40_{-0.23}^{+0.19}$ & 10.58$\pm$0.02 &  2.65$\pm$0.01 & SFG, MIPS24\\
nircam1-5666   &  214.94574593 &  +52.94394633 & 26.47$\pm$0.06 &  1.79$\pm$0.08 & $ 3.84_{-0.13}^{+0.09}$ &  9.18$\pm$0.02 & -1.26$\pm$0.03 & QG ($UVJ$-SFG)\\
nircam1-5800   &  214.97244326 &  +52.96219591 & 26.74$\pm$0.05 &  3.14$\pm$0.19 & $ 5.61_{-0.22}^{+0.25}$ &  9.42$\pm$0.24 &  2.09$\pm$0.03 & SFG\\
nircam1-5988   &  215.00455757 &  +52.98352355 & 23.67$\pm$0.03 &  3.02$\pm$0.05 & $ 3.71_{-0.18}^{+0.16}$ & 10.76$\pm$0.02 &  1.14$\pm$0.01 & SFG\\
nircam1-5996   &  215.03028056 &  +53.00187923 & 27.45$\pm$0.06 &  1.55$\pm$0.10 & $ 5.26_{-0.30}^{+0.17}$ &  8.84$\pm$0.14 & -0.53$\pm$0.01 & SFG\\
nircam1-6588   &  215.02153586 &  +52.99129837 & 22.52$\pm$0.03 &  2.19$\pm$0.04 & $ 2.86_{-0.18}^{+0.31}$ & 10.70$\pm$0.01 &  2.80$\pm$0.02 & SFG, MIPS24\\
nircam1-6842   &  215.03905021 &  +53.00277847 & 23.61$\pm$0.03 &  2.75$\pm$0.05 & $ 4.11_{-0.12}^{+0.12}$ & 10.51$\pm$0.01 &  0.06$\pm$0.04 & QG\\
nircam1-7369   &  215.00848883 &  +52.97797283 & 26.28$\pm$0.04 &  1.66$\pm$0.06 & $ 6.51_{-0.12}^{+1.30}$ &  9.59$\pm$0.05 &  0.82$\pm$0.02 & XELG-$z6$\\
nircam1-7723   &  214.97746629 &  +52.95349959 & 23.20$\pm$0.03 &  2.01$\pm$0.04 & $ 3.02_{-0.27}^{+0.20}$ & 10.63$\pm$0.02 &  1.16$\pm$0.04 & SFG, MIPS24\\
nircam1-7915   &  214.98303471 &  +52.95600639 & 24.50$\pm$0.03 &  1.54$\pm$0.04 & $ 7.32_{-0.09}^{+0.12}$ & 10.56$\pm$0.04 &  3.05$\pm$0.06 & XELG-$z6$\\
nircam1-8059   &  215.03681713 &  +52.99349730 & 23.70$\pm$0.03 &  1.44$\pm$0.04 & $ 2.21_{-0.15}^{+0.08}$ &  9.82$\pm$0.02 &  0.89$\pm$0.05 & SFG\\
nircam1-8677   &  215.02290766 &  +52.98006507 & 24.15$\pm$0.03 &  1.88$\pm$0.05 & $ 3.46_{-0.06}^{+0.09}$ & 10.17$\pm$0.04 &  1.37$\pm$0.04 & SFG\\
nircam1-8955   &  214.97926146 &  +52.94752785 & 24.90$\pm$0.03 &  1.71$\pm$0.05 & $ 4.70_{-0.06}^{+0.07}$ & 10.27$\pm$0.08 &  1.03$\pm$0.02 & SFG\\
nircam2-214    &  214.85208109 &  +52.90975593 & 22.87$\pm$0.03 &  1.56$\pm$0.04 & $ 2.27_{-0.14}^{+0.11}$ & 10.30$\pm$0.01 &  1.58$\pm$0.01 & SFG, MIPS24\\
nircam2-284    &  214.90311270 &  +52.94573042 & 23.99$\pm$0.03 &  1.63$\pm$0.04 & $ 5.21_{-0.02}^{+0.02}$ & 10.49$\pm$0.02 &  2.33$\pm$0.05 & SFG\\
nircam2-561    &  214.83510724 &  +52.89512113 & 25.48$\pm$0.04 &  1.70$\pm$0.06 & $ 3.97_{-0.05}^{+0.07}$ &  9.81$\pm$0.05 &  0.19$\pm$0.06 & SFG\\
nircam2-686    &  214.82930597 &  +52.89392478 & 21.87$\pm$0.03 &  2.33$\pm$0.04 & $ 2.67_{-0.17}^{+0.45}$ & 11.10$\pm$0.01 &  1.19$\pm$0.07 & SFG, MIPS24\\
nircam2-956    &  214.89189908 &  +52.93386575 & 25.07$\pm$0.03 &  2.48$\pm$0.05 & $ 4.29_{-0.41}^{+0.30}$ &  9.80$\pm$0.02 &  2.26$\pm$0.06 & SFG, SMG\\
nircam2-1196   &  214.91555380 &  +52.94901642 & 22.79$\pm$0.03 &  2.38$\pm$0.04 & $ 4.54_{-0.11}^{+0.08}$ & 10.87$\pm$0.05 &  0.62$\pm$0.02 & QG\\
nircam2-1199   &  214.91534933 &  +52.94867041 & 26.60$\pm$0.06 &  2.29$\pm$0.11 & $ 4.01_{-0.19}^{+0.07}$ &  9.20$\pm$0.41 & -0.89$\pm$0.06 & QG ($UVJ$-SFG)\\
nircam2-1423   &  214.91613181 &  +52.95189811 & 25.87$\pm$0.03 &  4.57$\pm$0.34 & $ 5.10_{-0.16}^{+0.11}$ & 10.17$\pm$0.10 &  1.14$\pm$0.04 & SFG ($UVJ$-QG), SMG\\
nircam2-2014   &  214.91379957 &  +52.94301951 & 22.97$\pm$0.03 &  1.43$\pm$0.04 & $ 2.88_{-0.50}^{+0.58}$ & 10.25$\pm$0.03 &  1.74$\pm$0.05 & SFG\\
nircam2-2134   &  214.92321780 &  +52.94911492 & 22.36$\pm$0.03 &  1.72$\pm$0.04 & $ 2.38_{-0.07}^{+0.14}$ & 10.40$\pm$0.01 &  2.08$\pm$0.03 & SFG\\
nircam2-2159   &  214.91454789 &  +52.94302256 & 26.55$\pm$0.05 &  2.12$\pm$0.11 & $ 4.59_{-0.03}^{+0.03}$ &  9.39$\pm$0.06 & -1.86$\pm$0.04 & QG\\
nircam2-2225   &  214.90485307 &  +52.93534900 & 23.00$\pm$0.03 &  1.72$\pm$0.04 & $ 3.27_{-0.18}^{+0.08}$ & 10.47$\pm$0.02 &  0.01$\pm$0.06 & QG\\
nircam2-2420   &  214.87518851 &  +52.91347925 & 24.22$\pm$0.03 &  1.63$\pm$0.04 & $ 3.53_{-0.34}^{+0.14}$ & 10.11$\pm$0.03 &  0.23$\pm$0.06 & SFG, X-ray\\
nircam2-2494   &  214.90911350 &  +52.93720417 & 26.42$\pm$0.04 &  4.16$\pm$0.40 & $ 5.22_{-0.13}^{+0.16}$ &  9.90$\pm$0.08 &  2.29$\pm$0.02 & SFG, SMG\\
nircam2-2624   &  214.90403232 &  +52.93270268 & 22.68$\pm$0.03 &  2.09$\pm$0.04 & $ 2.52_{-0.06}^{+0.12}$ & 10.63$\pm$0.01 &  0.49$\pm$0.02 & QG ($UVJ$-SFG)\\
nircam2-2998   &  214.92575938 &  +52.94566152 & 27.02$\pm$0.05 &  2.43$\pm$0.19 & $ 6.64_{-0.05}^{+0.04}$ &  9.95$\pm$0.13 &  2.54$\pm$0.06 & XELG-$z6$\\
nircam2-3078   &  214.88067652 &  +52.91296239 & 23.30$\pm$0.03 &  2.43$\pm$0.04 & $ 3.82_{-0.20}^{+0.20}$ & 10.66$\pm$0.02 &  1.09$\pm$0.01 & SFG, SMG\\
nircam2-3340   &  214.91838475 &  +52.93789089 & 24.29$\pm$0.03 &  2.30$\pm$0.05 & $ 3.59_{-0.37}^{+0.30}$ & 10.24$\pm$0.02 &  0.27$\pm$0.03 & SFG\\
nircam2-3545   &  214.85845902 &  +52.89259341 & 22.42$\pm$0.03 &  1.56$\pm$0.04 & $ 2.15_{-0.18}^{+0.19}$ & 10.23$\pm$0.01 &  1.89$\pm$0.01 & SFG\\
nircam2-3880   &  214.84828097 &  +52.88477584 & 26.18$\pm$0.04 &  2.02$\pm$0.06 & $ 3.66_{-0.22}^{+0.24}$ &  9.37$\pm$0.04 & -0.77$\pm$0.04 & QG ($UVJ$-SFG)\\
nircam2-3882   &  214.84854191 &  +52.88475632 & 27.14$\pm$0.06 &  2.35$\pm$0.09 & $ 4.08_{-0.08}^{+0.09}$ &  9.10$\pm$0.02 &  1.09$\pm$0.06 & SFG, MIPS24\\
nircam2-4231   &  214.90496370 &  +52.92240948 & 22.47$\pm$0.03 &  1.95$\pm$0.04 & $ 3.45_{-0.18}^{+0.14}$ & 10.94$\pm$0.03 &  1.29$\pm$0.01 & SFG, SMG\\
nircam2-4241   &  214.93696001 &  +52.94546314 & 22.54$\pm$0.03 &  1.85$\pm$0.04 & $ 2.40_{-0.12}^{+0.15}$ & 10.45$\pm$0.01 &  2.24$\pm$0.04 & SFG\\
nircam2-4830   &  214.93239676 &  +52.93825496 & 24.75$\pm$0.03 &  1.27$\pm$0.05 & $ 3.37_{-0.37}^{+0.27}$ &  9.77$\pm$0.05 &  0.47$\pm$0.01 & SFG\\
nircam2-5021   &  214.92724232 &  +52.93389331 & 25.55$\pm$0.03 &  1.17$\pm$0.05 & $ 6.36_{-0.12}^{+0.11}$ &  9.71$\pm$0.05 &  1.69$\pm$0.05 & XELG-$z6$\\
nircam2-5835   &  214.92337531 &  +52.92558836 & 26.21$\pm$0.03 &  1.43$\pm$0.05 & $ 5.59_{-0.00}^{+0.00}$ &  9.56$\pm$0.19 &  2.38$\pm$0.06 & SFG\\
nircam2-5870   &  214.86604164 &  +52.88408333 & 22.85$\pm$0.03 &  2.16$\pm$0.04 & $ 3.45_{-0.03}^{+0.03}$ & 10.62$\pm$0.02 &  0.03$\pm$0.05 & QG\\
nircam2-5966   &  214.86577570 &  +52.88341678 & 24.69$\pm$0.03 &  1.77$\pm$0.05 & $ 3.23_{-0.14}^{+0.15}$ &  9.88$\pm$0.02 &  0.58$\pm$0.05 & SFG\\
nircam2-6108   &  214.86704198 &  +52.88327391 & 23.47$\pm$0.03 &  1.61$\pm$0.04 & $ 3.40_{-0.25}^{+0.10}$ & 10.36$\pm$0.03 &  0.72$\pm$0.04 & SFG\\
nircam2-6296   &  214.86015469 &  +52.87742886 & 25.26$\pm$0.03 &  1.57$\pm$0.05 & $ 6.43_{-0.04}^{+0.03}$ &  9.25$\pm$0.08 & -0.08$\pm$0.02 & XELG-$z6$\\
nircam2-6729   &  214.87909561 &  +52.88805949 & 23.58$\pm$0.03 &  1.79$\pm$0.04 & $ 3.40_{-0.09}^{+0.05}$ & 10.28$\pm$0.03 & -0.56$\pm$0.01 & QG\\
nircam2-6980   &  214.92458649 &  +52.91868782 & 24.50$\pm$0.03 &  1.45$\pm$0.05 & $ 6.31_{-0.16}^{+0.09}$ & 10.03$\pm$0.08 &  0.49$\pm$0.05 & SFG\\
nircam2-7077   &  214.88906695 &  +52.89261021 & 23.96$\pm$0.03 &  2.92$\pm$0.05 & $ 3.65_{-0.12}^{+0.06}$ & 10.47$\pm$0.03 &  1.74$\pm$0.01 & SFG, SMG\\
nircam2-7122   &  214.92575863 &  +52.91852482 & 23.87$\pm$0.03 &  1.96$\pm$0.04 & $ 3.70_{-0.15}^{+0.11}$ & 10.50$\pm$0.02 &  1.44$\pm$0.01 & SFG\\
nircam2-7245   &  214.93157042 &  +52.92100293 & 25.50$\pm$0.03 &  3.07$\pm$0.08 & $ 5.02_{-0.40}^{+0.35}$ & 10.08$\pm$0.06 &  1.45$\pm$0.05 & SFG\\
nircam2-7343   &  214.94404404 &  +52.92973616 & 25.58$\pm$0.03 &  2.35$\pm$0.05 & $ 5.87_{-0.28}^{+0.15}$ & 10.02$\pm$0.04 &  0.37$\pm$0.06 & SFG, MIPS24\\
nircam2-7884   &  214.92542772 &  +52.91339370 & 23.77$\pm$0.03 &  1.51$\pm$0.04 & $ 2.48_{-0.05}^{+0.13}$ &  9.94$\pm$0.01 &  0.50$\pm$0.02 & SFG\\
nircam2-8436   &  214.93750634 &  +52.91828492 & 24.19$\pm$0.03 &  2.08$\pm$0.05 & $ 3.27_{-0.46}^{+0.99}$ & 10.12$\pm$0.06 &  0.21$\pm$0.04 & SFG\\
nircam3-53     &  214.74342038 &  +52.83687143 & 24.64$\pm$0.03 &  1.48$\pm$0.04 & $ 2.53_{-0.25}^{+0.07}$ &  9.59$\pm$0.01 & -0.01$\pm$0.04 & SFG\\
nircam3-54     &  214.74373107 &  +52.83682239 & 22.09$\pm$0.03 &  0.98$\pm$0.04 & $ 2.53_{-0.31}^{+0.48}$ & 10.73$\pm$0.03 &  0.91$\pm$0.01 & SFG\\
nircam3-371    &  214.82802002 &  +52.89411029 & 22.60$\pm$0.03 &  1.42$\pm$0.04 & $ 3.06_{-0.44}^{+0.67}$ & 10.58$\pm$0.02 &  1.20$\pm$0.01 & SFG ($UVJ$-QG)\\
nircam3-447    &  214.76339852 &  +52.84779402 & 23.66$\pm$0.03 &  2.05$\pm$0.04 & $ 4.09_{-0.41}^{+0.10}$ & 10.53$\pm$0.07 &  0.76$\pm$0.01 & SFG\\
nircam3-448    &  214.76324045 &  +52.84758887 & 27.21$\pm$0.04 &  1.31$\pm$0.06 & $ 3.31_{-0.06}^{+0.09}$ &  8.82$\pm$0.09 & -2.24$\pm$0.06 & QG ($UVJ$-SFG)\\
nircam3-938    &  214.81987894 &  +52.88487041 & 22.55$\pm$0.03 &  1.42$\pm$0.04 & $ 2.35_{-0.18}^{+0.21}$ & 10.25$\pm$0.02 &  2.12$\pm$0.01 & SFG\\
nircam3-1208   &  214.76316750 &  +52.84287412 & 23.18$\pm$0.03 &  1.35$\pm$0.04 & $ 2.23_{-0.33}^{+0.40}$ &  9.72$\pm$0.02 &  2.41$\pm$0.03 & SFG\\
nircam3-2288   &  214.81194745 &  +52.87090095 & 21.39$\pm$0.03 &  2.02$\pm$0.04 & $ 2.63_{-0.07}^{+0.15}$ & 11.19$\pm$0.01 &  1.99$\pm$0.02 & SFG, MIPS24\\
nircam3-2460   &  214.75981826 &  +52.83341487 & 26.24$\pm$0.04 & $>$ 3.54 & $ 4.35_{-0.03}^{+0.07}$ &  9.49$\pm$0.09 & -0.93$\pm$0.01 & QG\\
nircam3-2477   &  214.83281702 &  +52.88538218 & 27.25$\pm$0.05 &  2.02$\pm$0.13 & $ 6.11_{-0.02}^{+0.06}$ &  8.34$\pm$0.22 & -2.13$\pm$0.03 & XELG-$z6$\\
nircam3-2484   &  214.80914162 &  +52.86848352 & 27.53$\pm$0.06 &  2.39$\pm$0.13 & $ 5.72_{-0.34}^{+0.50}$ &  9.14$\pm$0.07 &  1.33$\pm$0.05 & SFG\\
nircam3-2514   &  214.77551871 &  +52.84408284 & 22.66$\pm$0.03 &  1.56$\pm$0.04 & $ 1.81_{-0.02}^{+0.08}$ &  9.79$\pm$0.02 &  1.73$\pm$0.01 & SFG\\
nircam3-3009   &  214.75584820 &  +52.82740774 & 26.91$\pm$0.12 &  0.92$\pm$0.13 & $ 4.39_{-0.33}^{+0.35}$ &  9.10$\pm$0.12 &  0.36$\pm$0.01 & SFG ($UVJ$-QG)\\
nircam3-3140   &  214.83841666 &  +52.88518348 & 25.62$\pm$0.03 &  2.06$\pm$0.05 & $ 5.83_{-0.03}^{+0.07}$ &  9.89$\pm$0.06 &  1.28$\pm$0.05 & SFG\\
nircam3-3141   &  214.83855378 &  +52.88520163 & 25.44$\pm$0.03 &  1.72$\pm$0.05 & $ 4.65_{-0.03}^{+0.02}$ &  9.83$\pm$0.04 & -0.21$\pm$0.02 & SFG\\
nircam3-3534   &  214.76657456 &  +52.83152404 & 25.88$\pm$0.03 &  3.17$\pm$0.13 & $ 4.39_{-0.05}^{+0.07}$ &  9.83$\pm$0.14 & -1.49$\pm$0.05 & QG\\
nircam3-3596   &  214.84318104 &  +52.88575245 & 27.48$\pm$0.09 &  3.20$\pm$0.36 & $ 3.30_{-0.04}^{+0.07}$ &  8.68$\pm$0.08 & -2.16$\pm$0.06 & QG ($UVJ$-SFG)\\
nircam3-4118   &  214.77922118 &  +52.83692199 & 27.59$\pm$0.05 &  2.09$\pm$0.16 & $ 4.36_{-0.10}^{+0.10}$ &  8.56$\pm$0.12 &  1.23$\pm$0.01 & SFG\\
nircam3-4206   &  214.81093951 &  +52.85892847 & 25.39$\pm$0.04 &  1.39$\pm$0.05 & $ 2.57_{-0.10}^{+0.10}$ &  9.56$\pm$0.08 & -2.59$\pm$0.04 & QG\\
nircam3-4209   &  214.81118070 &  +52.85864777 & 23.42$\pm$0.03 &  1.87$\pm$0.04 & $ 3.62_{-0.15}^{+0.04}$ & 10.69$\pm$0.05 &  1.26$\pm$0.05 & SFG, MIPS24\\
nircam3-4383   &  214.83570899 &  +52.87530939 & 22.97$\pm$0.03 &  1.55$\pm$0.04 & $ 2.71_{-0.24}^{+0.35}$ & 10.45$\pm$0.01 &  0.18$\pm$0.05 & QG ($UVJ$-SFG)\\
nircam3-4452   &  214.81315440 &  +52.85891894 & 23.85$\pm$0.03 &  1.75$\pm$0.04 & $ 3.77_{-0.13}^{+0.27}$ & 10.26$\pm$0.05 &  2.16$\pm$0.02 & SFG\\
nircam3-4584   &  214.79037102 &  +52.84192419 & 23.73$\pm$0.03 &  1.74$\pm$0.04 & $ 4.09_{-0.09}^{+0.07}$ & 10.37$\pm$0.06 &  1.13$\pm$0.01 & SFG\\
nircam3-4833   &  214.83685539 &  +52.87344971 & 22.91$\pm$0.03 &  1.58$\pm$0.04 & $ 3.65_{-0.29}^{+0.06}$ & 10.58$\pm$0.03 &  0.43$\pm$0.06 & QG\\
nircam3-5269   &  214.79149020 &  +52.83803602 & 22.08$\pm$0.03 &  2.39$\pm$0.04 & $ 2.77_{-0.11}^{+0.14}$ & 10.98$\pm$0.03 &  0.84$\pm$0.01 & SFG, MIPS24\\
nircam3-5415   &  214.83969255 &  +52.87172622 & 24.75$\pm$0.03 &  1.71$\pm$0.05 & $ 3.67_{-0.23}^{+0.05}$ &  9.86$\pm$0.05 & -0.64$\pm$0.01 & QG ($UVJ$-SFG)\\
nircam3-5755   &  214.79439771 &  +52.83740251 & 25.82$\pm$0.03 &  1.36$\pm$0.05 & $ 5.17_{-0.01}^{+0.00}$ &  9.38$\pm$0.05 &  2.12$\pm$0.06 & SFG\\
nircam3-5772   &  214.76722852 &  +52.81771089 & 23.43$\pm$0.03 &  1.76$\pm$0.04 & $ 3.63_{-0.12}^{+0.03}$ & 10.42$\pm$0.08 & -0.51$\pm$0.06 & QG\\
nircam3-6044   &  214.76785862 &  +52.81627127 & 25.89$\pm$0.03 &  2.54$\pm$0.06 & $ 3.70_{-0.04}^{+0.11}$ &  9.51$\pm$0.06 &  0.60$\pm$0.01 & SFG\\
nircam3-6045   &  214.76812426 &  +52.81651073 & 23.32$\pm$0.03 &  2.19$\pm$0.05 & $ 3.65_{-0.03}^{+0.08}$ & 10.52$\pm$0.03 &  2.20$\pm$0.02 & SFG, MIPS24\\
nircam3-6046   &  214.76802062 &  +52.81639750 & 23.24$\pm$0.03 &  2.18$\pm$0.05 & $ 3.68_{-0.05}^{+0.12}$ & 10.59$\pm$0.03 &  2.04$\pm$0.03 & SFG, MIPS24\\
nircam3-6606   &  214.78568915 &  +52.82581893 & 23.03$\pm$0.03 &  1.99$\pm$0.04 & $ 3.96_{-0.29}^{+0.18}$ & 10.81$\pm$0.04 &  1.54$\pm$0.04 & SFG, MIPS24\\
nircam3-7014   &  214.85908668 &  +52.87659995 & 24.25$\pm$0.03 &  1.58$\pm$0.04 & $ 4.85_{-0.42}^{+0.86}$ & 10.28$\pm$0.05 &  1.43$\pm$0.03 & SFG\\
nircam3-7254   &  214.84004223 &  +52.86064444 & 25.34$\pm$0.03 &  1.78$\pm$0.05 & $ 4.76_{-0.01}^{+0.00}$ &  9.67$\pm$0.05 &  0.97$\pm$0.04 & SFG\\
nircam3-7357   &  214.85057802 &  +52.86602095 & 23.19$\pm$0.03 &  2.54$\pm$0.04 & $ 4.21_{-0.04}^{+0.04}$ & 10.63$\pm$0.06 &  1.02$\pm$0.05 & SFG ($UVJ$-QG), MIPS24, X-ray\\
nircam3-8070   &  214.80816330 &  +52.83221638 & 24.85$\pm$0.03 &  2.74$\pm$0.05 & $ 4.42_{-0.12}^{+0.13}$ & 10.20$\pm$0.12 & -1.01$\pm$0.03 & QG\\
nircam3-8344   &  214.86450533 &  +52.87096418 & 25.50$\pm$0.04 &  2.08$\pm$0.06 & $ 5.19_{-0.08}^{+0.04}$ &  9.85$\pm$0.03 &  2.13$\pm$0.04 & SFG\\
nircam3-8492   &  214.76280838 &  +52.85127861 & 24.86$\pm$0.03 &  1.95$\pm$0.05 & $ 3.75_{-0.12}^{+0.21}$ &  9.84$\pm$0.09 & -0.60$\pm$0.01 & QG\\
nircam3-8553   &  214.80012546 &  +52.82321302 & 25.22$\pm$0.03 &  1.65$\pm$0.04 & $ 3.57_{-0.12}^{+0.03}$ &  9.67$\pm$0.04 & -0.13$\pm$0.04 & SFG, MIPS24\\
nircam3-8554   &  214.80023743 &  +52.82344678 & 22.19$\pm$0.03 &  1.41$\pm$0.04 & $ 3.09_{-0.40}^{+0.55}$ & 10.80$\pm$0.05 &  2.05$\pm$0.05 & SFG\\
nircam3-8577   &  214.85389989 &  +52.86135553 & 22.09$\pm$0.03 &  3.03$\pm$0.04 & $ 3.91_{-0.29}^{+0.19}$ & 11.13$\pm$0.02 &  1.44$\pm$0.04 & SFG ($UVJ$-QG), MIPS24, X-ray\\
nircam3-8604   &  214.80007515 &  +52.82337289 & 23.80$\pm$0.03 &  1.59$\pm$0.04 & $ 3.64_{-0.12}^{+0.01}$ & 10.31$\pm$0.05 &  2.17$\pm$0.03 & SFG\\
nircam3-8745   &  214.79996808 &  +52.82209154 & 24.06$\pm$0.03 &  1.70$\pm$0.04 & $ 3.66_{-0.03}^{+0.06}$ & 10.11$\pm$0.06 &  0.14$\pm$0.05 & SFG\\
nircam6-254    &  214.84759938 &  +52.85338126 & 25.71$\pm$0.03 &  3.66$\pm$0.19 & $ 5.46_{-0.21}^{+0.24}$ & 10.32$\pm$0.08 &  2.06$\pm$0.04 & SFG, SMG\\
nircam6-382    &  214.85883785 &  +52.86039354 & 24.89$\pm$0.03 &  2.58$\pm$0.05 & $ 4.34_{-0.14}^{+0.27}$ & 10.10$\pm$0.03 &  1.85$\pm$0.01 & SFG, SMG\\
nircam6-456    &  214.85982893 &  +52.86065303 & 22.88$\pm$0.03 &  2.06$\pm$0.04 & $ 3.45_{-0.20}^{+0.18}$ & 10.85$\pm$0.02 &  1.75$\pm$0.01 & SFG, MIPS24\\
nircam6-971    &  214.85629956 &  +52.85463226 & 25.89$\pm$0.04 &  2.73$\pm$0.08 & $ 3.52_{-0.10}^{+0.10}$ &  9.37$\pm$0.03 & -0.53$\pm$0.02 & SFG ($UVJ$-QG)\\
nircam6-972    &  214.85573017 &  +52.85461870 & 26.33$\pm$0.04 &  2.23$\pm$0.06 & $ 3.34_{-0.04}^{+0.03}$ &  9.20$\pm$0.01 & -0.76$\pm$0.01 & SFG ($UVJ$-QG), MIPS24\\
nircam6-973    &  214.85589007 &  +52.85467042 & 23.58$\pm$0.03 &  3.48$\pm$0.06 & $ 3.70_{-0.17}^{+0.13}$ & 10.57$\pm$0.02 &  0.97$\pm$0.02 & SFG ($UVJ$-QG), MIPS24\\
nircam6-981    &  214.85604193 &  +52.85467234 & 23.58$\pm$0.03 &  3.54$\pm$0.07 & $ 3.69_{-0.18}^{+0.13}$ & 10.60$\pm$0.03 &  0.86$\pm$0.01 & SFG, MIPS24\\
nircam6-1802   &  214.85546356 &  +52.84878066 & 22.20$\pm$0.03 &  2.05$\pm$0.04 & $ 3.29_{-0.18}^{+0.20}$ & 11.04$\pm$0.02 &  1.43$\pm$0.05 & SFG, SMG\\
nircam6-1822   &  214.85079817 &  +52.84544028 & 22.29$\pm$0.03 &  1.73$\pm$0.04 & $ 2.49_{-0.16}^{+0.10}$ & 10.62$\pm$0.01 &  0.90$\pm$0.01 & SFG\\
nircam6-2101   &  214.83488079 &  +52.83239114 & 23.04$\pm$0.03 &  1.44$\pm$0.04 & $ 2.45_{-0.11}^{+0.16}$ & 10.16$\pm$0.02 &  1.58$\pm$0.02 & SFG\\
nircam6-2426   &  214.87693944 &  +52.86038901 & 22.74$\pm$0.03 &  2.04$\pm$0.04 & $ 2.67_{-0.06}^{+0.07}$ & 10.58$\pm$0.01 &  0.96$\pm$0.05 & SFG, MIPS24\\
nircam6-2696   &  214.82773439 &  +52.82376844 & 22.61$\pm$0.03 &  1.45$\pm$0.04 & $ 2.72_{-0.13}^{+0.13}$ & 10.56$\pm$0.03 & -0.41$\pm$0.07 & QG\\
nircam6-2884   &  214.80996584 &  +52.80975670 & 22.99$\pm$0.03 &  2.13$\pm$0.04 & $ 3.53_{-0.30}^{+0.13}$ & 10.86$\pm$0.05 &  2.16$\pm$0.03 & SFG\\
nircam6-3383   &  214.82965538 &  +52.82078225 & 24.08$\pm$0.03 &  2.14$\pm$0.05 & $ 3.95_{-0.20}^{+0.13}$ & 10.47$\pm$0.06 &  1.82$\pm$0.06 & SFG\\
nircam6-3588   &  214.86242636 &  +52.84290647 & 23.19$\pm$0.03 &  1.89$\pm$0.04 & $ 2.59_{-0.08}^{+0.06}$ & 10.36$\pm$0.03 &  0.92$\pm$0.06 & SFG, MIPS24\\
nircam6-3646   &  214.87628882 &  +52.85234566 & 22.08$\pm$0.03 &  2.80$\pm$0.04 & $ 3.16_{-0.18}^{+0.21}$ & 11.17$\pm$0.02 &  1.59$\pm$0.02 & SFG, SMG\\
nircam6-3974   &  214.84033934 &  +52.82495509 & 22.65$\pm$0.03 &  2.39$\pm$0.04 & $ 2.97_{-0.35}^{+0.34}$ & 10.83$\pm$0.01 &  0.91$\pm$0.06 & SFG, MIPS24\\
nircam6-4083   &  214.87066778 &  +52.84610677 & 24.43$\pm$0.03 &  2.65$\pm$0.05 & $ 3.71_{-0.06}^{+0.07}$ & 10.17$\pm$0.03 &  1.59$\pm$0.02 & SFG\\
nircam6-4429   &  214.83771989 &  +52.82004573 & 24.16$\pm$0.03 &  6.58$\pm$0.40 & $ 4.75_{-0.04}^{+0.06}$ & 10.81$\pm$0.26 &  0.27$\pm$0.05 & SFG\\
nircam6-4469   &  214.88679853 &  +52.85537678 & 25.02$\pm$0.03 &  2.17$\pm$0.05 & $ 6.84_{-0.03}^{+0.01}$ & 10.62$\pm$0.12 &  2.48$\pm$0.03 & XELG-$z6$\\
nircam6-4912   &  214.89247872 &  +52.85688618 & 25.63$\pm$0.03 &  1.57$\pm$0.05 & $ 6.55_{-0.09}^{+0.04}$ &  9.75$\pm$0.07 &  1.58$\pm$0.04 & XELG-$z6$\\
nircam6-5237   &  214.84054178 &  +52.81794349 & 27.11$\pm$0.05 &  0.86$\pm$0.07 & $ 6.87_{-0.28}^{+0.14}$ &  9.33$\pm$0.42 &  2.25$\pm$0.03 & XELG-$z6$\\
nircam6-6199   &  214.88712291 &  +52.84537265 & 26.00$\pm$0.04 &  2.24$\pm$0.09 & $ 3.59_{-0.26}^{+0.06}$ &  9.44$\pm$0.04 &  0.42$\pm$0.02 & SFG\\
nircam6-6454   &  214.87176632 &  +52.83316727 & 26.37$\pm$0.05 &  1.38$\pm$0.06 & $ 6.28_{-0.15}^{+0.15}$ &  8.58$\pm$0.16 &  1.58$\pm$0.01 & XELG-$z6$\\
nircam6-6547   &  214.89670433 &  +52.84979232 & 22.79$\pm$0.03 &  1.55$\pm$0.04 & $ 2.47_{-0.06}^{+0.05}$ & 10.42$\pm$0.02 &  0.24$\pm$0.01 & QG ($UVJ$-SFG)\\
nircam6-6895   &  214.84937747 &  +52.81183310 & 25.76$\pm$0.04 &  1.67$\pm$0.05 & $ 6.20_{-0.08}^{+0.12}$ &  9.95$\pm$0.08 &  1.10$\pm$0.03 & XELG-$z6$\\
nircam6-7698   &  214.85507239 &  +52.81304211 & 25.15$\pm$0.03 &  1.95$\pm$0.05 & $ 3.96_{-0.09}^{+0.05}$ &  9.76$\pm$0.03 & -0.18$\pm$0.04 & SFG ($UVJ$-QG)\\
nircam6-7824   &  214.85021414 &  +52.80899072 & 21.32$\pm$0.03 &  1.95$\pm$0.04 & $ 1.91_{-0.02}^{+0.02}$ & 10.81$\pm$0.01 &  2.43$\pm$0.05 & SFG, MIPS24\\
nircam6-7981   &  214.84025146 &  +52.80112019 & 24.53$\pm$0.03 &  1.97$\pm$0.05 & $ 3.75_{-0.09}^{+0.31}$ & 10.21$\pm$0.06 &  0.28$\pm$0.01 & SFG\\
nircam6-8976   &  214.90114269 &  +52.83810288 & 22.68$\pm$0.03 &  1.83$\pm$0.04 & $ 2.57_{-0.11}^{+0.15}$ & 10.62$\pm$0.03 &  1.54$\pm$0.01 & SFG, MIPS24\\
\enddata
\tablecomments{\label{tab:selection}Table with basic information about the sample of galaxies in this paper: ID, coordinates, magnitude and colors used in the selection, redshift, stellar mass, SFR (averaged in the last 10~Myr) and comments (including galaxy activity type based on SED-based sSFR estimations  (see text for details; we also mark quiescence results based on the $UVJ$ diagram), and detection by MIPS, sub-mm, and/or X-ray surveys).}
\end{deluxetable*}

\startlongtable
\centerwidetable
\begin{deluxetable*}{lcrrrrccccc}
\tabletypesize{\scriptsize}
\tablehead{\colhead{Sample} & \colhead{$z$} & \colhead{$\log\mathrm{M\!_\star}$} & \colhead{$\log\,\mathrm{SFR(5~Myr)}$} & \colhead{$\log\,\mathrm{SFR(25~Myr)}$} & \colhead{$\log\,\mathrm{SFR(100~Myr)}$} & \colhead{$\log\,\mathrm{SFR(E-budget)}$} & \colhead{A(V)} & \colhead{Age$_\mathrm{m-w}$} & \colhead{$z_{5\%}$} & \colhead{$\log\Sigma_\mathrm{M}\!_\star$}\\
\colhead{} & \colhead{} & \colhead{[M$_{\odot}$]} & \colhead{[$\mathrm{M}_{\odot}\,\mathrm{yr}^{-1}$]} & \colhead{[$\mathrm{M}_{\odot}\,\mathrm{yr}^{-1}$]} & \colhead{[$\mathrm{M}_{\odot}\,\mathrm{yr}^{-1}$]} & \colhead{[$\mathrm{M}_{\odot}\,\mathrm{yr}^{-1}$]} & \colhead{[mag]} & \colhead{[Gyr]} & \colhead{} & \colhead{[M$_{\odot}\,\mathrm{kpc}^{-2}$]}}
\startdata
ALL          & $3.7_{3.1}^{4.6}$ & $10.2_{9.7}^{10.6}$ & $1.4_{0.3}^{2.1}$ & $0.7_{-0.1}^{1.4}$ & $0.5_{-0.1}^{1.2}$ & $1.5_{0.8}^{2.1}$ & $0.7_{0.2}^{1.5}$ & $0.77_{0.39}^{1.04}$ & $12.0_{9.0}^{15.2}$ & $8.4_{8.0}^{8.8}$\\
HST-faint    & $3.3_{2.5}^{3.7}$ & $10.4_{10.2}^{10.7}$ & $1.3_{0.6}^{2.1}$ & $0.8_{0.4}^{1.4}$ & $0.7_{0.3}^{1.3}$ & $1.6_{1.1}^{2.2}$ & $0.7_{0.2}^{1.4}$ & $0.77_{0.46}^{1.11}$ & $10.5_{7.8}^{13.5}$ & $8.4_{7.9}^{8.8}$\\
HST-dark     & $4.4_{3.7}^{5.7}$ &  $9.8_{9.4}^{10.2}$ & $1.4_{0.2}^{2.1}$ & $0.3_{-0.8}^{1.2}$ & $0.2_{-0.7}^{0.9}$ & $1.4_{0.5}^{1.9}$ & $0.9_{0.3}^{1.7}$ & $0.65_{0.25}^{0.88}$ & $13.5_{10.1}^{16.0}$ & $8.5_{8.2}^{8.9}$\\
SFG          & $3.6_{2.9}^{4.3}$ & $10.2_{9.8}^{10.6}$ & $1.5_{0.9}^{2.1}$ & $0.9_{0.4}^{1.5}$ & $0.7_{0.3}^{1.3}$ & $1.7_{1.2}^{2.2}$ & $0.9_{0.3}^{1.6}$ & $0.71_{0.33}^{1.04}$ & $12.1_{9.0}^{15.1}$ & $8.4_{8.0}^{8.8}$\\
SFG@$z<4$    & $3.3_{2.6}^{3.7}$ & $10.4_{9.9}^{10.7}$ & $1.4_{0.7}^{1.9}$ & $1.0_{0.6}^{1.5}$ & $0.9_{0.5}^{1.4}$ & $1.7_{1.1}^{2.1}$ & $0.9_{0.3}^{1.6}$ & $0.74_{0.35}^{1.11}$ & $11.7_{8.8}^{14.8}$ & $8.3_{8.0}^{8.8}$\\
SFG@$z>4$    & $5.0_{4.4}^{5.3}$ &  $9.9_{9.5}^{10.3}$ & $1.8_{1.1}^{2.5}$ & $0.6_{-0.0}^{1.2}$ & $0.4_{-0.0}^{0.8}$ & $1.8_{1.3}^{2.3}$ & $1.0_{0.2}^{2.1}$ & $0.48_{0.01}^{0.77}$ & $13.5_{9.5}^{15.1}$ & $8.8_{8.4}^{9.1}$\\
QG           & $3.6_{3.3}^{4.0}$ & $10.3_{9.5}^{10.5}$ & $-0.6_{-1.0}^{0.2}$ & $-0.6_{-1.0}^{0.1}$ & $-0.5_{-0.8}^{0.1}$ & $0.4_{-0.4}^{0.8}$ & $0.2_{0.0}^{0.5}$ & $0.82_{0.77}^{1.04}$ & $11.7_{9.2}^{13.6}$ & $8.4_{8.0}^{8.9}$\\
QG@$z<3.5$   & $3.2_{2.6}^{3.3}$ & $10.4_{9.9}^{10.6}$ & $0.0_{-1.4}^{0.2}$ & $-0.1_{-1.3}^{0.1}$ & $-0.0_{-1.2}^{0.1}$ & $0.7_{-0.1}^{0.9}$ & $0.2_{0.1}^{0.6}$ & $0.82_{0.71}^{1.23}$ & $9.7_{8.6}^{13.9}$ & $8.2_{7.9}^{8.7}$\\
QG@$z>3.5$   & $3.9_{3.7}^{4.4}$ &  $9.8_{9.4}^{10.4}$ & $-0.8_{-1.0}^{-0.0}$ & $-0.8_{-1.0}^{-0.5}$ & $-0.7_{-0.8}^{-0.4}$ & $0.3_{-0.4}^{0.7}$ & $0.1_{0.0}^{0.5}$ & $0.82_{0.77}^{0.82}$ & $12.0_{11.0}^{13.5}$ & $8.6_{8.4}^{9.0}$\\
XELG-$z6$    & $6.5_{6.3}^{6.6}$ &  $9.6_{9.0}^{9.8}$ & $2.2_{1.9}^{2.8}$ & $0.5_{-1.0}^{1.3}$ & $-0.4_{-1.0}^{0.8}$ & $1.8_{1.5}^{2.4}$ & $1.4_{0.8}^{2.4}$ & $0.003_{0.002}^{0.400}$ & $14.4_{8.6}^{17.8}$ & $9.0_{8.5}^{9.4}$\\
\enddata
\tablecomments{\label{tab:stats}Table with median and quartiles for relevant properties of the different galaxy subsamples presented in the paper, namely, all galaxies, HST-faint, HST-dark, star-forming galaxies SFGs (all, and divided in 2 redshift regimes), quiescent galaxies QGs (all, and divided in 2 redshift regimes), and extreme emission-line galaxies at $z\sim6$ XELG-$z6$. These properties are: stellar mass, SED-based SFRs averaged in the last 5, 25, and 100~Myr, SFR obtained from the energy budget calculations presented in Section~6.2.1, $V$-band attenuation, mass-weighted age, redshift at which the galaxies formed 5\% of their current mass, and stellar mass surface density (in a pixel-by-pixel basis).}
\end{deluxetable*}

\bibliography{ms_perezgonzalez_etal_ceers}

\begin{thebibliography}{}
\expandafter\ifx\csname natexlab\endcsname\relax\def\natexlab#1{#1}\fi
\providecommand{\url}[1]{\href{#1}{#1}}
\providecommand{\dodoi}[1]{doi:~\href{http://doi.org/#1}{\nolinkurl{#1}}}
\providecommand{\doeprint}[1]{\href{http://ascl.net/#1}{\nolinkurl{http://ascl.net/#1}}}
\providecommand{\doarXiv}[1]{\href{https://arxiv.org/abs/#1}{\nolinkurl{https://arxiv.org/abs/#1}}}

\bibitem[{{Abraham} {et~al.}(1999){Abraham}, {Ellis}, {Fabian}, {Tanvir}, \&
  {Glazebrook}}]{1999MNRAS.303..641A}
{Abraham}, R.~G., {Ellis}, R.~S., {Fabian}, A.~C., {Tanvir}, N.~R., \&
  {Glazebrook}, K. 1999, \mnras, 303, 641,
  \dodoi{10.1046/j.1365-8711.1999.02059.x}

\bibitem[{{Adams} {et~al.}(2022){Adams}, {Conselice}, {Ferreira}, {Austin},
  {Trussler}, {Juod{\v{z}}balis}, {Wilkins}, {Caruana}, {Dayal}, {Verma}, \&
  {Vijayan}}]{2022arXiv220711217A}
{Adams}, N.~J., {Conselice}, C.~J., {Ferreira}, L., {et~al.} 2022, arXiv
  e-prints, arXiv:2207.11217.
\newblock \doarXiv{2207.11217}

\bibitem[{{Alcalde Pampliega} {et~al.}(2019){Alcalde Pampliega},
  {P{\'e}rez-Gonz{\'a}lez}, {Barro}, {Dom{\'\i}nguez S{\'a}nchez},
  {Eliche-Moral}, {Cardiel}, {Hern{\'a}n-Caballero}, {Rodriguez-Mu{\~n}oz},
  {S{\'a}nchez Bl{\'a}zquez}, \& {Esquej}}]{2019ApJ...876..135A}
{Alcalde Pampliega}, B., {P{\'e}rez-Gonz{\'a}lez}, P.~G., {Barro}, G., {et~al.}
  2019, \apj, 876, 135, \dodoi{10.3847/1538-4357/ab14f2}

\bibitem[{{Astropy Collaboration} {et~al.}(2022){Astropy Collaboration},
  {Price-Whelan}, {Lim}, {Earl}, {Starkman}, {Bradley}, {Shupe}, {Patil},
  {Corrales}, {Brasseur}, {N{\"o}the}, {Donath}, {Tollerud}, {Morris},
  {Ginsburg}, {Vaher}, {Weaver}, {Tocknell}, {Jamieson}, {van Kerkwijk},
  {Robitaille}, {Merry}, {Bachetti}, {G{\"u}nther}, {Aldcroft},
  {Alvarado-Montes}, {Archibald}, {B{\'o}di}, {Bapat}, {Barentsen},
  {Baz{\'a}n}, {Biswas}, {Boquien}, {Burke}, {Cara}, {Cara}, {Conroy},
  {Conseil}, {Craig}, {Cross}, {Cruz}, {D'Eugenio}, {Dencheva}, {Devillepoix},
  {Dietrich}, {Eigenbrot}, {Erben}, {Ferreira}, {Foreman-Mackey}, {Fox},
  {Freij}, {Garg}, {Geda}, {Glattly}, {Gondhalekar}, {Gordon}, {Grant},
  {Greenfield}, {Groener}, {Guest}, {Gurovich}, {Handberg}, {Hart},
  {Hatfield-Dodds}, {Homeier}, {Hosseinzadeh}, {Jenness}, {Jones}, {Joseph},
  {Kalmbach}, {Karamehmetoglu}, {Ka{\l}uszy{\'n}ski}, {Kelley}, {Kern},
  {Kerzendorf}, {Koch}, {Kulumani}, {Lee}, {Ly}, {Ma}, {MacBride}, {Maljaars},
  {Muna}, {Murphy}, {Norman}, {O'Steen}, {Oman}, {Pacifici}, {Pascual},
  {Pascual-Granado}, {Patil}, {Perren}, {Pickering}, {Rastogi}, {Roulston},
  {Ryan}, {Rykoff}, {Sabater}, {Sakurikar}, {Salgado}, {Sanghi}, {Saunders},
  {Savchenko}, {Schwardt}, {Seifert-Eckert}, {Shih}, {Jain}, {Shukla}, {Sick},
  {Simpson}, {Singanamalla}, {Singer}, {Singhal}, {Sinha}, {Sip{\H{o}}cz},
  {Spitler}, {Stansby}, {Streicher}, {{\v{S}}umak}, {Swinbank}, {Taranu},
  {Tewary}, {Tremblay}, {Val-Borro}, {Van Kooten}, {Vasovi{\'c}}, {Verma}, {de
  Miranda Cardoso}, {Williams}, {Wilson}, {Winkel}, {Wood-Vasey}, {Xue},
  {Yoachim}, {Zhang}, {Zonca}, \& {Astropy Project
  Contributors}}]{2022ApJ...935..167A}
{Astropy Collaboration}, {Price-Whelan}, A.~M., {Lim}, P.~L., {et~al.} 2022,
  \apj, 935, 167, \dodoi{10.3847/1538-4357/ac7c74}

\bibitem[{{Atek} {et~al.}(2022){Atek}, {Furtak}, {Oesch}, {van Dokkum},
  {Reddy}, {Contini}, {Illingworth}, \& {Wilkins}}]{2022MNRAS.511.4464A}
{Atek}, H., {Furtak}, L.~J., {Oesch}, P., {et~al.} 2022, \mnras, 511, 4464,
  \dodoi{10.1093/mnras/stac360}

\bibitem[{{Bagley} {et~al.}(2022){Bagley}, {Finkelstein}, {Koekemoer},
  {Ferguson}, {Arrabal Haro}, {Dickinson}, {Kartaltepe}, {Papovich},
  {P{\'e}rez-Gonz{\'a}lez}, {Pirzkal}, {Somerville}, {Willmer}, {Yang}, {Yung},
  {Fontana}, {Grazian}, {Grogin}, {Hirschmann}, {Kewley}, {Kirkpatrick},
  {Kocevski}, {Lotz}, {Medrano}, {Morales}, {Pentericci}, {Ravindranath},
  {Trump}, {Wilkins}, {Calabr{\`o}}, {Cooper}, {Costantin}, {de la Vega},
  {Hutchison}, {Lucas}, {McGrath}, {Wang}, \& {Wuyts}}]{2022arXiv221102495B}
{Bagley}, M.~B., {Finkelstein}, S.~L., {Koekemoer}, A.~M., {et~al.} 2022, arXiv
  e-prints, arXiv:2211.02495.
\newblock \doarXiv{2211.02495}

\bibitem[{{Barro} {et~al.}(2011){Barro}, {P{\'e}rez-Gonz{\'a}lez}, {Gallego},
  {Ashby}, {Kajisawa}, {Miyazaki}, {Villar}, {Yamada}, \&
  {Zamorano}}]{2011ApJS..193...30B}
{Barro}, G., {P{\'e}rez-Gonz{\'a}lez}, P.~G., {Gallego}, J., {et~al.} 2011,
  \apjs, 193, 30, \dodoi{10.1088/0067-0049/193/2/30}

\bibitem[{{Barro} {et~al.}(2019){Barro}, {P{\'e}rez-Gonz{\'a}lez}, {Cava},
  {Brammer}, {Pandya}, {Eliche Moral}, {Esquej}, {Dom{\'\i}nguez-S{\'a}nchez},
  {Alcalde Pampliega}, {Guo}, {Koekemoer}, {Trump}, {Ashby}, {Cardiel},
  {Castellano}, {Conselice}, {Dickinson}, {Dolch}, {Donley}, {Espino Briones},
  {Faber}, {Fazio}, {Ferguson}, {Finkelstein}, {Fontana}, {Galametz},
  {Gardner}, {Gawiser}, {Giavalisco}, {Grazian}, {Grogin}, {Hathi}, {Hemmati},
  {Hern{\'a}n-Caballero}, {Kocevski}, {Koo}, {Kodra}, {Lee}, {Lin}, {Lucas},
  {Mobasher}, {McGrath}, {Nandra}, {Nayyeri}, {Newman}, {Pforr}, {Peth},
  {Rafelski}, {Rodr{\'\i}guez-Munoz}, {Salvato}, {Stefanon}, {van der Wel},
  {Willner}, {Wiklind}, \& {Wuyts}}]{2019ApJS..243...22B}
{Barro}, G., {P{\'e}rez-Gonz{\'a}lez}, P.~G., {Cava}, A., {et~al.} 2019, \apjs,
  243, 22, \dodoi{10.3847/1538-4365/ab23f2}

\bibitem[{{Barrufet} {et~al.}(2022){Barrufet}, {Oesch}, {Weibel}, {Brammer},
  {Bezanson}, {Bouwens}, {Fudamoto}, {Gonzalez}, {Illingworth}, {Heintz},
  {Holden}, {Labbe}, {Magee}, {Naidu}, {Nelson}, {Stefanon}, {Smit}, {van
  Dokkum}, {Weaver}, \& {Williams}}]{2022arXiv220714733B}
{Barrufet}, L., {Oesch}, P.~A., {Weibel}, A., {et~al.} 2022, arXiv e-prints,
  arXiv:2207.14733.
\newblock \doarXiv{2207.14733}

\bibitem[{{Bertin} \& {Arnouts}(1996)}]{1996A&AS..117..393B}
{Bertin}, E., \& {Arnouts}, S. 1996, \aaps, 117, 393,
  \dodoi{10.1051/aas:1996164}

\bibitem[{{Bowler} {et~al.}(2018){Bowler}, {Bourne}, {Dunlop}, {McLure}, \&
  {McLeod}}]{2018MNRAS.481.1631B}
{Bowler}, R.~A.~A., {Bourne}, N., {Dunlop}, J.~S., {McLure}, R.~J., \&
  {McLeod}, D.~J. 2018, \mnras, 481, 1631, \dodoi{10.1093/mnras/sty2368}

\bibitem[{{Bradley} {et~al.}(2022){Bradley}, {Sip{\H{o}}cz}, {Robitaille},
  {Tollerud}, {Vin{\'\i}cius}, {Deil}, {Barbary}, {Wilson}, {Busko}, {Donath},
  {G{\"u}nther}, {Cara}, {Lim}, {Me{\ss}linger}, {Conseil}, {Bostroem},
  {Droettboom}, {Bray}, {Andersen Bratholm}, {Barentsen}, {Craig}, {Rathi},
  {Pascual}, {Perren}, {Georgiev}, {De Val-Borro}, {Kerzendorf}, {Bach},
  {Quint}, \& {Souchereau}}]{2022zndo...6825092B}
{Bradley}, L., {Sip{\H{o}}cz}, B., {Robitaille}, T., {et~al.} 2022,
  {astropy/photutils: 1.5.0}, 1.5.0, Zenodo,  Zenodo,
  \dodoi{10.5281/zenodo.6825092}

\bibitem[{{Brammer} {et~al.}(2008){Brammer}, {van Dokkum}, \&
  {Coppi}}]{2008ApJ...686.1503B}
{Brammer}, G.~B., {van Dokkum}, P.~G., \& {Coppi}, P. 2008, \apj, 686, 1503,
  \dodoi{10.1086/591786}

\bibitem[{{Bruzual} \& {Charlot}(2003)}]{2003MNRAS.344.1000B}
{Bruzual}, G., \& {Charlot}, S. 2003, \mnras, 344, 1000,
  \dodoi{10.1046/j.1365-8711.2003.06897.x}

\bibitem[{{Buat} {et~al.}(2005){Buat}, {Iglesias-P{\'a}ramo}, {Seibert},
  {Burgarella}, {Charlot}, {Martin}, {Xu}, {Heckman}, {Boissier}, {Boselli},
  {Barlow}, {Bianchi}, {Byun}, {Donas}, {Forster}, {Friedman}, {Jelinski},
  {Lee}, {Madore}, {Malina}, {Milliard}, {Morissey}, {Neff}, {Rich},
  {Schiminovitch}, {Siegmund}, {Small}, {Szalay}, {Welsh}, \&
  {Wyder}}]{2005ApJ...619L..51B}
{Buat}, V., {Iglesias-P{\'a}ramo}, J., {Seibert}, M., {et~al.} 2005, \apjl,
  619, L51, \dodoi{10.1086/423241}

\bibitem[{{Burgarella} {et~al.}(2022){Burgarella}, {Bogdanoska}, {Nanni},
  {Bardelli}, {B{\'e}thermin}, {Boquien}, {Buat}, {Faisst},
  {Dessauges-Zavadsky}, {Fudamoto}, {Fujimoto}, {Giavalisco}, {Ginolfi},
  {Gruppioni}, {Hathi}, {Ibar}, {Jones}, {Koekemoer}, {Kohno}, {Lemaux},
  {Narayanan}, {Oesch}, {Ouchi}, {Riechers}, {Pozzi}, {Romano}, {Schaerer},
  {Talia}, {Theul{\'e}}, {Vergani}, {Zamorani}, {Zucca}, {Cassata}, \& {ALPINE
  Team}}]{2022A&A...664A..73B}
{Burgarella}, D., {Bogdanoska}, J., {Nanni}, A., {et~al.} 2022, \aap, 664, A73,
  \dodoi{10.1051/0004-6361/202142554}

\bibitem[{{Calzetti} {et~al.}(2000){Calzetti}, {Armus}, {Bohlin}, {Kinney},
  {Koornneef}, \& {Storchi-Bergmann}}]{2000ApJ...533..682C}
{Calzetti}, D., {Armus}, L., {Bohlin}, R.~C., {et~al.} 2000, \apj, 533, 682,
  \dodoi{10.1086/308692}

\bibitem[{{Calzetti} {et~al.}(2007){Calzetti}, {Kennicutt}, {Engelbracht},
  {Leitherer}, {Draine}, {Kewley}, {Moustakas}, {Sosey}, {Dale}, {Gordon},
  {Helou}, {Hollenbach}, {Armus}, {Bendo}, {Bot}, {Buckalew}, {Jarrett}, {Li},
  {Meyer}, {Murphy}, {Prescott}, {Regan}, {Rieke}, {Roussel}, {Sheth}, {Smith},
  {Thornley}, \& {Walter}}]{2007ApJ...666..870C}
{Calzetti}, D., {Kennicutt}, R.~C., {Engelbracht}, C.~W., {et~al.} 2007, \apj,
  666, 870, \dodoi{10.1086/520082}

\bibitem[{{Caputi} {et~al.}(2012){Caputi}, {Dunlop}, {McLure}, {Huang},
  {Fazio}, {Ashby}, {Castellano}, {Fontana}, {Cirasuolo}, {Almaini}, {Bell},
  {Dickinson}, {Donley}, {Faber}, {Ferguson}, {Giavalisco}, {Grogin},
  {Kocevski}, {Koekemoer}, {Koo}, {Lai}, {Newman}, \&
  {Somerville}}]{2012ApJ...750L..20C}
{Caputi}, K.~I., {Dunlop}, J.~S., {McLure}, R.~J., {et~al.} 2012, \apjl, 750,
  L20, \dodoi{10.1088/2041-8205/750/1/L20}

\bibitem[{{Carnall} {et~al.}(2022){Carnall}, {McLeod}, {McLure}, {Dunlop},
  {Begley}, {Cullen}, {Donnan}, {Hamadouche}, {Jewell}, {Jones}, {Pollock}, \&
  {Wild}}]{2022arXiv220800986C}
{Carnall}, A.~C., {McLeod}, D.~J., {McLure}, R.~J., {et~al.} 2022, arXiv
  e-prints, arXiv:2208.00986.
\newblock \doarXiv{2208.00986}

\bibitem[{{Castellano} {et~al.}(2022){Castellano}, {Fontana}, {Treu},
  {Santini}, {Merlin}, {Leethochawalit}, {Trenti}, {Mestric}, {Vanzella},
  {Bonchi}, {Belfiori}, {Nonino}, {Paris}, {Polenta}, {Roberts-Borsani},
  {Boyett}, {Bradac}, {Calabro}, {Glazebrook}, {Grillo}, {Mascia}, {Mason},
  {Mercurio}, {Morishita}, {Nanayakkara}, {Pentericci}, {Rosati}, {Vulcani},
  {Wang}, \& {Yang}}]{2022arXiv220709436C}
{Castellano}, M., {Fontana}, A., {Treu}, T., {et~al.} 2022, arXiv e-prints,
  arXiv:2207.09436.
\newblock \doarXiv{2207.09436}

\bibitem[{{Chabrier}(2003)}]{2003PASP..115..763C}
{Chabrier}, G. 2003, \pasp, 115, 763, \dodoi{10.1086/376392}

\bibitem[{{Chapman} {et~al.}(2005){Chapman}, {Blain}, {Smail}, \&
  {Ivison}}]{2005ApJ...622..772C}
{Chapman}, S.~C., {Blain}, A.~W., {Smail}, I., \& {Ivison}, R.~J. 2005, \apj,
  622, 772, \dodoi{10.1086/428082}

\bibitem[{{Chary} \& {Elbaz}(2001)}]{2001ApJ...556..562C}
{Chary}, R., \& {Elbaz}, D. 2001, \apj, 556, 562, \dodoi{10.1086/321609}

\bibitem[{{Costantin} {et~al.}(2020){Costantin}, {M{\'e}ndez-Abreu}, {Corsini},
  {Morelli}, {de Lorenzo-C{\'a}ceres}, {Pagotto}, {Cuomo}, {Aguerri}, \&
  {Rubino}}]{2020ApJ...889L...3C}
{Costantin}, L., {M{\'e}ndez-Abreu}, J., {Corsini}, E.~M., {et~al.} 2020,
  \apjl, 889, L3, \dodoi{10.3847/2041-8213/ab6459}

\bibitem[{{Daddi} {et~al.}(2004){Daddi}, {Cimatti}, {Renzini}, {Fontana},
  {Mignoli}, {Pozzetti}, {Tozzi}, \& {Zamorani}}]{2004ApJ...617..746D}
{Daddi}, E., {Cimatti}, A., {Renzini}, A., {et~al.} 2004, \apj, 617, 746,
  \dodoi{10.1086/425569}

\bibitem[{{Dahlen} {et~al.}(2013){Dahlen}, {Mobasher}, {Faber}, {Ferguson},
  {Barro}, {Finkelstein}, {Finlator}, {Fontana}, {Gruetzbauch}, {Johnson},
  {Pforr}, {Salvato}, {Wiklind}, {Wuyts}, {Acquaviva}, {Dickinson}, {Guo},
  {Huang}, {Huang}, {Newman}, {Bell}, {Conselice}, {Galametz}, {Gawiser},
  {Giavalisco}, {Grogin}, {Hathi}, {Kocevski}, {Koekemoer}, {Koo}, {Lee},
  {McGrath}, {Papovich}, {Peth}, {Ryan}, {Somerville}, {Weiner}, \&
  {Wilson}}]{2013ApJ...775...93D}
{Dahlen}, T., {Mobasher}, B., {Faber}, S.~M., {et~al.} 2013, \apj, 775, 93,
  \dodoi{10.1088/0004-637X/775/2/93}

\bibitem[{{Dannerbauer} {et~al.}(2008){Dannerbauer}, {Walter}, \&
  {Morrison}}]{2008ApJ...673L.127D}
{Dannerbauer}, H., {Walter}, F., \& {Morrison}, G. 2008, \apjl, 673, L127,
  \dodoi{10.1086/528794}

\bibitem[{{Davis} {et~al.}(2007){Davis}, {Guhathakurta}, {Konidaris}, {Newman},
  {Ashby}, {Biggs}, {Barmby}, {Bundy}, {Chapman}, {Coil}, {Conselice},
  {Cooper}, {Croton}, {Eisenhardt}, {Ellis}, {Faber}, {Fang}, {Fazio},
  {Georgakakis}, {Gerke}, {Goss}, {Gwyn}, {Harker}, {Hopkins}, {Huang},
  {Ivison}, {Kassin}, {Kirby}, {Koekemoer}, {Koo}, {Laird}, {Le Floc'h}, {Lin},
  {Lotz}, {Marshall}, {Martin}, {Metevier}, {Moustakas}, {Nandra}, {Noeske},
  {Papovich}, {Phillips}, {Rich}, {Rieke}, {Rigopoulou}, {Salim},
  {Schiminovich}, {Simard}, {Smail}, {Small}, {Weiner}, {Willmer}, {Willner},
  {Wilson}, {Wright}, \& {Yan}}]{2007ApJ...660L...1D}
{Davis}, M., {Guhathakurta}, P., {Konidaris}, N.~P., {et~al.} 2007, \apjl, 660,
  L1, \dodoi{10.1086/517931}

\bibitem[{{Dom{\'\i}nguez S{\'a}nchez} {et~al.}(2016){Dom{\'\i}nguez
  S{\'a}nchez}, {P{\'e}rez-Gonz{\'a}lez}, {Esquej}, {Eliche-Moral}, {Barro},
  {Cava}, {Koekemoer}, {Alcalde Pampliega}, {Alonso Herrero}, {Bruzual},
  {Cardiel}, {Cenarro}, {Ceverino}, {Charlot}, \& {Hern{\'a}n
  Caballero}}]{2016MNRAS.457.3743D}
{Dom{\'\i}nguez S{\'a}nchez}, H., {P{\'e}rez-Gonz{\'a}lez}, P.~G., {Esquej},
  P., {et~al.} 2016, \mnras, 457, 3743, \dodoi{10.1093/mnras/stw201}

\bibitem[{{Donnan} {et~al.}(2022){Donnan}, {McLeod}, {Dunlop}, {McLure},
  {Carnall}, {Begley}, {Cullen}, {Hamadouche}, {Bowler}, {McCracken},
  {Milvang-Jensen}, {Moneti}, \& {Targett}}]{2022arXiv220712356D}
{Donnan}, C.~T., {McLeod}, D.~J., {Dunlop}, J.~S., {et~al.} 2022, arXiv
  e-prints, arXiv:2207.12356.
\newblock \doarXiv{2207.12356}

\bibitem[{{Donnari} {et~al.}(2019){Donnari}, {Pillepich}, {Nelson},
  {Vogelsberger}, {Genel}, {Weinberger}, {Marinacci}, {Springel}, \&
  {Hernquist}}]{2019MNRAS.485.4817D}
{Donnari}, M., {Pillepich}, A., {Nelson}, D., {et~al.} 2019, \mnras, 485, 4817,
  \dodoi{10.1093/mnras/stz712}

\bibitem[{{Elbaz} {et~al.}(2011){Elbaz}, {Dickinson}, {Hwang},
  {D{\'\i}az-Santos}, {Magdis}, {Magnelli}, {Le Borgne}, {Galliano},
  {Pannella}, {Chanial}, {Armus}, {Charmandaris}, {Daddi}, {Aussel}, {Popesso},
  {Kartaltepe}, {Altieri}, {Valtchanov}, {Coia}, {Dannerbauer}, {Dasyra},
  {Leiton}, {Mazzarella}, {Alexander}, {Buat}, {Burgarella}, {Chary}, {Gilli},
  {Ivison}, {Juneau}, {Le Floc'h}, {Lutz}, {Morrison}, {Mullaney}, {Murphy},
  {Pope}, {Scott}, {Brodwin}, {Calzetti}, {Cesarsky}, {Charlot}, {Dole},
  {Eisenhardt}, {Ferguson}, {F{\"o}rster Schreiber}, {Frayer}, {Giavalisco},
  {Huynh}, {Koekemoer}, {Papovich}, {Reddy}, {Surace}, {Teplitz}, {Yun}, \&
  {Wilson}}]{2011A&A...533A.119E}
{Elbaz}, D., {Dickinson}, M., {Hwang}, H.~S., {et~al.} 2011, \aap, 533, A119,
  \dodoi{10.1051/0004-6361/201117239}

\bibitem[{{Elston} {et~al.}(1988){Elston}, {Rieke}, \&
  {Rieke}}]{1988ApJ...331L..77E}
{Elston}, R., {Rieke}, G.~H., \& {Rieke}, M.~J. 1988, \apjl, 331, L77,
  \dodoi{10.1086/185239}

\bibitem[{{Endsley} {et~al.}(2022){Endsley}, {Stark}, {Whitler}, {Topping},
  {Chen}, {Plat}, {Chisholm}, \& {Charlot}}]{2022arXiv220814999E}
{Endsley}, R., {Stark}, D.~P., {Whitler}, L., {et~al.} 2022, arXiv e-prints,
  arXiv:2208.14999.
\newblock \doarXiv{2208.14999}

\bibitem[{{Fang} {et~al.}(2018){Fang}, {Faber}, {Koo}, {Rodr{\'\i}guez-Puebla},
  {Guo}, {Barro}, {Behroozi}, {Brammer}, {Chen}, {Dekel}, {Ferguson},
  {Gawiser}, {Giavalisco}, {Kartaltepe}, {Kocevski}, {Koekemoer}, {McGrath},
  {McIntosh}, {Newman}, {Pacifici}, {Pandya}, {P{\'e}rez-Gonz{\'a}lez},
  {Primack}, {Salmon}, {Trump}, {Weiner}, {Willner}, {Acquaviva}, {Dahlen},
  {Finkelstein}, {Finlator}, {Fontana}, {Galametz}, {Grogin}, {Gruetzbauch},
  {Johnson}, {Mobasher}, {Papovich}, {Pforr}, {Salvato}, {Santini}, {van der
  Wel}, {Wiklind}, \& {Wuyts}}]{2018ApJ...858..100F}
{Fang}, J.~J., {Faber}, S.~M., {Koo}, D.~C., {et~al.} 2018, \apj, 858, 100,
  \dodoi{10.3847/1538-4357/aabcba}

\bibitem[{{Finkelstein} {et~al.}(2017){Finkelstein}, {Dickinson}, {Ferguson},
  {Grazian}, {Grogin}, {Kartaltepe}, {Kewley}, {Kocevski}, {Koekemoer}, {Lotz},
  {Papovich}, {Pentericci}, {Perez-Gonzalez}, {Pirzkal}, {Ravindranath},
  {Somerville}, {Trump}, \& {Wilkins}}]{2017jwst.prop.1345F}
{Finkelstein}, S.~L., {Dickinson}, M., {Ferguson}, H.~C., {et~al.} 2017, {The
  Cosmic Evolution Early Release Science (CEERS) Survey}, JWST Proposal ID
  1345. Cycle 0 Early Release Science

\bibitem[{{Finkelstein} {et~al.}(2022){Finkelstein}, {Bagley}, {Arrabal Haro},
  {Dickinson}, {Ferguson}, {Kartaltepe}, {Papovich}, {Burgarella}, {Kocevski},
  {Huertas-Company}, {Iyer}, {Larson}, {P{\'e}rez-Gonz{\'a}lez}, {Rose},
  {Tacchella}, {Wilkins}, {Chworowsky}, {Medrano}, {Morales}, {Somerville},
  {Yung}, {Fontana}, {Giavalisco}, {Grazian}, {Grogin}, {Kewley}, {Koekemoer},
  {Kirkpatrick}, {Kurczynski}, {Lotz}, {Pentericci}, {Pirzkal}, {Ravindranath},
  {Ryan}, {Trump}, {Yang}, {Almaini}, {Amor{\'\i}n}, {Annunziatella},
  {Backhaus}, {Barro}, {Behroozi}, {Bell}, {Bhatawdekar}, {Bisigello}, {Bromm},
  {Buat}, {Buitrago}, {Calabr{\'o}}, {Casey}, {Castellano}, {Ch{\'a}vez Ortiz},
  {Ciesla}, {Cleri}, {Cohen}, {Cole}, {Cooke}, {Cooper}, {Cooray}, {Costantin},
  {Cox}, {Croton}, {Daddi}, {Dav{\'e}}, {de la Vega}, {Dekel}, {Elbaz},
  {Estrada-Carpenter}, {Faber}, {Fern{\'a}ndez}, {Finkelstein}, {Freundlich},
  {Fujimoto}, {Garc{\'\i}a-Argum{\'a}nez}, {Gardner}, {Gawiser},
  {G{\'o}mez-Guijarro}, {Guo}, {Hamilton}, {Hathi}, {Holwerda}, {Hirschmann},
  {Hutchison}, {Jaskot}, {Jha}, {Jogee}, {Juneau}, {Jung}, {Kassin}, {Le Bail},
  {Leung}, {Lucas}, {Magnelli}, {Mantha}, {Matharu}, {McGrath}, {McIntosh},
  {Merlin}, {Mobasher}, {Newman}, {Nicholls}, {Pandya}, {Rafelski}, {Ronayne},
  {Santini}, {Seill{\'e}}, {Shah}, {Shen}, {Simons}, {Snyder}, {Stanway},
  {Straughn}, {Teplitz}, {Vanderhoof}, {Vega-Ferrero}, {Wang}, {Weiner},
  {Willmer}, {Wuyts}, \& {Zavala}}]{2022arXiv220712474F}
{Finkelstein}, S.~L., {Bagley}, M.~B., {Arrabal Haro}, P., {et~al.} 2022, arXiv
  e-prints, arXiv:2207.12474.
\newblock \doarXiv{2207.12474}

\bibitem[{{Franco} {et~al.}(2018){Franco}, {Elbaz}, {B{\'e}thermin},
  {Magnelli}, {Schreiber}, {Ciesla}, {Dickinson}, {Nagar}, {Silverman},
  {Daddi}, {Alexander}, {Wang}, {Pannella}, {Le Floc'h}, {Pope}, {Giavalisco},
  {Maury}, {Bournaud}, {Chary}, {Demarco}, {Ferguson}, {Finkelstein}, {Inami},
  {Iono}, {Juneau}, {Lagache}, {Leiton}, {Lin}, {Magdis}, {Messias},
  {Motohara}, {Mullaney}, {Okumura}, {Papovich}, {Pforr}, {Rujopakarn},
  {Sargent}, {Shu}, \& {Zhou}}]{2018A&A...620A.152F}
{Franco}, M., {Elbaz}, D., {B{\'e}thermin}, M., {et~al.} 2018, \aap, 620, A152,
  \dodoi{10.1051/0004-6361/201832928}

\bibitem[{{Franx} {et~al.}(2003){Franx}, {Labb{\'e}}, {Rudnick}, {van Dokkum},
  {Daddi}, {F{\"o}rster Schreiber}, {Moorwood}, {Rix}, {R{\"o}ttgering}, {van
  der Wel}, {van der Werf}, \& {van Starkenburg}}]{2003ApJ...587L..79F}
{Franx}, M., {Labb{\'e}}, I., {Rudnick}, G., {et~al.} 2003, \apjl, 587, L79,
  \dodoi{10.1086/375155}

\bibitem[{{Freitas} {et~al.}(2021){Freitas}, {Chies-Santos}, {Furlanetto}, \&
  {Ferrari}}]{2021IAUS..359..431F}
{Freitas}, R.~F., {Chies-Santos}, A.~L., {Furlanetto}, C., \& {Ferrari}, F.
  2021, in Galaxy Evolution and Feedback across Different Environments, ed.
  T.~{Storchi Bergmann}, W.~{Forman}, R.~{Overzier}, \& R.~{Riffel}, Vol. 359,
  431--432, \dodoi{10.1017/S1743921320002069}

\bibitem[{{Fumagalli} {et~al.}(2014){Fumagalli}, {Labb{\'e}}, {Patel}, {Franx},
  {van Dokkum}, {Brammer}, {da Cunha}, {F{\"o}rster Schreiber}, {Kriek},
  {Quadri}, {Rix}, {Wake}, {Whitaker}, {Lundgren}, {Marchesini}, {Maseda},
  {Momcheva}, {Nelson}, {Pacifici}, \& {Skelton}}]{2014ApJ...796...35F}
{Fumagalli}, M., {Labb{\'e}}, I., {Patel}, S.~G., {et~al.} 2014, \apj, 796, 35,
  \dodoi{10.1088/0004-637X/796/1/35}

\bibitem[{{Gaia Collaboration} {et~al.}(2016{\natexlab{a}}){Gaia
  Collaboration}, {Prusti}, {de Bruijne}, {Brown}, {Vallenari}, {Babusiaux},
  {Bailer-Jones}, {Bastian}, {Biermann}, {Evans}, {Eyer}, {Jansen}, {Jordi},
  {Klioner}, {Lammers}, {Lindegren}, {Luri}, {Mignard}, {Milligan}, {Panem},
  {Poinsignon}, {Pourbaix}, {Randich}, {Sarri}, {Sartoretti}, {Siddiqui},
  {Soubiran}, {Valette}, {van Leeuwen}, {Walton}, {Aerts}, {Arenou}, {Cropper},
  {Drimmel}, {H{\o}g}, {Katz}, {Lattanzi}, {O'Mullane}, {Grebel}, {Holland},
  {Huc}, {Passot}, {Bramante}, {Cacciari}, {Casta{\~n}eda}, {Chaoul}, {Cheek},
  {De Angeli}, {Fabricius}, {Guerra}, {Hern{\'a}ndez}, {Jean-Antoine-Piccolo},
  {Masana}, {Messineo}, {Mowlavi}, {Nienartowicz}, {Ord{\'o}{\~n}ez-Blanco},
  {Panuzzo}, {Portell}, {Richards}, {Riello}, {Seabroke}, {Tanga},
  {Th{\'e}venin}, {Torra}, {Els}, {Gracia-Abril}, {Comoretto},
  {Garcia-Reinaldos}, {Lock}, {Mercier}, {Altmann}, {Andrae}, {Astraatmadja},
  {Bellas-Velidis}, {Benson}, {Berthier}, {Blomme}, {Busso}, {Carry},
  {Cellino}, {Clementini}, {Cowell}, {Creevey}, {Cuypers}, {Davidson}, {De
  Ridder}, {de Torres}, {Delchambre}, {Dell'Oro}, {Ducourant}, {Fr{\'e}mat},
  {Garc{\'\i}a-Torres}, {Gosset}, {Halbwachs}, {Hambly}, {Harrison}, {Hauser},
  {Hestroffer}, {Hodgkin}, {Huckle}, {Hutton}, {Jasniewicz}, {Jordan},
  {Kontizas}, {Korn}, {Lanzafame}, {Manteiga}, {Moitinho}, {Muinonen},
  {Osinde}, {Pancino}, {Pauwels}, {Petit}, {Recio-Blanco}, {Robin}, {Sarro},
  {Siopis}, {Smith}, {Smith}, {Sozzetti}, {Thuillot}, {van Reeven}, {Viala},
  {Abbas}, {Abreu Aramburu}, {Accart}, {Aguado}, {Allan}, {Allasia},
  {Altavilla}, {{\'A}lvarez}, {Alves}, {Anderson}, {Andrei}, {Anglada Varela},
  {Antiche}, {Antoja}, {Ant{\'o}n}, {Arcay}, {Atzei}, {Ayache}, {Bach},
  {Baker}, {Balaguer-N{\'u}{\~n}ez}, {Barache}, {Barata}, {Barbier}, {Barblan},
  {Baroni}, {Barrado y Navascu{\'e}s}, {Barros}, {Barstow}, {Becciani},
  {Bellazzini}, {Bellei}, {Bello Garc{\'\i}a}, {Belokurov}, {Bendjoya},
  {Berihuete}, {Bianchi}, {Bienaym{\'e}}, {Billebaud}, {Blagorodnova},
  {Blanco-Cuaresma}, {Boch}, {Bombrun}, {Borrachero}, {Bouquillon}, {Bourda},
  {Bouy}, {Bragaglia}, {Breddels}, {Brouillet}, {Br{\"u}semeister},
  {Bucciarelli}, {Budnik}, {Burgess}, {Burgon}, {Burlacu}, {Busonero}, {Buzzi},
  {Caffau}, {Cambras}, {Campbell}, {Cancelliere}, {Cantat-Gaudin}, {Carlucci},
  {Carrasco}, {Castellani}, {Charlot}, {Charnas}, {Charvet}, {Chassat},
  {Chiavassa}, {Clotet}, {Cocozza}, {Collins}, {Collins}, {Costigan}, {Crifo},
  {Cross}, {Crosta}, {Crowley}, {Dafonte}, {Damerdji}, {Dapergolas}, {David},
  {David}, {De Cat}, {de Felice}, {de Laverny}, {De Luise}, {De March}, {de
  Martino}, {de Souza}, {Debosscher}, {del Pozo}, {Delbo}, {Delgado},
  {Delgado}, {di Marco}, {Di Matteo}, {Diakite}, {Distefano}, {Dolding}, {Dos
  Anjos}, {Drazinos}, {Dur{\'a}n}, {Dzigan}, {Ecale}, {Edvardsson}, {Enke},
  {Erdmann}, {Escolar}, {Espina}, {Evans}, {Eynard Bontemps}, {Fabre},
  {Fabrizio}, {Faigler}, {Falc{\~a}o}, {Farr{\`a}s Casas}, {Faye}, {Federici},
  {Fedorets}, {Fern{\'a}ndez-Hern{\'a}ndez}, {Fernique}, {Fienga}, {Figueras},
  {Filippi}, {Findeisen}, {Fonti}, {Fouesneau}, {Fraile}, {Fraser}, {Fuchs},
  {Furnell}, {Gai}, {Galleti}, {Galluccio}, {Garabato}, {Garc{\'\i}a-Sedano},
  {Gar{\'e}}, {Garofalo}, {Garralda}, {Gavras}, {Gerssen}, {Geyer}, {Gilmore},
  {Girona}, {Giuffrida}, {Gomes}, {Gonz{\'a}lez-Marcos},
  {Gonz{\'a}lez-N{\'u}{\~n}ez}, {Gonz{\'a}lez-Vidal}, {Granvik}, {Guerrier},
  {Guillout}, {Guiraud}, {G{\'u}rpide}, {Guti{\'e}rrez-S{\'a}nchez}, {Guy},
  {Haigron}, {Hatzidimitriou}, {Haywood}, {Heiter}, {Helmi}, {Hobbs},
  {Hofmann}, {Holl}, {Holland}, {Hunt}, {Hypki}, {Icardi}, {Irwin}, {Jevardat
  de Fombelle}, {Jofr{\'e}}, {Jonker}, {Jorissen}, {Julbe}, {Karampelas},
  {Kochoska}, {Kohley}, {Kolenberg}, {Kontizas}, {Koposov}, {Kordopatis},
  {Koubsky}, {Kowalczyk}, {Krone-Martins}, {Kudryashova}, {Kull}, {Bachchan},
  {Lacoste-Seris}, {Lanza}, {Lavigne}, {Le Poncin-Lafitte}, {Lebreton},
  {Lebzelter}, {Leccia}, {Leclerc}, {Lecoeur-Taibi}, {Lemaitre}, {Lenhardt},
  {Leroux}, {Liao}, {Licata}, {Lindstr{\o}m}, {Lister}, {Livanou}, {Lobel},
  {L{\"o}ffler}, {L{\'o}pez}, {Lopez-Lozano}, {Lorenz}, {Loureiro},
  {MacDonald}, {Magalh{\~a}es Fernandes}, {Managau}, {Mann}, {Mantelet},
  {Marchal}, {Marchant}, {Marconi}, {Marie}, {Marinoni}, {Marrese},
  {Marschalk{\'o}}, {Marshall}, {Mart{\'\i}n-Fleitas}, {Martino}, {Mary},
  {Matijevi{\v{c}}}, {Mazeh}, {McMillan}, {Messina}, {Mestre}, {Michalik},
  {Millar}, {Miranda}, {Molina}, {Molinaro}, {Molinaro}, {Moln{\'a}r},
  {Moniez}, {Montegriffo}, {Monteiro}, {Mor}, {Mora}, {Morbidelli}, {Morel},
  {Morgenthaler}, {Morley}, {Morris}, {Mulone}, {Muraveva}, {Musella},
  {Narbonne}, {Nelemans}, {Nicastro}, {Noval}, {Ord{\'e}novic},
  {Ordieres-Mer{\'e}}, {Osborne}, {Pagani}, {Pagano}, {Pailler}, {Palacin},
  {Palaversa}, {Parsons}, {Paulsen}, {Pecoraro}, {Pedrosa}, {Pentik{\"a}inen},
  {Pereira}, {Pichon}, {Piersimoni}, {Pineau}, {Plachy}, {Plum}, {Poujoulet},
  {Pr{\v{s}}a}, {Pulone}, {Ragaini}, {Rago}, {Rambaux}, {Ramos-Lerate},
  {Ranalli}, {Rauw}, {Read}, {Regibo}, {Renk}, {Reyl{\'e}}, {Ribeiro},
  {Rimoldini}, {Ripepi}, {Riva}, {Rixon}, {Roelens}, {Romero-G{\'o}mez},
  {Rowell}, {Royer}, {Rudolph}, {Ruiz-Dern}, {Sadowski}, {Sagrist{\`a}
  Sell{\'e}s}, {Sahlmann}, {Salgado}, {Salguero}, {Sarasso}, {Savietto},
  {Schnorhk}, {Schultheis}, {Sciacca}, {Segol}, {Segovia}, {Segransan},
  {Serpell}, {Shih}, {Smareglia}, {Smart}, {Smith}, {Solano}, {Solitro},
  {Sordo}, {Soria Nieto}, {Souchay}, {Spagna}, {Spoto}, {Stampa}, {Steele},
  {Steidelm{\"u}ller}, {Stephenson}, {Stoev}, {Suess}, {S{\"u}veges}, {Surdej},
  {Szabados}, {Szegedi-Elek}, {Tapiador}, {Taris}, {Tauran}, {Taylor},
  {Teixeira}, {Terrett}, {Tingley}, {Trager}, {Turon}, {Ulla}, {Utrilla},
  {Valentini}, {van Elteren}, {Van Hemelryck}, {van Leeuwen}, {Varadi},
  {Vecchiato}, {Veljanoski}, {Via}, {Vicente}, {Vogt}, {Voss}, {Votruba},
  {Voutsinas}, {Walmsley}, {Weiler}, {Weingrill}, {Werner}, {Wevers},
  {Whitehead}, {Wyrzykowski}, {Yoldas}, {{\v{Z}}erjal}, {Zucker}, {Zurbach},
  {Zwitter}, {Alecu}, {Allen}, {Allende Prieto}, {Amorim},
  {Anglada-Escud{\'e}}, {Arsenijevic}, {Azaz}, {Balm}, {Beck}, {Bernstein},
  {Bigot}, {Bijaoui}, {Blasco}, {Bonfigli}, {Bono}, {Boudreault}, {Bressan},
  {Brown}, {Brunet}, {Bunclark}, {Buonanno}, {Butkevich}, {Carret}, {Carrion},
  {Chemin}, {Ch{\'e}reau}, {Corcione}, {Darmigny}, {de Boer}, {de Teodoro}, {de
  Zeeuw}, {Delle Luche}, {Domingues}, {Dubath}, {Fodor}, {Fr{\'e}zouls},
  {Fries}, {Fustes}, {Fyfe}, {Gallardo}, {Gallegos}, {Gardiol}, {Gebran},
  {Gomboc}, {G{\'o}mez}, {Grux}, {Gueguen}, {Heyrovsky}, {Hoar}, {Iannicola},
  {Isasi Parache}, {Janotto}, {Joliet}, {Jonckheere}, {Keil}, {Kim},
  {Klagyivik}, {Klar}, {Knude}, {Kochukhov}, {Kolka}, {Kos}, {Kutka}, {Lainey},
  {LeBouquin}, {Liu}, {Loreggia}, {Makarov}, {Marseille}, {Martayan},
  {Martinez-Rubi}, {Massart}, {Meynadier}, {Mignot}, {Munari}, {Nguyen},
  {Nordlander}, {Ocvirk}, {O'Flaherty}, {Olias Sanz}, {Ortiz}, {Osorio},
  {Oszkiewicz}, {Ouzounis}, {Palmer}, {Park}, {Pasquato}, {Peltzer}, {Peralta},
  {P{\'e}turaud}, {Pieniluoma}, {Pigozzi}, {Poels}, {Prat}, {Prod'homme},
  {Raison}, {Rebordao}, {Risquez}, {Rocca-Volmerange}, {Rosen}, {Ruiz-Fuertes},
  {Russo}, {Sembay}, {Serraller Vizcaino}, {Short}, {Siebert}, {Silva},
  {Sinachopoulos}, {Slezak}, {Soffel}, {Sosnowska}, {Strai{\v{z}}ys}, {ter
  Linden}, {Terrell}, {Theil}, {Tiede}, {Troisi}, {Tsalmantza}, {Tur},
  {Vaccari}, {Vachier}, {Valles}, {Van Hamme}, {Veltz}, {Virtanen}, {Wallut},
  {Wichmann}, {Wilkinson}, {Ziaeepour}, \& {Zschocke}}]{2016A&A...595A...1G}
{Gaia Collaboration}, {Prusti}, T., {de Bruijne}, J.~H.~J., {et~al.}
  2016{\natexlab{a}}, \aap, 595, A1, \dodoi{10.1051/0004-6361/201629272}

\bibitem[{{Gaia Collaboration} {et~al.}(2016{\natexlab{b}}){Gaia
  Collaboration}, {Brown}, {Vallenari}, {Prusti}, {de Bruijne}, {Mignard},
  {Drimmel}, {Babusiaux}, {Bailer-Jones}, {Bastian}, {Biermann}, {Evans},
  {Eyer}, {Jansen}, {Jordi}, {Katz}, {Klioner}, {Lammers}, {Lindegren}, {Luri},
  {O'Mullane}, {Panem}, {Pourbaix}, {Randich}, {Sartoretti}, {Siddiqui},
  {Soubiran}, {Valette}, {van Leeuwen}, {Walton}, {Aerts}, {Arenou}, {Cropper},
  {H{\o}g}, {Lattanzi}, {Grebel}, {Holland}, {Huc}, {Passot}, {Perryman},
  {Bramante}, {Cacciari}, {Casta{\~n}eda}, {Chaoul}, {Cheek}, {De Angeli},
  {Fabricius}, {Guerra}, {Hern{\'a}ndez}, {Jean-Antoine-Piccolo}, {Masana},
  {Messineo}, {Mowlavi}, {Nienartowicz}, {Ord{\'o}{\~n}ez-Blanco}, {Panuzzo},
  {Portell}, {Richards}, {Riello}, {Seabroke}, {Tanga}, {Th{\'e}venin},
  {Torra}, {Els}, {Gracia-Abril}, {Comoretto}, {Garcia-Reinaldos}, {Lock},
  {Mercier}, {Altmann}, {Andrae}, {Astraatmadja}, {Bellas-Velidis}, {Benson},
  {Berthier}, {Blomme}, {Busso}, {Carry}, {Cellino}, {Clementini}, {Cowell},
  {Creevey}, {Cuypers}, {Davidson}, {De Ridder}, {de Torres}, {Delchambre},
  {Dell'Oro}, {Ducourant}, {Fr{\'e}mat}, {Garc{\'\i}a-Torres}, {Gosset},
  {Halbwachs}, {Hambly}, {Harrison}, {Hauser}, {Hestroffer}, {Hodgkin},
  {Huckle}, {Hutton}, {Jasniewicz}, {Jordan}, {Kontizas}, {Korn}, {Lanzafame},
  {Manteiga}, {Moitinho}, {Muinonen}, {Osinde}, {Pancino}, {Pauwels}, {Petit},
  {Recio-Blanco}, {Robin}, {Sarro}, {Siopis}, {Smith}, {Smith}, {Sozzetti},
  {Thuillot}, {van Reeven}, {Viala}, {Abbas}, {Abreu Aramburu}, {Accart},
  {Aguado}, {Allan}, {Allasia}, {Altavilla}, {{\'A}lvarez}, {Alves},
  {Anderson}, {Andrei}, {Anglada Varela}, {Antiche}, {Antoja}, {Ant{\'o}n},
  {Arcay}, {Bach}, {Baker}, {Balaguer-N{\'u}{\~n}ez}, {Barache}, {Barata},
  {Barbier}, {Barblan}, {Barrado y Navascu{\'e}s}, {Barros}, {Barstow},
  {Becciani}, {Bellazzini}, {Bello Garc{\'\i}a}, {Belokurov}, {Bendjoya},
  {Berihuete}, {Bianchi}, {Bienaym{\'e}}, {Billebaud}, {Blagorodnova},
  {Blanco-Cuaresma}, {Boch}, {Bombrun}, {Borrachero}, {Bouquillon}, {Bourda},
  {Bouy}, {Bragaglia}, {Breddels}, {Brouillet}, {Br{\"u}semeister},
  {Bucciarelli}, {Burgess}, {Burgon}, {Burlacu}, {Busonero}, {Buzzi}, {Caffau},
  {Cambras}, {Campbell}, {Cancelliere}, {Cantat-Gaudin}, {Carlucci},
  {Carrasco}, {Castellani}, {Charlot}, {Charnas}, {Chiavassa}, {Clotet},
  {Cocozza}, {Collins}, {Costigan}, {Crifo}, {Cross}, {Crosta}, {Crowley},
  {Dafonte}, {Damerdji}, {Dapergolas}, {David}, {David}, {De Cat}, {de Felice},
  {de Laverny}, {De Luise}, {De March}, {de Martino}, {de Souza}, {Debosscher},
  {del Pozo}, {Delbo}, {Delgado}, {Delgado}, {Di Matteo}, {Diakite},
  {Distefano}, {Dolding}, {Dos Anjos}, {Drazinos}, {Duran}, {Dzigan},
  {Edvardsson}, {Enke}, {Evans}, {Eynard Bontemps}, {Fabre}, {Fabrizio},
  {Faigler}, {Falc{\~a}o}, {Farr{\`a}s Casas}, {Federici}, {Fedorets},
  {Fern{\'a}ndez-Hern{\'a}ndez}, {Fernique}, {Fienga}, {Figueras}, {Filippi},
  {Findeisen}, {Fonti}, {Fouesneau}, {Fraile}, {Fraser}, {Fuchs}, {Gai},
  {Galleti}, {Galluccio}, {Garabato}, {Garc{\'\i}a-Sedano}, {Garofalo},
  {Garralda}, {Gavras}, {Gerssen}, {Geyer}, {Gilmore}, {Girona}, {Giuffrida},
  {Gomes}, {Gonz{\'a}lez-Marcos}, {Gonz{\'a}lez-N{\'u}{\~n}ez},
  {Gonz{\'a}lez-Vidal}, {Granvik}, {Guerrier}, {Guillout}, {Guiraud},
  {G{\'u}rpide}, {Guti{\'e}rrez-S{\'a}nchez}, {Guy}, {Haigron},
  {Hatzidimitriou}, {Haywood}, {Heiter}, {Helmi}, {Hobbs}, {Hofmann}, {Holl},
  {Holland}, {Hunt}, {Hypki}, {Icardi}, {Irwin}, {Jevardat de Fombelle},
  {Jofr{\'e}}, {Jonker}, {Jorissen}, {Julbe}, {Karampelas}, {Kochoska},
  {Kohley}, {Kolenberg}, {Kontizas}, {Koposov}, {Kordopatis}, {Koubsky},
  {Krone-Martins}, {Kudryashova}, {Kull}, {Bachchan}, {Lacoste-Seris}, {Lanza},
  {Lavigne}, {Le Poncin-Lafitte}, {Lebreton}, {Lebzelter}, {Leccia}, {Leclerc},
  {Lecoeur-Taibi}, {Lemaitre}, {Lenhardt}, {Leroux}, {Liao}, {Licata},
  {Lindstr{\o}m}, {Lister}, {Livanou}, {Lobel}, {L{\"o}ffler}, {L{\'o}pez},
  {Lorenz}, {MacDonald}, {Magalh{\~a}es Fernandes}, {Managau}, {Mann},
  {Mantelet}, {Marchal}, {Marchant}, {Marconi}, {Marinoni}, {Marrese},
  {Marschalk{\'o}}, {Marshall}, {Mart{\'\i}n-Fleitas}, {Martino}, {Mary},
  {Matijevi{\v{c}}}, {Mazeh}, {McMillan}, {Messina}, {Michalik}, {Millar},
  {Miranda}, {Molina}, {Molinaro}, {Molinaro}, {Moln{\'a}r}, {Moniez},
  {Montegriffo}, {Mor}, {Mora}, {Morbidelli}, {Morel}, {Morgenthaler},
  {Morris}, {Mulone}, {Muraveva}, {Musella}, {Narbonne}, {Nelemans},
  {Nicastro}, {Noval}, {Ord{\'e}novic}, {Ordieres-Mer{\'e}}, {Osborne},
  {Pagani}, {Pagano}, {Pailler}, {Palacin}, {Palaversa}, {Parsons}, {Pecoraro},
  {Pedrosa}, {Pentik{\"a}inen}, {Pichon}, {Piersimoni}, {Pineau}, {Plachy},
  {Plum}, {Poujoulet}, {Pr{\v{s}}a}, {Pulone}, {Ragaini}, {Rago}, {Rambaux},
  {Ramos-Lerate}, {Ranalli}, {Rauw}, {Read}, {Regibo}, {Reyl{\'e}}, {Ribeiro},
  {Rimoldini}, {Ripepi}, {Riva}, {Rixon}, {Roelens}, {Romero-G{\'o}mez},
  {Rowell}, {Royer}, {Ruiz-Dern}, {Sadowski}, {Sagrist{\`a} Sell{\'e}s},
  {Sahlmann}, {Salgado}, {Salguero}, {Sarasso}, {Savietto}, {Schultheis},
  {Sciacca}, {Segol}, {Segovia}, {Segransan}, {Shih}, {Smareglia}, {Smart},
  {Solano}, {Solitro}, {Sordo}, {Soria Nieto}, {Souchay}, {Spagna}, {Spoto},
  {Stampa}, {Steele}, {Steidelm{\"u}ller}, {Stephenson}, {Stoev}, {Suess},
  {S{\"u}veges}, {Surdej}, {Szabados}, {Szegedi-Elek}, {Tapiador}, {Taris},
  {Tauran}, {Taylor}, {Teixeira}, {Terrett}, {Tingley}, {Trager}, {Turon},
  {Ulla}, {Utrilla}, {Valentini}, {van Elteren}, {Van Hemelryck}, {van
  Leeuwen}, {Varadi}, {Vecchiato}, {Veljanoski}, {Via}, {Vicente}, {Vogt},
  {Voss}, {Votruba}, {Voutsinas}, {Walmsley}, {Weiler}, {Weingrill}, {Wevers},
  {Wyrzykowski}, {Yoldas}, {{\v{Z}}erjal}, {Zucker}, {Zurbach}, {Zwitter},
  {Alecu}, {Allen}, {Allende Prieto}, {Amorim}, {Anglada-Escud{\'e}},
  {Arsenijevic}, {Azaz}, {Balm}, {Beck}, {Bernstein}, {Bigot}, {Bijaoui},
  {Blasco}, {Bonfigli}, {Bono}, {Boudreault}, {Bressan}, {Brown}, {Brunet},
  {Bunclark}, {Buonanno}, {Butkevich}, {Carret}, {Carrion}, {Chemin},
  {Ch{\'e}reau}, {Corcione}, {Darmigny}, {de Boer}, {de Teodoro}, {de Zeeuw},
  {Delle Luche}, {Domingues}, {Dubath}, {Fodor}, {Fr{\'e}zouls}, {Fries},
  {Fustes}, {Fyfe}, {Gallardo}, {Gallegos}, {Gardiol}, {Gebran}, {Gomboc},
  {G{\'o}mez}, {Grux}, {Gueguen}, {Heyrovsky}, {Hoar}, {Iannicola}, {Isasi
  Parache}, {Janotto}, {Joliet}, {Jonckheere}, {Keil}, {Kim}, {Klagyivik},
  {Klar}, {Knude}, {Kochukhov}, {Kolka}, {Kos}, {Kutka}, {Lainey}, {LeBouquin},
  {Liu}, {Loreggia}, {Makarov}, {Marseille}, {Martayan}, {Martinez-Rubi},
  {Massart}, {Meynadier}, {Mignot}, {Munari}, {Nguyen}, {Nordlander}, {Ocvirk},
  {O'Flaherty}, {Olias Sanz}, {Ortiz}, {Osorio}, {Oszkiewicz}, {Ouzounis},
  {Palmer}, {Park}, {Pasquato}, {Peltzer}, {Peralta}, {P{\'e}turaud},
  {Pieniluoma}, {Pigozzi}, {Poels}, {Prat}, {Prod'homme}, {Raison}, {Rebordao},
  {Risquez}, {Rocca-Volmerange}, {Rosen}, {Ruiz-Fuertes}, {Russo}, {Sembay},
  {Serraller Vizcaino}, {Short}, {Siebert}, {Silva}, {Sinachopoulos}, {Slezak},
  {Soffel}, {Sosnowska}, {Strai{\v{z}}ys}, {ter Linden}, {Terrell}, {Theil},
  {Tiede}, {Troisi}, {Tsalmantza}, {Tur}, {Vaccari}, {Vachier}, {Valles}, {Van
  Hamme}, {Veltz}, {Virtanen}, {Wallut}, {Wichmann}, {Wilkinson}, {Ziaeepour},
  \& {Zschocke}}]{2016A&A...595A...2G}
{Gaia Collaboration}, {Brown}, A.~G.~A., {Vallenari}, A., {et~al.}
  2016{\natexlab{b}}, \aap, 595, A2, \dodoi{10.1051/0004-6361/201629512}

\bibitem[{{Garc{\'\i}a-Argum{\'a}nez}
  {et~al.}(2022){Garc{\'\i}a-Argum{\'a}nez}, {P{\'e}rez-Gonz{\'a}lez}, {Gil de
  Paz}, {Snyder}, {Arrabal Haro}, {Bagley}, {Finkelstein}, {Kartaltepe},
  {Koekemoer}, {Papovich}, {Pirzkal}, {Ferguson}, {Yung}, {Annunziatella},
  {Cleri}, {Cooper}, {Costantin}, {Holwerda}, {Mar{\'\i}a M{\'e}rida
  Gonz{\'a}lez}, {Rose}, {Giavalisco}, {Grogin}, \&
  {Kocevski}}]{2022arXiv220714062G}
{Garc{\'\i}a-Argum{\'a}nez}, {\'A}., {P{\'e}rez-Gonz{\'a}lez}, P.~G., {Gil de
  Paz}, A., {et~al.} 2022, arXiv e-prints, arXiv:2207.14062.
\newblock \doarXiv{2207.14062}

\bibitem[{{Grogin} {et~al.}(2011){Grogin}, {Kocevski}, {Faber}, {Ferguson},
  {Koekemoer}, {Riess}, {Acquaviva}, {Alexander}, {Almaini}, {Ashby}, {Barden},
  {Bell}, {Bournaud}, {Brown}, {Caputi}, {Casertano}, {Cassata}, {Castellano},
  {Challis}, {Chary}, {Cheung}, {Cirasuolo}, {Conselice}, {Roshan Cooray},
  {Croton}, {Daddi}, {Dahlen}, {Dav{\'e}}, {de Mello}, {Dekel}, {Dickinson},
  {Dolch}, {Donley}, {Dunlop}, {Dutton}, {Elbaz}, {Fazio}, {Filippenko},
  {Finkelstein}, {Fontana}, {Gardner}, {Garnavich}, {Gawiser}, {Giavalisco},
  {Grazian}, {Guo}, {Hathi}, {H{\"a}ussler}, {Hopkins}, {Huang}, {Huang},
  {Jha}, {Kartaltepe}, {Kirshner}, {Koo}, {Lai}, {Lee}, {Li}, {Lotz}, {Lucas},
  {Madau}, {McCarthy}, {McGrath}, {McIntosh}, {McLure}, {Mobasher},
  {Moustakas}, {Mozena}, {Nandra}, {Newman}, {Niemi}, {Noeske}, {Papovich},
  {Pentericci}, {Pope}, {Primack}, {Rajan}, {Ravindranath}, {Reddy}, {Renzini},
  {Rix}, {Robaina}, {Rodney}, {Rosario}, {Rosati}, {Salimbeni}, {Scarlata},
  {Siana}, {Simard}, {Smidt}, {Somerville}, {Spinrad}, {Straughn}, {Strolger},
  {Telford}, {Teplitz}, {Trump}, {van der Wel}, {Villforth}, {Wechsler},
  {Weiner}, {Wiklind}, {Wild}, {Wilson}, {Wuyts}, {Yan}, \&
  {Yun}}]{2011ApJS..197...35G}
{Grogin}, N.~A., {Kocevski}, D.~D., {Faber}, S.~M., {et~al.} 2011, \apjs, 197,
  35, \dodoi{10.1088/0067-0049/197/2/35}

\bibitem[{{Gruppioni} {et~al.}(2013){Gruppioni}, {Pozzi}, {Rodighiero},
  {Delvecchio}, {Berta}, {Pozzetti}, {Zamorani}, {Andreani}, {Cimatti},
  {Ilbert}, {Le Floc'h}, {Lutz}, {Magnelli}, {Marchetti}, {Monaco}, {Nordon},
  {Oliver}, {Popesso}, {Riguccini}, {Roseboom}, {Rosario}, {Sargent},
  {Vaccari}, {Altieri}, {Aussel}, {Bongiovanni}, {Cepa}, {Daddi},
  {Dom{\'\i}nguez-S{\'a}nchez}, {Elbaz}, {F{\"o}rster Schreiber}, {Genzel},
  {Iribarrem}, {Magliocchetti}, {Maiolino}, {Poglitsch}, {P{\'e}rez
  Garc{\'\i}a}, {Sanchez-Portal}, {Sturm}, {Tacconi}, {Valtchanov}, {Amblard},
  {Arumugam}, {Bethermin}, {Bock}, {Boselli}, {Buat}, {Burgarella},
  {Castro-Rodr{\'\i}guez}, {Cava}, {Chanial}, {Clements}, {Conley}, {Cooray},
  {Dowell}, {Dwek}, {Eales}, {Franceschini}, {Glenn}, {Griffin},
  {Hatziminaoglou}, {Ibar}, {Isaak}, {Ivison}, {Lagache}, {Levenson}, {Lu},
  {Madden}, {Maffei}, {Mainetti}, {Nguyen}, {O'Halloran}, {Page}, {Panuzzo},
  {Papageorgiou}, {Pearson}, {P{\'e}rez-Fournon}, {Pohlen}, {Rigopoulou},
  {Rowan-Robinson}, {Schulz}, {Scott}, {Seymour}, {Shupe}, {Smith}, {Stevens},
  {Symeonidis}, {Trichas}, {Tugwell}, {Vigroux}, {Wang}, {Wright}, {Xu},
  {Zemcov}, {Bardelli}, {Carollo}, {Contini}, {Le F{\'e}vre}, {Lilly},
  {Mainieri}, {Renzini}, {Scodeggio}, \& {Zucca}}]{2013MNRAS.432...23G}
{Gruppioni}, C., {Pozzi}, F., {Rodighiero}, G., {et~al.} 2013, \mnras, 432, 23,
  \dodoi{10.1093/mnras/stt308}

\bibitem[{{Guo} {et~al.}(2013){Guo}, {Ferguson}, {Giavalisco}, {Barro},
  {Willner}, {Ashby}, {Dahlen}, {Donley}, {Faber}, {Fontana}, {Galametz},
  {Grazian}, {Huang}, {Kocevski}, {Koekemoer}, {Koo}, {McGrath}, {Peth},
  {Salvato}, {Wuyts}, {Castellano}, {Cooray}, {Dickinson}, {Dunlop}, {Fazio},
  {Gardner}, {Gawiser}, {Grogin}, {Hathi}, {Hsu}, {Lee}, {Lucas}, {Mobasher},
  {Nandra}, {Newman}, \& {van der Wel}}]{2013ApJS..207...24G}
{Guo}, Y., {Ferguson}, H.~C., {Giavalisco}, M., {et~al.} 2013, \apjs, 207, 24,
  \dodoi{10.1088/0067-0049/207/2/24}

\bibitem[{{Harikane} {et~al.}(2022){Harikane}, {Ouchi}, {Oguri}, {Ono},
  {Nakajima}, {Isobe}, {Umeda}, {Mawatari}, \& {Zhang}}]{2022arXiv220801612H}
{Harikane}, Y., {Ouchi}, M., {Oguri}, M., {et~al.} 2022, arXiv e-prints,
  arXiv:2208.01612.
\newblock \doarXiv{2208.01612}

\bibitem[{{Huang} {et~al.}(2011){Huang}, {Zheng}, {Rigopoulou}, {Magdis},
  {Fazio}, \& {Wang}}]{2011ApJ...742L..13H}
{Huang}, J.~S., {Zheng}, X.~Z., {Rigopoulou}, D., {et~al.} 2011, \apjl, 742,
  L13, \dodoi{10.1088/2041-8205/742/1/L13}

\bibitem[{{Hunter}(2007)}]{2007CSE.....9...90H}
{Hunter}, J.~D. 2007, Computing in Science and Engineering, 9, 90,
  \dodoi{10.1109/MCSE.2007.55}

\bibitem[{{Ivison} {et~al.}(2007){Ivison}, {Chapman}, {Faber}, {Smail},
  {Biggs}, {Conselice}, {Wilson}, {Salim}, {Huang}, \&
  {Willner}}]{2007ApJ...660L..77I}
{Ivison}, R.~J., {Chapman}, S.~C., {Faber}, S.~M., {et~al.} 2007, \apjl, 660,
  L77, \dodoi{10.1086/517917}

\bibitem[{{Ji} \& {Giavalisco}(2022)}]{2022arXiv220804325J}
{Ji}, Z., \& {Giavalisco}, M. 2022, arXiv e-prints, arXiv:2208.04325.
\newblock \doarXiv{2208.04325}

\bibitem[{{Johnson} {et~al.}(2021){Johnson}, {Leja}, {Conroy}, \&
  {Speagle}}]{2021ApJS..254...22J}
{Johnson}, B.~D., {Leja}, J., {Conroy}, C., \& {Speagle}, J.~S. 2021, \apjs,
  254, 22, \dodoi{10.3847/1538-4365/abef67}

\bibitem[{{Kennicutt}(1998)}]{1998ARA&A..36..189K}
{Kennicutt}, Robert~C., J. 1998, \araa, 36, 189,
  \dodoi{10.1146/annurev.astro.36.1.189}

\bibitem[{{Koekemoer} {et~al.}(2011){Koekemoer}, {Faber}, {Ferguson}, {Grogin},
  {Kocevski}, {Koo}, {Lai}, {Lotz}, {Lucas}, {McGrath}, {Ogaz}, {Rajan},
  {Riess}, {Rodney}, {Strolger}, {Casertano}, {Castellano}, {Dahlen},
  {Dickinson}, {Dolch}, {Fontana}, {Giavalisco}, {Grazian}, {Guo}, {Hathi},
  {Huang}, {van der Wel}, {Yan}, {Acquaviva}, {Alexander}, {Almaini}, {Ashby},
  {Barden}, {Bell}, {Bournaud}, {Brown}, {Caputi}, {Cassata}, {Challis},
  {Chary}, {Cheung}, {Cirasuolo}, {Conselice}, {Roshan Cooray}, {Croton},
  {Daddi}, {Dav{\'e}}, {de Mello}, {de Ravel}, {Dekel}, {Donley}, {Dunlop},
  {Dutton}, {Elbaz}, {Fazio}, {Filippenko}, {Finkelstein}, {Frazer}, {Gardner},
  {Garnavich}, {Gawiser}, {Gruetzbauch}, {Hartley}, {H{\"a}ussler},
  {Herrington}, {Hopkins}, {Huang}, {Jha}, {Johnson}, {Kartaltepe},
  {Khostovan}, {Kirshner}, {Lani}, {Lee}, {Li}, {Madau}, {McCarthy},
  {McIntosh}, {McLure}, {McPartland}, {Mobasher}, {Moreira}, {Mortlock},
  {Moustakas}, {Mozena}, {Nandra}, {Newman}, {Nielsen}, {Niemi}, {Noeske},
  {Papovich}, {Pentericci}, {Pope}, {Primack}, {Ravindranath}, {Reddy},
  {Renzini}, {Rix}, {Robaina}, {Rosario}, {Rosati}, {Salimbeni}, {Scarlata},
  {Siana}, {Simard}, {Smidt}, {Snyder}, {Somerville}, {Spinrad}, {Straughn},
  {Telford}, {Teplitz}, {Trump}, {Vargas}, {Villforth}, {Wagner}, {Wandro},
  {Wechsler}, {Weiner}, {Wiklind}, {Wild}, {Wilson}, {Wuyts}, \&
  {Yun}}]{2011ApJS..197...36K}
{Koekemoer}, A.~M., {Faber}, S.~M., {Ferguson}, H.~C., {et~al.} 2011, \apjs,
  197, 36, \dodoi{10.1088/0067-0049/197/2/36}

\bibitem[{{Kokorev} {et~al.}(2022){Kokorev}, {Brammer}, {Fujimoto}, {Kohno},
  {Magdis}, {Valentino}, {Toft}, {Oesch}, {Bauer}, {Coe}, {Egami}, {Oguri},
  {Ouchi}, {Postman}, {Richard}, {Jolly}, {Knudsen}, {Sun}, {Weaver}, {Ao},
  {Baker}, {Caputi}, {Espada}, {Hatsukade}, {Koekemoer}, {Mu{\~n}oz Arancibia},
  {Shimasaku}, {Umehata}, {Wang}, \& {Wang}}]{2022arXiv220707125K}
{Kokorev}, V., {Brammer}, G., {Fujimoto}, S., {et~al.} 2022, arXiv e-prints,
  arXiv:2207.07125.
\newblock \doarXiv{2207.07125}

\bibitem[{{Kriek} {et~al.}(2015){Kriek}, {Shapley}, {Reddy}, {Siana}, {Coil},
  {Mobasher}, {Freeman}, {de Groot}, {Price}, {Sanders}, {Shivaei}, {Brammer},
  {Momcheva}, {Skelton}, {van Dokkum}, {Whitaker}, {Aird}, {Azadi}, {Kassis},
  {Bullock}, {Conroy}, {Dav{\'e}}, {Kere{\v{s}}}, \&
  {Krumholz}}]{2015ApJS..218...15K}
{Kriek}, M., {Shapley}, A.~E., {Reddy}, N.~A., {et~al.} 2015, \apjs, 218, 15,
  \dodoi{10.1088/0067-0049/218/2/15}

\bibitem[{{Kron}(1980)}]{1980ApJS...43..305K}
{Kron}, R.~G. 1980, \apjs, 43, 305, \dodoi{10.1086/190669}

\bibitem[{{Labb{\'e}} {et~al.}(2013){Labb{\'e}}, {Oesch}, {Bouwens},
  {Illingworth}, {Magee}, {Gonz{\'a}lez}, {Carollo}, {Franx}, {Trenti}, {van
  Dokkum}, \& {Stiavelli}}]{2013ApJ...777L..19L}
{Labb{\'e}}, I., {Oesch}, P.~A., {Bouwens}, R.~J., {et~al.} 2013, \apjl, 777,
  L19, \dodoi{10.1088/2041-8205/777/2/L19}

\bibitem[{{Labbe} {et~al.}(2022){Labbe}, {van Dokkum}, {Nelson}, {Bezanson},
  {Suess}, {Leja}, {Brammer}, {Whitaker}, {Mathews}, \&
  {Stefanon}}]{2022arXiv220712446L}
{Labbe}, I., {van Dokkum}, P., {Nelson}, E., {et~al.} 2022, arXiv e-prints,
  arXiv:2207.12446.
\newblock \doarXiv{2207.12446}

\bibitem[{{Laporte} {et~al.}(2022){Laporte}, {Ellis}, {Witten}, \&
  {Roberts-Borsani}}]{2022arXiv221205072L}
{Laporte}, N., {Ellis}, R.~S., {Witten}, C.~E.~C., \& {Roberts-Borsani}, G.
  2022, arXiv e-prints, arXiv:2212.05072.
\newblock \doarXiv{2212.05072}

\bibitem[{{Leja} {et~al.}(2019{\natexlab{a}}){Leja}, {Carnall}, {Johnson},
  {Conroy}, \& {Speagle}}]{2019ApJ...876....3L}
{Leja}, J., {Carnall}, A.~C., {Johnson}, B.~D., {Conroy}, C., \& {Speagle},
  J.~S. 2019{\natexlab{a}}, \apj, 876, 3, \dodoi{10.3847/1538-4357/ab133c}

\bibitem[{{Leja} {et~al.}(2019{\natexlab{b}}){Leja}, {Tacchella}, \&
  {Conroy}}]{2019ApJ...880L...9L}
{Leja}, J., {Tacchella}, S., \& {Conroy}, C. 2019{\natexlab{b}}, \apjl, 880,
  L9, \dodoi{10.3847/2041-8213/ab2f8c}

\bibitem[{{Lisiecki} {et~al.}(2022){Lisiecki}, {Ma{\l}ek}, {Siudek}, {Pollo},
  {Krywult}, {Karska}, \& {Junais}}]{2022arXiv220804601L}
{Lisiecki}, K., {Ma{\l}ek}, K., {Siudek}, M., {et~al.} 2022, arXiv e-prints,
  arXiv:2208.04601.
\newblock \doarXiv{2208.04601}

\bibitem[{{Lohmann} {et~al.}(2021){Lohmann}, {Schnorr-M{\"u}ller}, {Trevisan},
  {Riffel}, {Mallmann}, {Chies-Santos}, \& {Furlanetto}}]{2021IAUS..359..441L}
{Lohmann}, F.~S., {Schnorr-M{\"u}ller}, A., {Trevisan}, M., {et~al.} 2021, in
  Galaxy Evolution and Feedback across Different Environments, ed. T.~{Storchi
  Bergmann}, W.~{Forman}, R.~{Overzier}, \& R.~{Riffel}, Vol. 359, 441--443,
  \dodoi{10.1017/S1743921320002045}

\bibitem[{{Lovell} {et~al.}(2022){Lovell}, {Harrison}, {Harikane}, {Tacchella},
  \& {Wilkins}}]{2022arXiv220810479L}
{Lovell}, C.~C., {Harrison}, I., {Harikane}, Y., {Tacchella}, S., \& {Wilkins},
  S.~M. 2022, arXiv e-prints, arXiv:2208.10479.
\newblock \doarXiv{2208.10479}

\bibitem[{{Mancini} {et~al.}(2009){Mancini}, {Matute}, {Cimatti}, {Daddi},
  {Dickinson}, {Rodighiero}, {Bolzonella}, \& {Pozzetti}}]{2009A&A...500..705M}
{Mancini}, C., {Matute}, I., {Cimatti}, A., {et~al.} 2009, \aap, 500, 705,
  \dodoi{10.1051/0004-6361/200810630}

\bibitem[{{Manning} {et~al.}(2022){Manning}, {Casey}, {Zavala}, {Magdis},
  {Drew}, {Champagne}, {Aravena}, {B{\'e}thermin}, {Clements}, {Finkelstein},
  {Fujimoto}, {Hayward}, {Hodge}, {Ilbert}, {Kartaltepe}, {Knudsen},
  {Koekemoer}, {Man}, {Sanders}, {Sheth}, {Spilker}, {Staguhn}, {Talia},
  {Treister}, \& {Yun}}]{2022ApJ...925...23M}
{Manning}, S.~M., {Casey}, C.~M., {Zavala}, J.~A., {et~al.} 2022, \apj, 925,
  23, \dodoi{10.3847/1538-4357/ac366a}

\bibitem[{{Matthee} {et~al.}(2022){Matthee}, {Mackenzie}, {Simcoe}, {Kashino},
  {Lilly}, {Bordoloi}, \& {Eilers}}]{2022arXiv221108255M}
{Matthee}, J., {Mackenzie}, R., {Simcoe}, R.~A., {et~al.} 2022, arXiv e-prints,
  arXiv:2211.08255.
\newblock \doarXiv{2211.08255}

\bibitem[{{Miller} {et~al.}(2022){Miller}, {Whitaker}, {Nelson}, {van Dokkum},
  {Bezanson}, {Brammer}, {Heintz}, {Leja}, {Suess}, \&
  {Weaver}}]{2022arXiv220912954M}
{Miller}, T.~B., {Whitaker}, K.~E., {Nelson}, E.~J., {et~al.} 2022, arXiv
  e-prints, arXiv:2209.12954.
\newblock \doarXiv{2209.12954}

\bibitem[{{Muzzin} {et~al.}(2013){Muzzin}, {Marchesini}, {Stefanon}, {Franx},
  {McCracken}, {Milvang-Jensen}, {Dunlop}, {Fynbo}, {Brammer}, {Labb{\'e}}, \&
  {van Dokkum}}]{2013ApJ...777...18M}
{Muzzin}, A., {Marchesini}, D., {Stefanon}, M., {et~al.} 2013, \apj, 777, 18,
  \dodoi{10.1088/0004-637X/777/1/18}

\bibitem[{{Naidu} {et~al.}(2022){Naidu}, {Oesch}, {van Dokkum}, {Nelson},
  {Suess}, {Whitaker}, {Allen}, {Bezanson}, {Bouwens}, {Brammer}, {Conroy},
  {Illingworth}, {Labbe}, {Leja}, {Leonova}, {Matthee}, {Price}, {Setton},
  {Strait}, {Stefanon}, {Tacchella}, {Toft}, {Weaver}, \&
  {Weibel}}]{2022arXiv220709434N}
{Naidu}, R.~P., {Oesch}, P.~A., {van Dokkum}, P., {et~al.} 2022, arXiv
  e-prints, arXiv:2207.09434.
\newblock \doarXiv{2207.09434}

\bibitem[{{Nandra} {et~al.}(2015){Nandra}, {Laird}, {Aird}, {Salvato},
  {Georgakakis}, {Barro}, {Perez-Gonzalez}, {Barmby}, {Chary}, {Coil},
  {Cooper}, {Davis}, {Dickinson}, {Faber}, {Fazio}, {Guhathakurta}, {Gwyn},
  {Hsu}, {Huang}, {Ivison}, {Koo}, {Newman}, {Rangel}, {Yamada}, \&
  {Willmer}}]{2015ApJS..220...10N}
{Nandra}, K., {Laird}, E.~S., {Aird}, J.~A., {et~al.} 2015, \apjs, 220, 10,
  \dodoi{10.1088/0067-0049/220/1/10}

\bibitem[{{Nayyeri} {et~al.}(2014){Nayyeri}, {Mobasher}, {Hemmati}, {De
  Barros}, {Ferguson}, {Wiklind}, {Dahlen}, {Dickinson}, {Giavalisco},
  {Fontana}, {Ashby}, {Barro}, {Guo}, {Hathi}, {Kassin}, {Koekemoer},
  {Willner}, {Dunlop}, {Paris}, \& {Targett}}]{2014ApJ...794...68N}
{Nayyeri}, H., {Mobasher}, B., {Hemmati}, S., {et~al.} 2014, \apj, 794, 68,
  \dodoi{10.1088/0004-637X/794/1/68}

\bibitem[{{Nelson} {et~al.}(2022){Nelson}, {Suess}, {Bezanson}, {Price}, {van
  Dokkum}, {Leja}, {Whitaker}, {Labb{\'e}}, {Barrufet}, {Brammer},
  {Eisenstein}, {Heintz}, {Johnson}, {Mathews}, {Miller}, {Oesch}, {Sandles},
  {Setton}, {Speagle}, {Tacchella}, {Tadaki}, \&
  {Weaver}}]{2022arXiv220801630N}
{Nelson}, E.~J., {Suess}, K.~A., {Bezanson}, R., {et~al.} 2022, arXiv e-prints,
  arXiv:2208.01630.
\newblock \doarXiv{2208.01630}

\bibitem[{{Oke} \& {Gunn}(1983)}]{1983ApJ...266..713O}
{Oke}, J.~B., \& {Gunn}, J.~E. 1983, \apj, 266, 713, \dodoi{10.1086/160817}

\bibitem[{{Ono} {et~al.}(2022){Ono}, {Harikane}, {Ouchi}, {Yajima}, {Abe},
  {Isobe}, {Shibuya}, {Zhang}, {Nakajima}, \& {Umeda}}]{2022arXiv220813582O}
{Ono}, Y., {Harikane}, Y., {Ouchi}, M., {et~al.} 2022, arXiv e-prints,
  arXiv:2208.13582.
\newblock \doarXiv{2208.13582}

\bibitem[{{Pacifici} {et~al.}(2012){Pacifici}, {Charlot}, {Blaizot}, \&
  {Brinchmann}}]{2012MNRAS.421.2002P}
{Pacifici}, C., {Charlot}, S., {Blaizot}, J., \& {Brinchmann}, J. 2012, \mnras,
  421, 2002, \dodoi{10.1111/j.1365-2966.2012.20431.x}

\bibitem[{{Patel} {et~al.}(2012){Patel}, {Holden}, {Kelson}, {Franx}, {van der
  Wel}, \& {Illingworth}}]{2012ApJ...748L..27P}
{Patel}, S.~G., {Holden}, B.~P., {Kelson}, D.~D., {et~al.} 2012, \apjl, 748,
  L27, \dodoi{10.1088/2041-8205/748/2/L27}

\bibitem[{{Peng} {et~al.}(2002){Peng}, {Ho}, {Impey}, \&
  {Rix}}]{2002AJ....124..266P}
{Peng}, C.~Y., {Ho}, L.~C., {Impey}, C.~D., \& {Rix}, H.-W. 2002, \aj, 124,
  266, \dodoi{10.1086/340952}

\bibitem[{{P{\'e}rez-Gonz{\'a}lez} {et~al.}(2003){P{\'e}rez-Gonz{\'a}lez}, {Gil
  de Paz}, {Zamorano}, {Gallego}, {Alonso-Herrero}, \&
  {Arag{\'o}n-Salamanca}}]{2003MNRAS.338..508P}
{P{\'e}rez-Gonz{\'a}lez}, P.~G., {Gil de Paz}, A., {Zamorano}, J., {et~al.}
  2003, \mnras, 338, 508, \dodoi{10.1046/j.1365-8711.2003.06077.x}

\bibitem[{{P{\'e}rez-Gonz{\'a}lez}
  {et~al.}(2008{\natexlab{a}}){P{\'e}rez-Gonz{\'a}lez}, {Trujillo}, {Barro},
  {Gallego}, {Zamorano}, \& {Conselice}}]{2008ApJ...687...50P}
{P{\'e}rez-Gonz{\'a}lez}, P.~G., {Trujillo}, I., {Barro}, G., {et~al.}
  2008{\natexlab{a}}, \apj, 687, 50, \dodoi{10.1086/591843}

\bibitem[{{P{\'e}rez-Gonz{\'a}lez} {et~al.}(2005){P{\'e}rez-Gonz{\'a}lez},
  {Rieke}, {Egami}, {Alonso-Herrero}, {Dole}, {Papovich}, {Blaylock}, {Jones},
  {Rieke}, {Rigby}, {Barmby}, {Fazio}, {Huang}, \&
  {Martin}}]{2005ApJ...630...82P}
{P{\'e}rez-Gonz{\'a}lez}, P.~G., {Rieke}, G.~H., {Egami}, E., {et~al.} 2005,
  \apj, 630, 82, \dodoi{10.1086/431894}

\bibitem[{{P{\'e}rez-Gonz{\'a}lez}
  {et~al.}(2008{\natexlab{b}}){P{\'e}rez-Gonz{\'a}lez}, {Rieke}, {Villar},
  {Barro}, {Blaylock}, {Egami}, {Gallego}, {Gil de Paz}, {Pascual}, {Zamorano},
  \& {Donley}}]{2008ApJ...675..234P}
{P{\'e}rez-Gonz{\'a}lez}, P.~G., {Rieke}, G.~H., {Villar}, V., {et~al.}
  2008{\natexlab{b}}, \apj, 675, 234, \dodoi{10.1086/523690}

\bibitem[{{P{\'e}rez-Gonz{\'a}lez} {et~al.}(2010){P{\'e}rez-Gonz{\'a}lez},
  {Egami}, {Rex}, {Rawle}, {Kneib}, {Richard}, {Johansson}, {Altieri}, {Blain},
  {Bock}, {Boone}, {Bridge}, {Chung}, {Cl{\'e}ment}, {Clowe}, {Combes}, {Cuby},
  {Dessauges-Zavadsky}, {Dowell}, {Espino-Briones}, {Fadda}, {Fiedler},
  {Gonzalez}, {Horellou}, {Ilbert}, {Ivison}, {Jauzac}, {Lutz}, {Pell{\'o}},
  {Pereira}, {Rieke}, {Rodighiero}, {Schaerer}, {Smith}, {Valtchanov}, {Walth},
  {van der Werf}, {Werner}, \& {Zemcov}}]{2010A&A...518L..15P}
{P{\'e}rez-Gonz{\'a}lez}, P.~G., {Egami}, E., {Rex}, M., {et~al.} 2010, \aap,
  518, L15, \dodoi{10.1051/0004-6361/201014593}

\bibitem[{{Petrosian}(1976)}]{1976ApJ...209L...1P}
{Petrosian}, V. 1976, \apjl, 210, L53, \dodoi{10.1086/182301}

\bibitem[{{Rawle} {et~al.}(2016){Rawle}, {Altieri}, {Egami},
  {P{\'e}rez-Gonz{\'a}lez}, {Boone}, {Clement}, {Ivison}, {Richard},
  {Rujopakarn}, {Valtchanov}, {Walth}, {Weiner}, {Blain}, {Dessauges-Zavadsky},
  {Kneib}, {Lutz}, {Rodighiero}, {Schaerer}, \& {Smail}}]{2016MNRAS.459.1626R}
{Rawle}, T.~D., {Altieri}, B., {Egami}, E., {et~al.} 2016, \mnras, 459, 1626,
  \dodoi{10.1093/mnras/stw712}

\bibitem[{{Reddy} {et~al.}(2006){Reddy}, {Steidel}, {Fadda}, {Yan}, {Pettini},
  {Shapley}, {Erb}, \& {Adelberger}}]{2006ApJ...644..792R}
{Reddy}, N.~A., {Steidel}, C.~C., {Fadda}, D., {et~al.} 2006, \apj, 644, 792,
  \dodoi{10.1086/503739}

\bibitem[{{Reddy} {et~al.}(2015){Reddy}, {Kriek}, {Shapley}, {Freeman},
  {Siana}, {Coil}, {Mobasher}, {Price}, {Sanders}, \&
  {Shivaei}}]{2015ApJ...806..259R}
{Reddy}, N.~A., {Kriek}, M., {Shapley}, A.~E., {et~al.} 2015, \apj, 806, 259,
  \dodoi{10.1088/0004-637X/806/2/259}

\bibitem[{{Rinaldi} {et~al.}(2022){Rinaldi}, {Caputi}, {van Mierlo}, {Ashby},
  {Caminha}, \& {Iani}}]{2022ApJ...930..128R}
{Rinaldi}, P., {Caputi}, K.~I., {van Mierlo}, S.~E., {et~al.} 2022, \apj, 930,
  128, \dodoi{10.3847/1538-4357/ac5d39}

\bibitem[{{Roberts-Borsani} {et~al.}(2016){Roberts-Borsani}, {Bouwens},
  {Oesch}, {Labbe}, {Smit}, {Illingworth}, {van Dokkum}, {Holden}, {Gonzalez},
  {Stefanon}, {Holwerda}, \& {Wilkins}}]{2016ApJ...823..143R}
{Roberts-Borsani}, G.~W., {Bouwens}, R.~J., {Oesch}, P.~A., {et~al.} 2016,
  \apj, 823, 143, \dodoi{10.3847/0004-637X/823/2/143}

\bibitem[{{Rodighiero} {et~al.}(2022){Rodighiero}, {Bisigello}, {Iani},
  {Marasco}, {Grazian}, {Sinigaglia}, {Cassata}, \&
  {Gruppioni}}]{2022arXiv220802825R}
{Rodighiero}, G., {Bisigello}, L., {Iani}, E., {et~al.} 2022, arXiv e-prints,
  arXiv:2208.02825.
\newblock \doarXiv{2208.02825}

\bibitem[{{Rodriguez-Gomez} {et~al.}(2019){Rodriguez-Gomez}, {Snyder}, {Lotz},
  {Nelson}, {Pillepich}, {Springel}, {Genel}, {Weinberger}, {Tacchella},
  {Pakmor}, {Torrey}, {Marinacci}, {Vogelsberger}, {Hernquist}, \&
  {Thilker}}]{2019MNRAS.483.4140R}
{Rodriguez-Gomez}, V., {Snyder}, G.~F., {Lotz}, J.~M., {et~al.} 2019, \mnras,
  483, 4140, \dodoi{10.1093/mnras/sty3345}

\bibitem[{{Rodr{\'\i}guez-Mu{\~n}oz} {et~al.}(2019){Rodr{\'\i}guez-Mu{\~n}oz},
  {Rodighiero}, {Mancini}, {P{\'e}rez-Gonz{\'a}lez}, {Rawle}, {Egami},
  {Mercurio}, {Rosati}, {Puglisi}, {Franceschini}, {Balestra}, {Baronchelli},
  {Biviano}, {Ebeling}, {Edge}, {Enia}, {Grillo}, {Haines}, {Iani}, {Jones},
  {Nonino}, {Valtchanov}, {Vulcani}, \& {Zemcov}}]{2019MNRAS.485..586R}
{Rodr{\'\i}guez-Mu{\~n}oz}, L., {Rodighiero}, G., {Mancini}, C., {et~al.} 2019,
  \mnras, 485, 586, \dodoi{10.1093/mnras/sty3335}

\bibitem[{{Sanders} \& {Mirabel}(1996)}]{1996ARA&A..34..749S}
{Sanders}, D.~B., \& {Mirabel}, I.~F. 1996, \araa, 34, 749,
  \dodoi{10.1146/annurev.astro.34.1.749}

\bibitem[{{Santini} {et~al.}(2017){Santini}, {Fontana}, {Castellano}, {Di
  Criscienzo}, {Merlin}, {Amorin}, {Cullen}, {Daddi}, {Dickinson}, {Dunlop},
  {Grazian}, {Lamastra}, {McLure}, {Micha{\l}owski}, {Pentericci}, \&
  {Shu}}]{2017ApJ...847...76S}
{Santini}, P., {Fontana}, A., {Castellano}, M., {et~al.} 2017, \apj, 847, 76,
  \dodoi{10.3847/1538-4357/aa8874}

\bibitem[{{Santini} {et~al.}(2022){Santini}, {Fontana}, {Castellano},
  {Leethochawalit}, {Trenti}, {Treu}, {Belfiori}, {Birrer}, {Bonchi}, {Merlin},
  {Mason}, {Morishita}, {Nonino}, {Paris}, {Polenta}, {Rosati}, {Yang},
  {Bradac}, {Calabr{\`o}}, {Dressler}, {Glazebrook}, {Marchesini}, {Mascia},
  {Nanayakkara}, {Pentericci}, {Roberts-Borsani}, {Scarlata}, {Vulcani}, \&
  {Wang}}]{2022arXiv220711379S}
---. 2022, arXiv e-prints, arXiv:2207.11379.
\newblock \doarXiv{2207.11379}

\bibitem[{{Schouws} {et~al.}(2022){Schouws}, {Stefanon}, {Bouwens}, {Smit},
  {Hodge}, {Labb{\'e}}, {Algera}, {Boogaard}, {Carniani}, {Fudamoto},
  {Holwerda}, {Illingworth}, {Maiolino}, {Maseda}, {Oesch}, \& {van der
  Werf}}]{2022ApJ...928...31S}
{Schouws}, S., {Stefanon}, M., {Bouwens}, R., {et~al.} 2022, \apj, 928, 31,
  \dodoi{10.3847/1538-4357/ac4605}

\bibitem[{{Schreiber} {et~al.}(2015){Schreiber}, {Pannella}, {Elbaz},
  {B{\'e}thermin}, {Inami}, {Dickinson}, {Magnelli}, {Wang}, {Aussel}, {Daddi},
  {Juneau}, {Shu}, {Sargent}, {Buat}, {Faber}, {Ferguson}, {Giavalisco},
  {Koekemoer}, {Magdis}, {Morrison}, {Papovich}, {Santini}, \&
  {Scott}}]{2015A&A...575A..74S}
{Schreiber}, C., {Pannella}, M., {Elbaz}, D., {et~al.} 2015, \aap, 575, A74,
  \dodoi{10.1051/0004-6361/201425017}

\bibitem[{{Scoville} {et~al.}(2016){Scoville}, {Sheth}, {Aussel}, {Vanden
  Bout}, {Capak}, {Bongiorno}, {Casey}, {Murchikova}, {Koda},
  {{\'A}lvarez-M{\'a}rquez}, {Lee}, {Laigle}, {McCracken}, {Ilbert}, {Pope},
  {Sanders}, {Chu}, {Toft}, {Ivison}, \& {Manohar}}]{2016ApJ...820...83S}
{Scoville}, N., {Sheth}, K., {Aussel}, H., {et~al.} 2016, \apj, 820, 83,
  \dodoi{10.3847/0004-637X/820/2/83}

\bibitem[{{Sersic}(1968)}]{1968adga.book.....S}
{Sersic}, J.~L. 1968, {Atlas de Galaxias Australes}

\bibitem[{{Simpson} {et~al.}(2014){Simpson}, {Swinbank}, {Smail}, {Alexander},
  {Brandt}, {Bertoldi}, {de Breuck}, {Chapman}, {Coppin}, {da Cunha},
  {Danielson}, {Dannerbauer}, {Greve}, {Hodge}, {Ivison}, {Karim}, {Knudsen},
  {Poggianti}, {Schinnerer}, {Thomson}, {Walter}, {Wardlow}, {Wei{\ss}}, \&
  {van der Werf}}]{2014ApJ...788..125S}
{Simpson}, J.~M., {Swinbank}, A.~M., {Smail}, I., {et~al.} 2014, \apj, 788,
  125, \dodoi{10.1088/0004-637X/788/2/125}

\bibitem[{{Skelton} {et~al.}(2014){Skelton}, {Whitaker}, {Momcheva}, {Brammer},
  {van Dokkum}, {Labb{\'e}}, {Franx}, {van der Wel}, {Bezanson}, {Da Cunha},
  {Fumagalli}, {F{\"o}rster Schreiber}, {Kriek}, {Leja}, {Lundgren}, {Magee},
  {Marchesini}, {Maseda}, {Nelson}, {Oesch}, {Pacifici}, {Patel}, {Price},
  {Rix}, {Tal}, {Wake}, \& {Wuyts}}]{2014ApJS..214...24S}
{Skelton}, R.~E., {Whitaker}, K.~E., {Momcheva}, I.~G., {et~al.} 2014, \apjs,
  214, 24, \dodoi{10.1088/0067-0049/214/2/24}

\bibitem[{{Smit} {et~al.}(2014){Smit}, {Bouwens}, {Labb{\'e}}, {Zheng},
  {Bradley}, {Donahue}, {Lemze}, {Moustakas}, {Umetsu}, {Zitrin}, {Coe},
  {Postman}, {Gonzalez}, {Bartelmann}, {Ben{\'\i}tez}, {Broadhurst}, {Ford},
  {Grillo}, {Infante}, {Jimenez-Teja}, {Jouvel}, {Kelson}, {Lahav}, {Maoz},
  {Medezinski}, {Melchior}, {Meneghetti}, {Merten}, {Molino}, {Moustakas},
  {Nonino}, {Rosati}, \& {Seitz}}]{2014ApJ...784...58S}
{Smit}, R., {Bouwens}, R.~J., {Labb{\'e}}, I., {et~al.} 2014, \apj, 784, 58,
  \dodoi{10.1088/0004-637X/784/1/58}

\bibitem[{{Smit} {et~al.}(2015){Smit}, {Bouwens}, {Franx}, {Oesch}, {Ashby},
  {Willner}, {Labb{\'e}}, {Holwerda}, {Fazio}, \&
  {Huang}}]{2015ApJ...801..122S}
{Smit}, R., {Bouwens}, R.~J., {Franx}, M., {et~al.} 2015, \apj, 801, 122,
  \dodoi{10.1088/0004-637X/801/2/122}

\bibitem[{{Sparre} {et~al.}(2015){Sparre}, {Hayward}, {Springel},
  {Vogelsberger}, {Genel}, {Torrey}, {Nelson}, {Sijacki}, \&
  {Hernquist}}]{2015MNRAS.447.3548S}
{Sparre}, M., {Hayward}, C.~C., {Springel}, V., {et~al.} 2015, \mnras, 447,
  3548, \dodoi{10.1093/mnras/stu2713}

\bibitem[{{Stefanon} {et~al.}(2015){Stefanon}, {Marchesini}, {Muzzin},
  {Brammer}, {Dunlop}, {Franx}, {Fynbo}, {Labb{\'e}}, {Milvang-Jensen}, \& {van
  Dokkum}}]{2015ApJ...803...11S}
{Stefanon}, M., {Marchesini}, D., {Muzzin}, A., {et~al.} 2015, \apj, 803, 11,
  \dodoi{10.1088/0004-637X/803/1/11}

\bibitem[{{Stefanon} {et~al.}(2017){Stefanon}, {Yan}, {Mobasher}, {Barro},
  {Donley}, {Fontana}, {Hemmati}, {Koekemoer}, {Lee}, {Lee}, {Nayyeri}, {Peth},
  {Pforr}, {Salvato}, {Wiklind}, {Wuyts}, {Ashby}, {Castellano}, {Conselice},
  {Cooper}, {Cooray}, {Dolch}, {Ferguson}, {Galametz}, {Giavalisco}, {Guo},
  {Willner}, {Dickinson}, {Faber}, {Fazio}, {Gardner}, {Gawiser}, {Grazian},
  {Grogin}, {Kocevski}, {Koo}, {Lee}, {Lucas}, {McGrath}, {Nandra}, {Newman},
  \& {van der Wel}}]{2017ApJS..229...32S}
{Stefanon}, M., {Yan}, H., {Mobasher}, B., {et~al.} 2017, \apjs, 229, 32,
  \dodoi{10.3847/1538-4365/aa66cb}

\bibitem[{{Suess} {et~al.}(2019){Suess}, {Kriek}, {Price}, \&
  {Barro}}]{2019ApJ...885L..22S}
{Suess}, K.~A., {Kriek}, M., {Price}, S.~H., \& {Barro}, G. 2019, \apjl, 885,
  L22, \dodoi{10.3847/2041-8213/ab4db3}

\bibitem[{{Sun} {et~al.}(2021){Sun}, {Egami}, {P{\'e}rez-Gonz{\'a}lez},
  {Smail}, {Caputi}, {Bauer}, {Rawle}, {Fujimoto}, {Kohno},
  {Dudzevi{\v{c}}i{\={u}}t{\.{e}}}, {Atek}, {Bianconi}, {Chapman}, {Combes},
  {Jauzac}, {Jolly}, {Koekemoer}, {Magdis}, {Rodighiero}, {Rujopakarn},
  {Schaerer}, {Steinhardt}, {Van der Werf}, {Walth}, \&
  {Weaver}}]{2021ApJ...922..114S}
{Sun}, F., {Egami}, E., {P{\'e}rez-Gonz{\'a}lez}, P.~G., {et~al.} 2021, \apj,
  922, 114, \dodoi{10.3847/1538-4357/ac2578}

\bibitem[{{Swinbank} {et~al.}(2012){Swinbank}, {Karim}, {Smail}, {Hodge},
  {Walter}, {Bertoldi}, {Biggs}, {de Breuck}, {Chapman}, {Coppin}, {Cox},
  {Danielson}, {Dannerbauer}, {Ivison}, {Greve}, {Knudsen}, {Menten},
  {Simpson}, {Schinnerer}, {Wardlow}, {Wei{\ss}}, \& {van der
  Werf}}]{2012MNRAS.427.1066S}
{Swinbank}, A.~M., {Karim}, A., {Smail}, I., {et~al.} 2012, \mnras, 427, 1066,
  \dodoi{10.1111/j.1365-2966.2012.22048.x}

\bibitem[{{Tacchella} {et~al.}(2018){Tacchella}, {Carollo}, {F{\"o}rster
  Schreiber}, {Renzini}, {Dekel}, {Genzel}, {Lang}, {Lilly}, {Mancini},
  {Onodera}, {Tacconi}, {Wuyts}, \& {Zamorani}}]{2018ApJ...859...56T}
{Tacchella}, S., {Carollo}, C.~M., {F{\"o}rster Schreiber}, N.~M., {et~al.}
  2018, \apj, 859, 56, \dodoi{10.3847/1538-4357/aabf8b}

\bibitem[{{Tacchella} {et~al.}(2022{\natexlab{a}}){Tacchella}, {Finkelstein},
  {Bagley}, {Dickinson}, {Ferguson}, {Giavalisco}, {Graziani}, {Grogin},
  {Hathi}, {Hutchison}, {Jung}, {Koekemoer}, {Larson}, {Papovich}, {Pirzkal},
  {Rojas-Ruiz}, {Song}, {Schneider}, {Somerville}, {Wilkins}, \&
  {Yung}}]{2022ApJ...927..170T}
{Tacchella}, S., {Finkelstein}, S.~L., {Bagley}, M., {et~al.}
  2022{\natexlab{a}}, \apj, 927, 170, \dodoi{10.3847/1538-4357/ac4cad}

\bibitem[{{Tacchella} {et~al.}(2022{\natexlab{b}}){Tacchella}, {Conroy},
  {Faber}, {Johnson}, {Leja}, {Barro}, {Cunningham}, {Deason}, {Guhathakurta},
  {Guo}, {Hernquist}, {Koo}, {McKinnon}, {Rockosi}, {Speagle}, {van Dokkum}, \&
  {Yesuf}}]{2022ApJ...926..134T}
{Tacchella}, S., {Conroy}, C., {Faber}, S.~M., {et~al.} 2022{\natexlab{b}},
  \apj, 926, 134, \dodoi{10.3847/1538-4357/ac449b}

\bibitem[{{Tadaki} {et~al.}(2020){Tadaki}, {Belli}, {Burkert}, {Dekel},
  {F{\"o}rster Schreiber}, {Genzel}, {Hayashi}, {Herrera-Camus}, {Kodama},
  {Kohno}, {Koyama}, {Lee}, {Lutz}, {Mowla}, {Nelson}, {Renzini}, {Suzuki},
  {Tacconi}, {{\"U}bler}, {Wisnioski}, \& {Wuyts}}]{2020ApJ...901...74T}
{Tadaki}, K.-i., {Belli}, S., {Burkert}, A., {et~al.} 2020, \apj, 901, 74,
  \dodoi{10.3847/1538-4357/abaf4a}

\bibitem[{{Talia} {et~al.}(2021){Talia}, {Cimatti}, {Giulietti}, {Zamorani},
  {Bethermin}, {Faisst}, {Le F{\`e}vre}, \&
  {Smol{\c{c}}i{\'c}}}]{2021ApJ...909...23T}
{Talia}, M., {Cimatti}, A., {Giulietti}, M., {et~al.} 2021, \apj, 909, 23,
  \dodoi{10.3847/1538-4357/abd6e3}

\bibitem[{{Tortora} {et~al.}(2020){Tortora}, {Napolitano}, {Radovich},
  {Spiniello}, {Hunt}, {Roy}, {Moscardini}, {Scognamiglio}, {Spavone},
  {Brescia}, {Cavuoti}, {D`Ago}, {Longo}, {Bellagamba}, {Maturi}, \&
  {Roncarelli}}]{2020A&A...638L..11T}
{Tortora}, C., {Napolitano}, N.~R., {Radovich}, M., {et~al.} 2020, \aap, 638,
  L11, \dodoi{10.1051/0004-6361/202038373}

\bibitem[{{van der Walt} {et~al.}(2011){van der Walt}, {Colbert}, \&
  {Varoquaux}}]{2011CSE....13b..22V}
{van der Walt}, S., {Colbert}, S.~C., \& {Varoquaux}, G. 2011, Computing in
  Science and Engineering, 13, 22, \dodoi{10.1109/MCSE.2011.37}

\bibitem[{{van der Wel} {et~al.}(2014){van der Wel}, {Franx}, {van Dokkum},
  {Skelton}, {Momcheva}, {Whitaker}, {Brammer}, {Bell}, {Rix}, {Wuyts},
  {Ferguson}, {Holden}, {Barro}, {Koekemoer}, {Chang}, {McGrath},
  {H{\"a}ussler}, {Dekel}, {Behroozi}, {Fumagalli}, {Leja}, {Lundgren},
  {Maseda}, {Nelson}, {Wake}, {Patel}, {Labb{\'e}}, {Faber}, {Grogin}, \&
  {Kocevski}}]{2014ApJ...788...28V}
{van der Wel}, A., {Franx}, M., {van Dokkum}, P.~G., {et~al.} 2014, \apj, 788,
  28, \dodoi{10.1088/0004-637X/788/1/28}

\bibitem[{{Virtanen} {et~al.}(2020){Virtanen}, {Gommers}, {Oliphant},
  {Haberland}, {Reddy}, {Cournapeau}, {Burovski}, {Peterson}, {Weckesser},
  {Bright}, {van der Walt}, {Brett}, {Wilson}, {Millman}, {Mayorov}, {Nelson},
  {Jones}, {Kern}, {Larson}, {Carey}, {Polat}, {Feng}, {Moore}, {VanderPlas},
  {Laxalde}, {Perktold}, {Cimrman}, {Henriksen}, {Quintero}, {Harris},
  {Archibald}, {Ribeiro}, {Pedregosa}, {van Mulbregt}, \& {SciPy 1. 0
  Contributors}}]{2020NatMe..17..261V}
{Virtanen}, P., {Gommers}, R., {Oliphant}, T.~E., {et~al.} 2020, Nature
  Methods, 17, 261, \dodoi{10.1038/s41592-019-0686-2}

\bibitem[{{Wang} {et~al.}(2021){Wang}, {Gao}, {Best}, {Duncan}, {Hardcastle},
  {Kondapally}, {Ma{\l}ek}, {McCheyne}, {Sabater}, {Shimwell}, {Tasse},
  {Bonato}, {Bondi}, {Cochrane}, {Farrah}, {G{\"u}rkan}, {Haskell}, {Pearson},
  {Prandoni}, {R{\"o}ttgering}, {Smith}, {Vaccari}, \&
  {Williams}}]{2021A&A...648A...8W}
{Wang}, L., {Gao}, F., {Best}, P.~N., {et~al.} 2021, \aap, 648, A8,
  \dodoi{10.1051/0004-6361/202038811}

\bibitem[{{Wang} {et~al.}(2016){Wang}, {Elbaz}, {Schreiber}, {Pannella}, {Shu},
  {Willner}, {Ashby}, {Huang}, {Fontana}, {Dekel}, {Daddi}, {Ferguson},
  {Dunlop}, {Ciesla}, {Koekemoer}, {Giavalisco}, {Boutsia}, {Finkelstein},
  {Juneau}, {Barro}, {Koo}, {Micha{\l}owski}, {Orellana}, {Lu}, {Castellano},
  {Bourne}, {Buitrago}, {Santini}, {Faber}, {Hathi}, {Lucas}, \&
  {P{\'e}rez-Gonz{\'a}lez}}]{2016ApJ...816...84W}
{Wang}, T., {Elbaz}, D., {Schreiber}, C., {et~al.} 2016, \apj, 816, 84,
  \dodoi{10.3847/0004-637X/816/2/84}

\bibitem[{{Wang} {et~al.}(2019){Wang}, {Schreiber}, {Elbaz}, {Yoshimura},
  {Kohno}, {Shu}, {Yamaguchi}, {Pannella}, {Franco}, {Huang}, {Lim}, \&
  {Wang}}]{2019Natur.572..211W}
{Wang}, T., {Schreiber}, C., {Elbaz}, D., {et~al.} 2019, \nat, 572, 211,
  \dodoi{10.1038/s41586-019-1452-4}

\bibitem[{{Wang} {et~al.}(2017){Wang}, {Faber}, {Liu}, {Guo}, {Pacifici},
  {Koo}, {Kassin}, {Mao}, {Fang}, {Chen}, {Koekemoer}, {Kocevski}, \&
  {Ashby}}]{2017MNRAS.469.4063W}
{Wang}, W., {Faber}, S.~M., {Liu}, F.~S., {et~al.} 2017, \mnras, 469, 4063,
  \dodoi{10.1093/mnras/stx1148}

\bibitem[{{Wang} {et~al.}(2018){Wang}, {Kassin}, {Pacifici}, {Barro}, {de la
  Vega}, {Simons}, {Faber}, {Salmon}, {Ferguson}, {P{\'e}rez-Gonz{\'a}lez},
  {Snyder}, {Gordon}, {Chen}, \& {Kodra}}]{2018ApJ...869..161W}
{Wang}, W., {Kassin}, S.~A., {Pacifici}, C., {et~al.} 2018, \apj, 869, 161,
  \dodoi{10.3847/1538-4357/aaef79}

\bibitem[{{Whitaker} {et~al.}(2010){Whitaker}, {van Dokkum}, {Brammer},
  {Kriek}, {Franx}, {Labb{\'e}}, {Marchesini}, {Quadri}, {Bezanson},
  {Illingworth}, {Lee}, {Muzzin}, {Rudnick}, \& {Wake}}]{2010ApJ...719.1715W}
{Whitaker}, K.~E., {van Dokkum}, P.~G., {Brammer}, G., {et~al.} 2010, \apj,
  719, 1715, \dodoi{10.1088/0004-637X/719/2/1715}

\bibitem[{{Williams} {et~al.}(2019){Williams}, {Labbe}, {Spilker}, {Stefanon},
  {Leja}, {Whitaker}, {Bezanson}, {Narayanan}, {Oesch}, \&
  {Weiner}}]{2019ApJ...884..154W}
{Williams}, C.~C., {Labbe}, I., {Spilker}, J., {et~al.} 2019, \apj, 884, 154,
  \dodoi{10.3847/1538-4357/ab44aa}

\bibitem[{{Williams} {et~al.}(2009){Williams}, {Quadri}, {Franx}, {van Dokkum},
  \& {Labb{\'e}}}]{2009ApJ...691.1879W}
{Williams}, R.~J., {Quadri}, R.~F., {Franx}, M., {van Dokkum}, P., \&
  {Labb{\'e}}, I. 2009, \apj, 691, 1879, \dodoi{10.1088/0004-637X/691/2/1879}

\bibitem[{{Woodrum} {et~al.}(2022){Woodrum}, {Williams}, {Rieke}, {Leja},
  {Johnson}, {Bezanson}, {Kennicutt}, {Spilker}, \&
  {Tacchella}}]{2022arXiv221003832W}
{Woodrum}, C., {Williams}, C.~C., {Rieke}, M., {et~al.} 2022, arXiv e-prints,
  arXiv:2210.03832.
\newblock \doarXiv{2210.03832}

\bibitem[{{Wuyts} {et~al.}(2012){Wuyts}, {F{\"o}rster Schreiber}, {Genzel},
  {Guo}, {Barro}, {Bell}, {Dekel}, {Faber}, {Ferguson}, {Giavalisco}, {Grogin},
  {Hathi}, {Huang}, {Kocevski}, {Koekemoer}, {Koo}, {Lotz}, {Lutz}, {McGrath},
  {Newman}, {Rosario}, {Saintonge}, {Tacconi}, {Weiner}, \& {van der
  Wel}}]{2012ApJ...753..114W}
{Wuyts}, S., {F{\"o}rster Schreiber}, N.~M., {Genzel}, R., {et~al.} 2012, \apj,
  753, 114, \dodoi{10.1088/0004-637X/753/2/114}

\bibitem[{{Xiao} {et~al.}(2022){Xiao}, {Elbaz}, {G{\'o}mez-Guijarro}, {Leroy},
  {Bing}, {Daddi}, {Magnelli}, {Franco}, {Zhou}, {Dickinson}, {Wang},
  {Rujopakarn}, {Magdis}, {Treister}, {Inami}, {Demarco}, {Sargent}, {Shu},
  {Kartaltepe}, {Alexander}, {B{\'e}thermin}, {Bournaud}, {Chary}, {Ciesla},
  {Ferguson}, {Finkelstein}, {Giavalisco}, {Gu}, {Iono}, {Juneau}, {Lagache},
  {Leiton}, {Messias}, {Motohara}, {Mullaney}, {Nagar}, {Pannella}, {Papovich},
  {Pope}, {Schreiber}, \& {Silverman}}]{2022arXiv221003135X}
{Xiao}, M., {Elbaz}, D., {G{\'o}mez-Guijarro}, C., {et~al.} 2022, arXiv
  e-prints, arXiv:2210.03135.
\newblock \doarXiv{2210.03135}

\bibitem[{{Yamaguchi} {et~al.}(2019){Yamaguchi}, {Kohno}, {Hatsukade}, {Wang},
  {Yoshimura}, {Ao}, {Caputi}, {Dunlop}, {Egami}, {Espada}, {Fujimoto},
  {Hayatsu}, {Ivison}, {Kodama}, {Kusakabe}, {Nagao}, {Ouchi}, {Rujopakarn},
  {Tadaki}, {Tamura}, {Ueda}, {Umehata}, {Wang}, \&
  {Yun}}]{2019ApJ...878...73Y}
{Yamaguchi}, Y., {Kohno}, K., {Hatsukade}, B., {et~al.} 2019, \apj, 878, 73,
  \dodoi{10.3847/1538-4357/ab0d22}

\bibitem[{{Zavala} {et~al.}(2017){Zavala}, {Aretxaga}, {Geach}, {Hughes},
  {Birkinshaw}, {Chapin}, {Chapman}, {Chen}, {Clements}, {Dunlop}, {Farrah},
  {Ivison}, {Jenness}, {Micha{\l}owski}, {Robson}, {Scott}, {Simpson},
  {Spaans}, \& {van der Werf}}]{2017MNRAS.464.3369Z}
{Zavala}, J.~A., {Aretxaga}, I., {Geach}, J.~E., {et~al.} 2017, \mnras, 464,
  3369, \dodoi{10.1093/mnras/stw2630}

\bibitem[{{Zavala} {et~al.}(2018){Zavala}, {Aretxaga}, {Dunlop},
  {Micha{\l}owski}, {Hughes}, {Bourne}, {Chapin}, {Cowley}, {Farrah}, {Lacey},
  {Targett}, \& {van der Werf}}]{2018MNRAS.475.5585Z}
{Zavala}, J.~A., {Aretxaga}, I., {Dunlop}, J.~S., {et~al.} 2018, \mnras, 475,
  5585, \dodoi{10.1093/mnras/sty217}

\bibitem[{{Zavala} {et~al.}(2022){Zavala}, {Buat}, {Casey}, {Burgarella},
  {Finkelstein}, {Bagley}, {Ciesla}, {Daddi}, {Dickinson}, {Ferguson},
  {Franco}, {Jim'enez-Andrade}, {Kartaltepe}, {Koekemoer}, {Le Bail}, {Murphy},
  {Papovich}, {Tacchella}, {Wilkins}, {Fontana}, {Giavalisco}, {Grazian},
  {Grogin}, {Kewley}, {Kocevski}, {Kirkpatrick}, {Lotz}, {Pentericci},
  {Perez-Gonzalez}, {Pirzkal}, {Ravindranath}, {Somerville}, {Trump}, {Yang},
  {Yung}, {Almaini}, {Amorin}, {Annunziatella}, {Arrabal Haro}, {Backhaus},
  {Barro}, {Behroozi}, {Bell}, {Bhatawdekar}, {Bisigello}, {Buitrago},
  {Calabro}, {Castellano}, {Chavez Ortiz}, {Chworowsky}, {Cleri}, {Cohen},
  {Cole}, {Cooke}, {Cooper}, {Cooray}, {Costantin}, {Cox}, {Croton}, {Dave},
  {de la Vega}, {Dekel}, {Elbaz}, {Estrada-Carpenter}, {Fern{\'a}ndez},
  {Finkelstein}, {Freundlich}, {Fujimoto}, {Garc{\'\i}a-Argum{\'a}nez},
  {Gardner}, {Gawiser}, {G{\'o}mez-Guijarro}, {Guo}, {Hamilton}, {Hathi},
  {Holwerda}, {Hirschmann}, {Huertas-Company}, {Hutchison}, {Iyer}, {Jaskot},
  {Jha}, {Jogee}, {Juneau}, {Jung}, {Kassin}, {Kurczynski}, {Larson}, {Leung},
  {Lucas}, {Magnelli}, {Mantha}, {Matharu}, {McGrath}, {McIntosh}, {Medrano},
  {Merlin}, {Mobasher}, {Morales}, {Newman}, {Nicholls}, {Pandya}, {Rafelski},
  {Ronayne}, {Rose}, {Ryan}, {Santini}, {Seill{\'e}}, {Shah}, {Shen}, {Simons},
  {Snyder}, {Stanway}, {Straughn}, {Teplitz}, {Vanderhoof}, {Vega-Ferrero},
  {Wang}, {Weiner}, {Willmer}, \& {Wuyts}}]{2022arXiv220801816Z}
{Zavala}, J.~A., {Buat}, V., {Casey}, C.~M., {et~al.} 2022, arXiv e-prints,
  arXiv:2208.01816.
\newblock \doarXiv{2208.01816}

\bibitem[{{Zhou} {et~al.}(2020){Zhou}, {Elbaz}, {Franco}, {Magnelli},
  {Schreiber}, {Wang}, {Ciesla}, {Daddi}, {Dickinson}, {Nagar}, {Magdis},
  {Alexander}, {B{\'e}thermin}, {Demarco}, {Mullaney}, {Bournaud}, {Ferguson},
  {Finkelstein}, {Giavalisco}, {Inami}, {Iono}, {Juneau}, {Lagache}, {Messias},
  {Motohara}, {Okumura}, {Pannella}, {Papovich}, {Pope}, {Rujopakarn}, {Shi},
  {Shu}, \& {Silverman}}]{2020A&A...642A.155Z}
{Zhou}, L., {Elbaz}, D., {Franco}, M., {et~al.} 2020, \aap, 642, A155,
  \dodoi{10.1051/0004-6361/202038059}

\bibitem[{{Zuckerman} {et~al.}(2021){Zuckerman}, {Belli}, {Leja}, \&
  {Tacchella}}]{2021ApJ...922L..32Z}
{Zuckerman}, L.~D., {Belli}, S., {Leja}, J., \& {Tacchella}, S. 2021, \apjl,
  922, L32, \dodoi{10.3847/2041-8213/ac3831}

\end{thebibliography}
\bibliographystyle{aasjournal}

\end{document}